\RequirePackage{lineno} 

\documentclass[twocolumn,showpacs,preprintnumbers,amsmath,amssymb,nofootinbib]{revtex4}

\usepackage{graphicx}
\usepackage{dcolumn}
\usepackage{bm}
 
\usepackage{epstopdf}

\def\bea{\begin{eqnarray}}
\def\eea{\end{eqnarray}}

\def\pp{\mbox{$p$-$p$}}
\def\pa{\mbox{$p$-A}}
\def\da{\mbox{$d$-A}}
\def\auau{\mbox{Au-Au}}

\def\aa{\mbox{A-A}}

\def\ee{\mbox{$e^+$-$e^-$}}
\def\ppbar{\mbox{$p$-$\bar p$}}

\def\pt{$p_t$}
\def\yt{$y_t$}
\def\nch{$n_{ch}$}

\begin{document} 

\setpagewiselinenumbers
\modulolinenumbers[5]

\preprint{Version 2.4}

\title{Predicting minimum-bias trigger-associated dijet correlations in p-p collisions
}

\author{Thomas A.\ Trainor}\affiliation{CENPA 354290, University of Washington, Seattle, Washington 98195}


\date{\today}

\begin{abstract}
A method is derived to predict the structure of dijet-related hard components of trigger-associated (TA) hadron correlations from \pp\ collisions based on measured fragmentation functions (FFs) from \ee\ or \ppbar\ collisions and a minimum-bias (MB) jet or scattered-parton spectrum from \ppbar\ collisions. The method is based on the probability chain rule and Bayes' theorem and relates a trigger-parton--associated-fragment system from reconstructed dijets to a trigger-hadron--associated hadron system from \pp\ data. The method is tested in this study by comparisons of FF TA structure with preliminary \pp\ TA data but can also be applied to \pa, \da\ and \aa\ collisions over a range of energies. Quantitative comparisons of measured TA correlations with FF-derived predictions may confirm a QCD MB dijet mechanism for certain spectrum and correlation structures whose origins are currently questioned and have been attributed by some to collective expansion (flows).
\end{abstract}


\maketitle

 \section{Introduction}
 
The role of QCD jets in high-energy nuclear collisions is currently strongly debated. In a high-energy physics (HEP) context dijet production has been long accepted as an important mechanism for hadron formation by parton fragmentation~\cite{hepjets,rick1,rick2}. Projectile-nucleon fragmentation (dissociation) is the other principal hadron production mechanism in elementary \pp\ collisions~\cite{atlas}. But in a heavy-ion context ``freezeout'' from a flowing bulk medium is assumed to be the nearly-exclusive hadron formation mechanism~\cite{bec}. Recent experimental results are interpreted to suggest that bulk-medium collectivity may even play a role in small \pp, $p$-A and $d$-A  systems~\cite{bron}. In effect, previously-accepted contributions from minimum-bias (MB) dijets to high-energy nuclear collisions are displaced by the claimed presence of strong collective motion (flows) in a thermalized bulk medium or quark-gluon plasma to explain spectrum and correlation structure, in small as well as large collision systems.
 
Claims of a flowing bulk medium, quark-gluon plasma (QGP) or ``perfect liquid''~\cite{perfliq1,perfliq2} are based in part on the {\it a priori} assumption that all hadrons with transverse momentum $p_t < 2$ GeV/c emerge from a thermalized bulk medium~\cite{2gev}, an interval that includes almost all final-state hadrons and more than 90\% of  minimum-bias (MB) jet fragments, consistent with direct jet measurements and QCD predictions~\cite{eeprd,hardspec,fragevo,jetspec}. Differential analysis of spectra and correlations in the same systems appears to confirm a dominant role for dijet production~\cite{ppprd,hardspec,fragevo,porter2,porter3,axialci,anomalous}. Resolution of the apparent contradiction requires establishment of a more comprehensive description of dijet manifestations in elementary \pp\ collisions that applies to all jet phenomenology (yields, spectra and correlations) over the complete hadron fragment momentum space down to the kinematic energy and momentum limits {\em where most jet fragments should appear}. 


In previous studies a jet-related {\em hard  component} was  isolated from the \pt\ spectrum of 200 GeV \pp\ collisions by means of its charge multiplicity \nch\ dependence, leading to a two-component (soft+hard) spectrum model (TCM)~\cite{ppprd}. A similar differential analysis of \auau\ spectra revealed that the spectrum hard component persists for all centralities, although the form changes quantitatively in more-central collisions~\cite{hardspec}. The spectrum hard components for \pp\ and \auau\ have been described quantitatively by a pQCD calculation~\cite{fragevo} based on a MB parton spectrum~\cite{jetspec2} and measured parton fragmentation functions (FFs)~\cite{eeprd}. Those results strongly suggest that 
(a) the spectrum and angular-correlation hard components in \pp\ collisions and \auau\ collisions for all centralities are jet related~\cite{hardspec,fragevo},  
(b) one third of  final-state hadrons in 200 GeV central \auau\ collisions are included within intact MB jets (minijets)~\cite{anomalous,jetspec}
and
(c) dijet production in \pp\ collisions increases as $n_{ch}^2$, a trend inconsistent with the eikonal approximation~\cite{ppprd,pptheory}.

The present study extends that program by introducing a method to predict two-dimensional (2D) trigger-associated (TA) hadron correlations arising from dijets produced in high-energy \pp\ collisions based on measured FFs and a large-angle-scattered parton spectrum. Comparisons with measured TA correlations may establish the kinematic lower limits on dijet energy and fragment momentum. Identification of unique dijet contributions to TA correlations in \pp\ collisions may then test claims of novel nonjet physics (bulk collectivity) in \mbox{$p$-A}, $d$-A and  heavy ion (\aa) collisions at the Relativistic Heavy-Ion Collider (RHIC) and the Large Hadron Collider (LHC).


The 1D \pt\ spectrum TCM describes a marginal projection of 2D TA correlations formed by pairing the highest-momentum hadron in each event (trigger particle) with any other hadron (associated particle)~\cite{pptrig}. Some fraction of those pairs (the TA hard  component, possibly jet related) can be isolated by subtracting a TA TCM soft-component model~\cite{pptrig}. The purpose of the present study is prediction of the 2D TA hard component obtained from \pp\ data using measured FFs and a MB jet spectrum to confirm a dijet mechanism for the hard component. The method consists in relating the QCD parton-fragment system to the hadronic trigger-associated system via the probability chain rule and Bayes' theorem.

This article is arranged as follows:
Section~\ref{anal} presents analysis methods.
Section~\ref{2comp} describes a two-component model for yields and momentum/rapidity spectra from 200 GeV \pp\ collisions.
Section~\ref{tatcm}  presents a TCM for \pp\ TA correlations.
Section~\ref{partition} describes a partition of parton fragmentation functions into trigger and associated components.
Section~\ref{fftrig} derives a QCD trigger spectrum from FFs.
Section~\ref{fftcm} presents a QCD prediction for 2D TA correlation hard components.
Section~\ref{datacomp} compares QCD TA predictions derived from FFs to preliminary TA data from 200 GeV \pp\ collisions. 
Section~\ref{syserr} discusses systematic uncertainties.
Sections~\ref{disc} and~\ref{summ} present discussion and summary.

 \section{Analysis methods} \label{anal}
 
A TCM for single-particle spectra and symmetrized two-particle correlations was applied previously to \pp\ and \auau\ collisions at the RHIC. The methods and results are described in Refs.~\cite{ppprd,hardspec,porter1,porter2,porter3,inverse,axialci,axialcd,anomalous,davidhq,davidhq2}.
In the present study we extend those techniques to relate scattered-parton FFs and a MB jet spectrum to hadron TA correlations from \pp\ collisions. We introduce the concept of conditional (asymmetric) TA correlations, based on a trigger hadron as the highest-momentum hadron in a dijet or \pp\ collision event, as the principal focus of study. We combine measured jet (scattered-parton) spectra from \ppbar\ collisions with parton fragmentation functions from \ppbar\ and \ee\ collisions to generate QCD predictions for TA correlations. To establish a quantitative relation we start with the concept of  a joint three-particle distribution on the parent parton energy and its trigger and associated fragment momenta. From projections onto various subspaces we derive a method to predict \pp\ TA correlations from jet spectra and FFs. 

\subsection{Kinematic variables and spaces} \label{kine}

High-energy nuclear collisions are described efficiently near mid-rapidity by a cylindrical coordinate system $(p_t,\eta,\phi)$, where $p_t$ is transverse momentum, $\phi$ is the azimuth angle from some reference direction and pseudorapidity $\eta = - \ln [\tan(\theta/2)] \approx \cos(\theta)$ is a measure of the polar angle, the approximation valid near $\eta = 0$. A finite detector angular acceptance is denoted by intervals ($\Delta \eta,\Delta \phi$) in the primary single-particle space $(\eta,\phi)$.

Although scalar momenta are directly measured by particle detectors, in this study we prefer to use alternative rapidity measures. To provide better visual access to low-momentum structure and to simplify the description of jet-related spectrum hard components and fragmentation functions (FFs) (both defined below) we present single-particle (SP) spectra and FFs in terms of three rapidity variables. Transverse rapidity $y_t = \ln[(m_t + p_t)/ m_h]$ with transverse mass $m_t = \sqrt{p_t^2 + m_h^2}$ and $m_h = m_\pi$ assumed for unidentified hadrons visualizes spectrum structure equally well at small or large \yt.  Total rapidity  $y = \ln[(E + p)/ m_h]$ or longitudinal rapidity $y_z = \ln[(E + p_z)/ m_t]$ can be used to describe FFs. Jet energies are then represented by $y_{max} \equiv \ln[2E_{jet}/ m_h]$. Trigger and associated hadrons may be described in the \pp\ collision system in terms of \yt\ or in the FF system in terms of $y$, in either case with corresponding subscripts: $y_{assoc}$ or $y_{trig}$ for FFs and $y_{ta}$ or $y_{tt}$ for \pp\ collisions.

\subsection{Probability chain rule}

This study relies on manipulation of compound probabilities via the chain rule. For a joint  distribution on two variables the chain rule is  $P(A\cap B) = P(B|A)P(A) = P(A|B)P(B)$, where for example $P(B|A)$ denotes the conditional probability of $B$ given $A$. Bayes' theorem $P(B|A) = P(A|B)P(B) / P(A)$ is a simple consequence. If $B$ is independent of $A$ the chain rule leads to factorization of $P(A \cap B)$. For arbitrary joint distribution $f(x,y)$ with marginal projections (integrals over $y$ or $x$) $f(x)$  or $f(y)$ the chain rule is $f(x,y) = \hat f(x|y) f(y)$ defining the unit-normal (on $x$) conditional distribution $\hat f(x|y)$, a caret signifying a unit-normal distribution. If $f(y)$ is also unit normal one has the equivalent of a joint probability system as described above. 
Generally, a joint distribution may be factorized as $f(x,y) = g(x|y) h(y)$, and $g(x|y)$ may not be unit normal on $x$. 

\subsection{Joint, conditional and marginal distributions} \label{math}

In this study we consider relationships among joint and conditional two-particle distributions on rapidity variables for several particle types denoted by $p$ (partons), $u$ (unidentified hadrons), $t$ or $trig$ (trigger hadrons) and $a$ or $assoc$ (associated [with a trigger] hadrons). Symbol $u$ for unidentified hadrons avoids confusion with $h$ representing distribution hard components (defined below).
Joint distributions are represented by $F_{\alpha \beta}(x,y)$ with subscripts $\alpha,\,\beta$ indicating two particle types. $D_\alpha(x|y)$ represents a distribution on $x$ conditional on $y$, $F_\alpha(x)$ is a marginal 1D projection onto $x$, $S_\alpha(y)$ denotes a particle spectrum and $\hat S_\alpha(y)$ denotes a unit-normal spectrum.

An ensemble of FFs denoted by $D_u(y|y_{max})$ and conditional on parton rapidity $y_{max}$ can be combined with a unit-normal parton spectrum $\hat S_p(y_{max})$ to form joint distribution $F_{up}(y,y_{max}) = D_u(y|y_{max}) \hat S_p(y_{max})$. Its marginal projections are the mean dijet fragment distribution $D_u(y)$ and the spectrum-weighted dijet multiplicity distribution $2n_{ch,j}(y_{max}) \hat S_p(y_{max})$. Integration of either marginal projection gives ensemble-mean dijet fragment number $2\bar n_{ch,j}$ (within $4\pi$ angular acceptance). 
Other joint distributions include trigger-fragment--parton distribution $F_{tp}$, whose marginals are trigger-fragment and parton spectra, associated-fragment--parton distribution $F_{ap}$ and associated-trigger two-fragment distribution $F_{at}$.  A principal goal of this study is establishment of a quantitative relation between measured $F_{at}(y_{ta},y_{tt})$ obtained from 200 GeV \pp\ collisions and $F_{up}(y,y_{max})$ derived from measured jet spectra from \pp\ and \ppbar\ collisions and FFs from \ee\ and \ppbar\ collisions.

 \subsection{Conditional trigger-associated correlations}

Symmetrized two-particle correlations on transverse mass $m_t \times m_t$~\cite{mtxmt} and on transverse rapidity $y_t \times y_t$~\cite{porter2,ytxyt} or angle differences  $(\eta_\Delta,\phi_\Delta)$~\cite{axialci,axialcd,porter2,porter3,anomalous} have been studied extensively and are well-described by a TCM consistent with the SP spectrum TCM and with expectations for jet-related  angular correlations. Conditional (asymmetric) TA correlations on $(y_{t,assoc},y_{t,trig})$ or $(y_{ta},y_{tt})$ (with $y_{ta} < y_{tt}$) are related to, but not equivalent to, symmetrized correlations on $y_t \times y_t$. Asymmetric TA correlations retain additional correlation information, and the TA hard component is directly comparable to measured FFs and pQCD parton (jet) spectrum predictions as described in this study.

One-dimensional trigger spectra and 2D TA correlations are formed as follows:  \pp\ events in a given $n_{ch}$ class are sorted into {\em trigger} classes ($y_{tt}$ bins) based on the highest $y_t$ in each event (trigger particle). The trigger hadron from each event is assumed (with some probability) to be the proxy for a scattered parton. The distribution of trigger particles (events) on $y_{tt}$ is the trigger spectrum.
Spectra for $n_{ch}-1$ {\em associated} hadrons (some may be jet fragments) distributed on $y_{ta}$ are accumulated for each event in trigger class $y_{tt}$ with trigger particles excluded (no self pairs). The resulting 2D TA distribution $F_{at}(y_{ta},y_{tt},n_{ch})$ can be factored according to the chain rule into a unit-normal trigger spectrum $\hat T(y_{tt},n_{ch})$ (comparable to a pQCD parton spectrum) and a 2D ensemble of conditional distributions $A(y_{ta}|y_{tt},n_{ch})$: associated-hadron spectra conditional on specific trigger $y_{tt}$ values (comparable to an ensemble of fragmentation functions conditional on parton rapidity $y_{max}$). All nontrivial combinatoric pairs from all events in a given $n_{ch}$ class are retained---no particles or pairs are excluded by $p_t$ cuts. 

\subsection{Modeling the TA hard component with FFs}

A 2D TCM for \pp\ TA correlations $F_{at}$ can be constructed from the 1D SP spectrum TCM~\cite{pptrig}.
From the TA TCM we obtain 1D and 2D soft-component models that can be subtracted from measured distributions $\hat T$ and $A$ to obtain jet-related trigger hard component $\hat T_{hh}$ and TA conditional hard component $R_h A_{hh}(y_{ta}|y_{tt},n_{ch})$. To provide a basis for direct comparison of measured FFs with TA hard components inferred from \pp\ collisions we decompose FFs into trigger and associated components by means of void probabilities (defined below). The trigger and associated FF components $\hat S_t(y_{trig}|y_{max})$ and $D_a(y_{assoc}|y_{max})$ are then combined to predict measured trigger $\hat T_{hh}(y_{tt})$ and conditional TA $A_{hh}(y_{ta}|y_{tt})$ hard components, the main goal of this study.


\section{$\bf p$-$\bf p$ Two-component model} \label{2comp}

We want to describe all aspects of minimum-bias jet contributions to \pp\ spectra and correlations, the fragments from all dijets produced in \pp\ collisions appearing within some detector acceptance. Here we define a TCM for yields and 1D SP spectra. In Sec.~\ref{tatcm} we derive the corresponding TCM for 2D TA correlations~\cite{pptrig}. In Sec.~\ref{fftcm} we derive a corresponding TA model or prediction from parametrizations of measured FFs and from measured jet spectra.  By combining those results we establish a quantitative correspondence between event-wise reconstructed dijets and jet-related fragments as they appear in \pp\ spectra and correlations with various conditions imposed.

\subsection{Soft and hard events and yield $\bf n_{ch}$ components} \label{tcmyield}

\pp\ collisions can be separated into soft and hard event types. Hard events include at least one minimum-bias dijet within the angular acceptance and therefore both soft and hard spectrum and correlation components. Soft events include no jet structure within the acceptance and therefore only soft components.  Soft and hard event types are distinguished from soft and hard components of ensemble-averaged (MB) yields, spectra and correlations. Events are separated into several multiplicity classes. For the purpose of illustration and to compare with preliminary data analysis we define seven classes indexed by $n \in [1,7]$ as $n_{ch} / \Delta \eta$ = 1.7, 3.4, 5.5,  7.6, 10.0, 13.7, 18.8.

Dijet production in \pp\ collisions scales approximately as $n_{ch}^2$~\cite{ppprd} and is more directly related to the multiplicity soft component $n_s$ (defined below). For a given $n_s$ the dijet number within some $\eta$ acceptance $\Delta \eta$ is $n_j(n_s) = \Delta \eta\, f(n_s)$, with dijet frequency $f(n_s)$ per unit $\eta$ scaled from non-single-diffractive (NSD) \pp\ collisions.  The $n_j(n_s)$ trend is described in Sec.~\ref{dijet}. The Poisson probabilities for soft and hard events are then respectively $P_s(n_s) = \exp(-n_j)$ and $P_h(n_s) = 1- P_s(n_s)$.

The yields $n_x$ and $n_{xy}$ defined below correspond to spectrum integrals within some angular acceptance $(2\pi,\Delta \eta)$. We then define 1D angular densities $\rho_x = n_x / \Delta \eta$. For each multiplicity class 
defined by some $n_{ch}$ interval we have $n_s + n_h = n_{ch}$ averaged over all events. For soft events $n_{ss} = n_{ch}$ and for hard events $n_{hs} + n_{hh} = n_{ch}$. We then obtain the  relations
\bea \label{nchdefs}
n_{ch} &=& n_s + n_h = P_s n_{ss} + P_h (n_{hs} + n_{hh}) \\ \nonumber
n_s &=& P_s n_{ch} + P_h n_{hs} ~~\text{and}~~ n_h = P_h n_{hh}.
\eea
The relation between hard components $n_h$ and $n_{hh}$ and dijet production is further described in Sec.~\ref{dijet}.

 \begin{figure}[h]
  \includegraphics[width=1.65in]{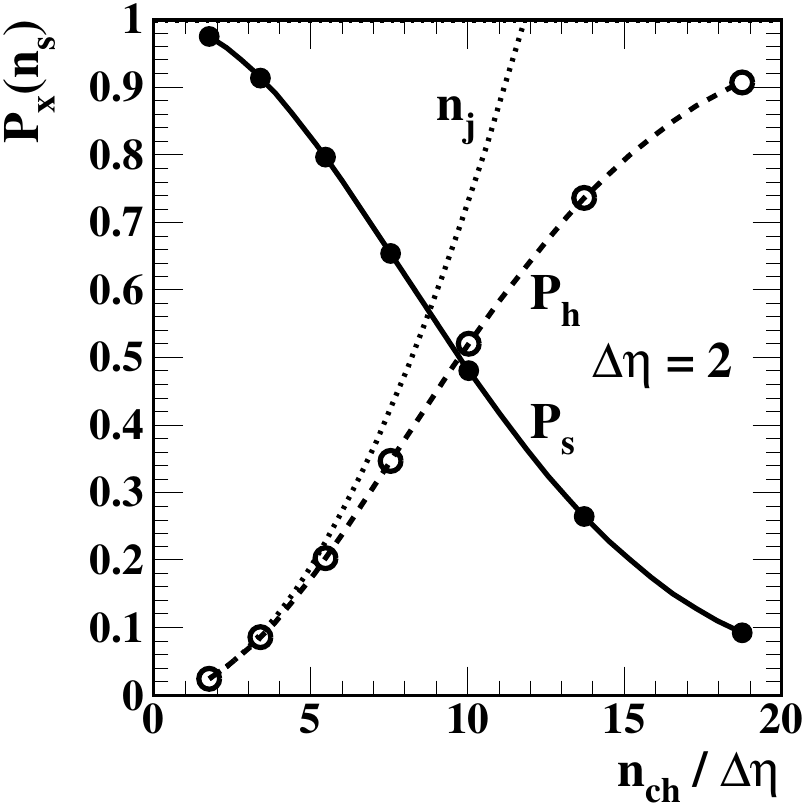}
  \includegraphics[width=1.65in]{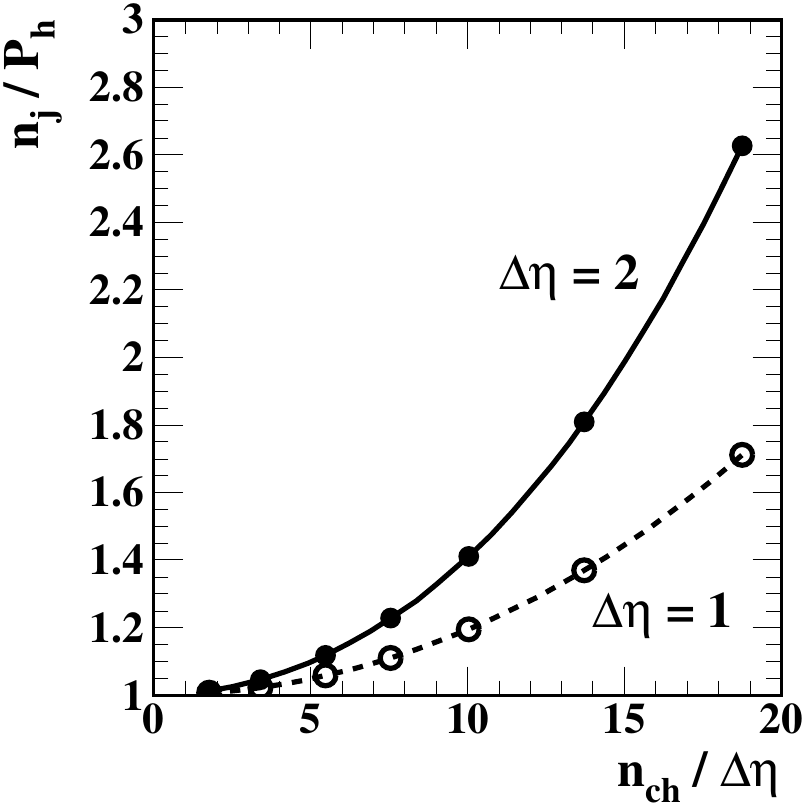}
\caption{\label{njph}
Left: Soft and hard event probabilities $P_s$ and $P_h$ and dijet number $n_j$ within acceptance $\Delta \eta = 2$ for seven multiplicity classes.
Right: The number of dijets per hard \pp\ collision for two acceptances, all for 200 GeV \pp\ collisions.
 }  
 \end{figure}

Figure~\ref{njph} (left) shows the variation of soft and hard event-type probabilities $P_x$ with event multiplicity $n_{ch}$ for acceptance $\Delta \eta = 2$. The mean number $n_j$ of dijets per event within some acceptance $\Delta \eta$ follows $P_h$ for smaller multiplicities but exceeds that quantity for larger multiplicities. Figure~\ref{njph} (right) shows ratio $n_j / P_h$ vs $n_{ch}$ representing the mean number of dijets per hard event for two acceptances. For convenience in this study we also define the dijet excess per hard event $\Delta n_j = n_j / P_h - 1$.

This treatment assumes that $n_{ch}$ is the total number of charged hadrons appearing in angular acceptance $(2\pi,\Delta \eta)$ integrated over all $y_t$. If the $y_t$ acceptance has a nonzero lower limit (detector acceptance) the affected quantities in the reduced acceptance are $\tilde n_{ch}$, $\tilde n_s$, $\tilde n_{ss}$ and $\tilde n_{hs}$ with $\tilde n_{ch} = \tilde n_s + n_h = \tilde n_{ss} = \tilde n_{hs} + n_{hh}$. The spectrum hard components are assumed to be fully included within typical detector $y_t$ acceptances. To simplify notation in what follows we introduce tildes explicitly only at points where the effect of partial \yt\ acceptance is relevant.


\subsection{p-p single-particle spectra} \label{ptspec} \label{ppspecc}

Single-particle spectra from 200 GeV \pp\ collisions plotted on $y_t$ for several charge multiplicity $n_{ch}$ classes (within some acceptance $\Delta \eta$) reveal a composite spectrum structure represented by two fixed functional forms  [unit-integral soft and hard components $\hat S_0(y_t)$ and $\hat H_0(y_t)$] with amplitudes scaling approximately as $n_{ch}$ and $n_{ch}^2$~\cite{ppprd}.  The TCM for SP $y_t$ spectra from \pp\ collisions is then described by
\bea \label{ppspec}
\rho(y_t,n_{ch}) \equiv \frac{d^2n_{ch}}{y_t dy_t d \eta} \hspace{-.05in}&=&\hspace{-.05in} S(y_t,n_{ch}) + H(y_t,n_{ch}) \\ \nonumber
&=&  \rho_s( n_{ch}) \hat S_0(y_t)  +  \rho_{h}( n_{ch}) \hat H_0(y_t),
\eea
where $\rho_s = n_s / \Delta \eta$ and $\rho_h = n_h / \Delta \eta$ are soft and hard angular densities, and $\rho_0(n_{ch}) = n_{ch} / \Delta \eta$ is the corresponding  total charge density. Soft component $\hat S_0(y_t)$ of the SP TCM is defined as the limiting form as $\rho_s \rightarrow 0$ of spectra normalized as $\rho / \rho_s$. Hard component $\hat H_0(y_t)$ models data hard components $H(y_{t},n_{ch})/\rho_s$ obtained by subtracting  model $\hat S_0(y_t)$ from those normalized spectra. 

Based on comparisons with theoretical models the soft component is interpreted to represent longitudinal projectile-nucleon fragmentation (dissociation) while the hard component represents transverse-scattered-parton fragmentation~\cite{hardspec}. The hard component of 1D SP spectra interpreted as a manifestation of MB dijet structure is consistent with jet-related two-particle correlations and in quantitative agreement with pQCD predictions~\cite{ppprd,hardspec,fragevo,jetspec}.  The \pp\ SP spectrum TCM serves in turn as the basis of a 2D TCM for \pp\ TA correlations.


SP spectra can also be expressed in terms of soft and hard event types
\bea \label{spevtype}
\rho(y_t,n_{ch}) &=& P_s(n_{ch}) S_s(y_t,n_{ch}) \\ \nonumber
&+& P_h(n_{ch}) [S_h(y_t,n_{ch}) + H_h(y_t,n_{ch})]
\eea
with corresponding TCM model elements $S_s = \rho_0 \hat S_0(y_t)$, $S_h = \rho_{hs} \hat S_0(y_t)$, $H_h = \rho_{hh} \hat H_0(y_t)$ and $H = P_h H_h$. 

In this analysis we compare fragment distributions derived from FFs defined on total momentum $p$ or total rapidity $y$ with a TCM based on \pp\ spectra defined on transverse momentum $p_t$ or transverse rapidity $y_t$.  Spectra are defined as $dn_{ch}/y_t dy_t$ whereas fragmentation  functions are defined  as $2dn_{ch,j}/dy$. The spectrum hard component in the form $y_tH(y_t)$ is compared with a fragment distribution $D_u(y)$ derived from FFs as in Ref.~\cite{fragevo}. 

\subsection{p-p minimum-bias dijet production} \label{dijet}

Equation~(\ref{ppspec}) integrated over some angular acceptance $(2\pi,\Delta \eta)$ becomes
\bea \label{ppspec2}
 F(y_t,n_{ch}) &=& \frac{d n_{ch}}{y_t dy_t} = n_s \hat S_0(y_t) + n_h \hat H_0(y_t),
\eea
with $F(y_t,n_{ch}) \equiv n_{ch} \hat F(y_t,n_{ch}) = \Delta \eta S + \Delta \eta H$. For hard  events $F_h = F_{hs} + F_{hh}$, and corresponding $n_{hs}$ and $n_{hh}$ from Eq.~(\ref{nchdefs}) are the integrals over $y_t$. The multiplicity trend for the extracted hard components reported in Ref.~\cite{ppprd} implies that $n_h/n_s = \alpha\, n_s/\Delta \eta$ or
\bea \label{freq}
\rho_h(n_{s}) &=& \alpha(\Delta \eta) \rho_s^2 \\ \nonumber
&\equiv& f(n_{s}) \epsilon(\Delta \eta) 2 \bar n_{ch,j},
\eea
where the second line represents the jet hypothesis from Ref.~\cite{ppprd} and defines dijet frequency $f = dn_j/ d\eta$ (mean dijet number per \pp\ event per unit $\eta$) with mean dijet fragment multiplicity $2\bar n_{ch,j}$ into $4\pi$ acceptance. The factor $\epsilon \in [0.5,1]$ represents the fraction of a dijet that appears within acceptance $\Delta \eta$ in hard events~\cite{fragevo}. 
Given $\rho_0 = \rho_s + \alpha \rho_s^2$ we have 
 \bea
 \rho_s(n_{ch}) &=& \frac{\sqrt{4 \alpha \rho_0 + 1} - 1}{2 \alpha}
 \eea
 as the soft component derived from charge density $\rho_0$.

Based on a universal model of jet spectra (see App.~\ref{parspec}) we derive an estimate of the jet frequency $f_{NSD}$ for 200 GeV non-single-diffractive (NSD) \pp\ collisions
\bea \label{fxxx}
f_{NSD} &=& \frac{1}{\sigma_{NSD}} \frac{d\sigma_j}{d\eta} = 0.025 \pm 0.005,
\eea
and an estimate of the MB dijet mean fragment multiplicity within $4 \pi$ acceptance
\bea \label{f}
 2\bar n_{ch,j} &=& \frac{\alpha(\Delta \eta) \,\rho_{s,NSD}^2}{f_{NSD}\epsilon(\Delta \eta)} = 2.2\pm 0.5
\eea
assuming $\rho_{s,NSD}  \approx 2.5$ for 200 GeV NSD \pp\ collisions, with $\alpha \approx 0.005$~\cite{ppprd} and $\epsilon(\Delta \eta) \approx 0.6$ for $\Delta \eta = 1$~\cite{fragevo,jetspec}.  We also obtain $\rho_{h,NSD} = 0.03 \pm 0.005$. 

From a TCM analysis of \pp\  spectrum \nch\ dependence we determine the model functions $\hat S_0(y_t)$ and $\hat H_0(y_t)$ and  the dijet frequency $f(n_{s}) = n_j(n_{s})/\Delta\eta = (\rho_s / 2.5)^2 f_{NSD}$.
In general, dividing a hard-component contribution to hard events (spectra or correlations) by factor $n_j(n_s) / P_h(n_s)$ (number of dijets per hard event in $\Delta \eta$) should produce a result directly comparable to the properties of single dijets.
We next employ the 1D SP spectrum TCM to derive a model for 2D TA correlations.

 \section{Trigger-associated TCM} \label{tatcm}
 

The TCM for 1D single-particle $y_t$ spectra in Eq.~(\ref{ppspec})  derived from 200 GeV \pp\ collisions can be used to define a TCM for 2D TA correlations in the form $F_{at}(y_{ta},y_{tt},n_{ch}) = \hat T(y_{tt},n_{ch}) A(y_{ta}|y_{tt},n_{ch})$ including soft and hard components as in Ref.~\cite{pptrig}. By subtracting a 2D TA soft-component model from the associated-particle conditional distribution $A(y_{ta}|y_{tt},n_{ch})$ we can isolate a conditional TA hard component of hard events $A_{hh}(y_{ta}|y_{tt},n_{ch})$ that may be compared directly with measured FF systematics and pQCD parton spectra. All TA results depend on a specific angular acceptance specified in this case as $2\pi$ azimuth and $\Delta \eta$.


\subsection{Trigger spectrum TCM} \label{trigtcm}

We first define the TCM for 1D unit-normal trigger spectrum $\hat T(y_{tt},n_{ch})$ based on the SP spectrum TCM from Sec.~\ref{ptspec}. 
Trigger particles arise from (sample) soft or hard events, and for hard events arise from either the soft or hard spectrum component. For a sequence of independent trials (collision events), each including $n_{ch}$ samples from a fixed unit-normal parent spectrum $\hat F_x(y_t)$ ($x$ denotes soft $s$ or hard $h$ event type), we sort events according to the maximum sample value $y_{tt}$ in each event. 
Parent distributions are denoted by  $\hat F_s(y_{tt}) = \hat S_0(y_{tt})$ for soft events and $\hat F_h(y_{tt},n_{ch}) = p_{hs} (n_{ch}) \hat S_0(y_{tt}) + p_{hh}(n_{ch}) \hat H_0(y_{tt})$ for hard events, with $p_{hs} = \tilde n_{hs} / \tilde n_{ch}$ and $p_{hh} = n_{hh} / \tilde n_{ch}$. 
%
We then define the trigger spectrum as
\bea \label{trigspec}
\hat T(y_{tt},n_{ch}) &\equiv& \frac{1}{N_{evt}(n_{ch})} \frac{dn_{trig}}{y_{tt}dy_{tt}} \\ \nonumber 
&& \hspace{-.3in} =  P_s(n_{ch})G_s(y_{tt},n_{ch})\, \tilde n_{ch} \hat F_s(y_{tt}) \\ \nonumber 
&& \hspace{-.27in} +      P_h(n_{ch})G_h(y_{tt},n_{ch})\, \tilde n_{ch} \hat F_h(y_{tt},n_{ch}) \\ \nonumber
&& \hspace{-.3in} = P_s(n_{ch}) \hat T_s(y_{tt},n_{ch}) + P_h(n_{ch})\hat T_h(y_{tt},n_{ch}). 
\eea
In each term factor $\hat F_x(y_{tt},n_{ch})$ is the probability of a sample at $y_{tt}$ from the parent spectrum, and factor $G_x(y_{tt},n_{ch})$ (defined below) is the probability of a void (no samples) above that $y_{tt}$. The $P_x$ are the event-type probabilities defined in Sec.~\ref{tcmyield}. The sum of products gives the probability of a trigger particle at $y_{tt}$ from either event type. The corresponding data trigger spectrum is unit normal by construction. 

The void probabilities $G_x(y_{tt},n_{ch})$ are defined as follows: For events with a trigger at $y_{tt}$ no samples can appear with $y_t > y_{tt}$ (void interval). The mean spectrum integral {\em above} $y_{tt}$ within acceptance $\Delta \eta$ is
\bea \label{nsum}
n_{x\Sigma}(y_{tt},n_{ch}) 
 &=& \int _{y_{tt}}^\infty dy_t y_t \tilde n_{ch} \hat F_x(y_{t})
\eea
separately for spectra $\hat F_x(y_t)$ from soft or hard events. The void probability for event type $x$ is $G_x(y_{tt},n_{ch}) = \exp[- \kappa_x n_{x\Sigma}(y_{tt},n_{ch})]$, where $O(1)$ factors $\kappa_x$ may account for non-Poisson correlations. $\kappa$ is the only adjustable parameter in the trigger-spectrum model. In Ref.~\cite{pptrig} ad hoc $O(1)$ correction factors $T_{x0}(n_{ch})$ were introduced. However, such factors are neither necessary nor permitted in the data description because $G_x \rightarrow 1$ for larger $y_{tt}$, and model $n_{hh} \hat H_0(y_{tt})$ describes the measured spectrum hard component accurately in that region.

For hard events the void probability factorizes as $G_h = G_{hs}G_{hh}$ because $n_{h\Sigma} = n_{hs\Sigma} + n_{hh\Sigma}$ has contributions from both the soft and hard components of $F_h(y_{tt},n_{ch}) = \tilde n_{ch} \hat F_h(y_{tt},n_{ch}) = \Delta \eta (S_h + H_h)$. The trigger spectrum for hard events is then decomposed as
\bea  \label{thhh}
\hat T_h &=& G_h \tilde n_{ch} \hat F_h
\\ \nonumber
&=&G_{hh} \hat T_{hs} + G_{hs} \hat T_{hh}
\eea
with $G_{hs} \hat T_{hh}$ representing all hard triggers from dijets in hard \pp\ collisions. Note that these definitions of $\hat T_{hs}$ and $\hat T_{hh}$ are different from those in Ref.~\cite{pptrig}.
Factor $G_{hs}$ is a form of inefficiency. What appears in a measured \pp\ trigger spectrum is the result of competition between soft and hard components to provide the trigger in a given hard \pp\ collision. Potential jet triggers are replaced by soft triggers in some fraction of  hard events. 

 \begin{figure}[h]
  \includegraphics[width=1.65in]{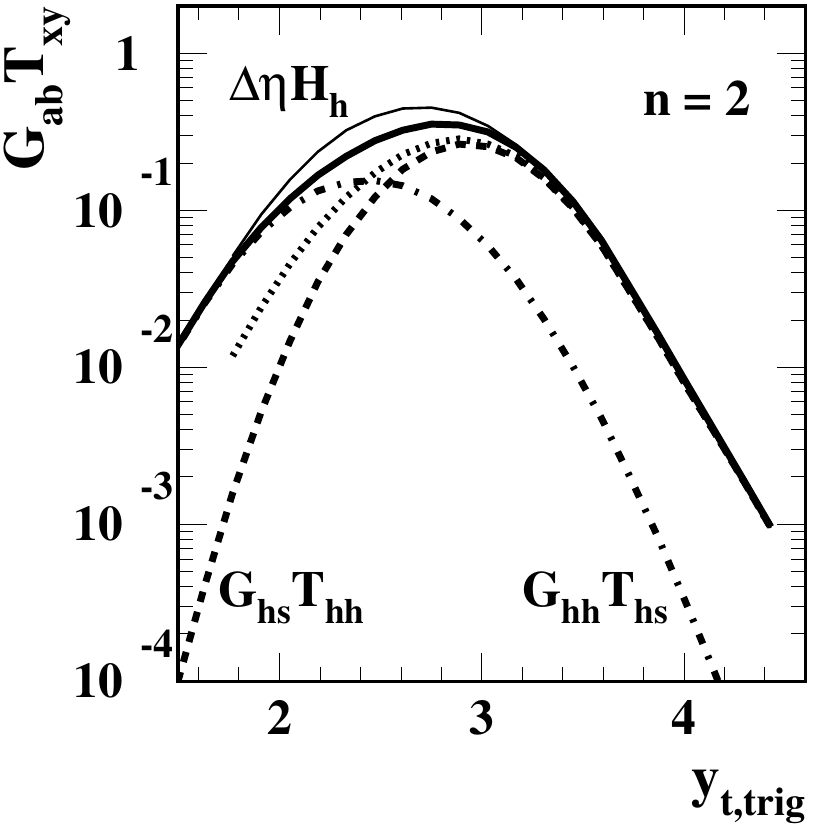}
  \includegraphics[width=1.65in]{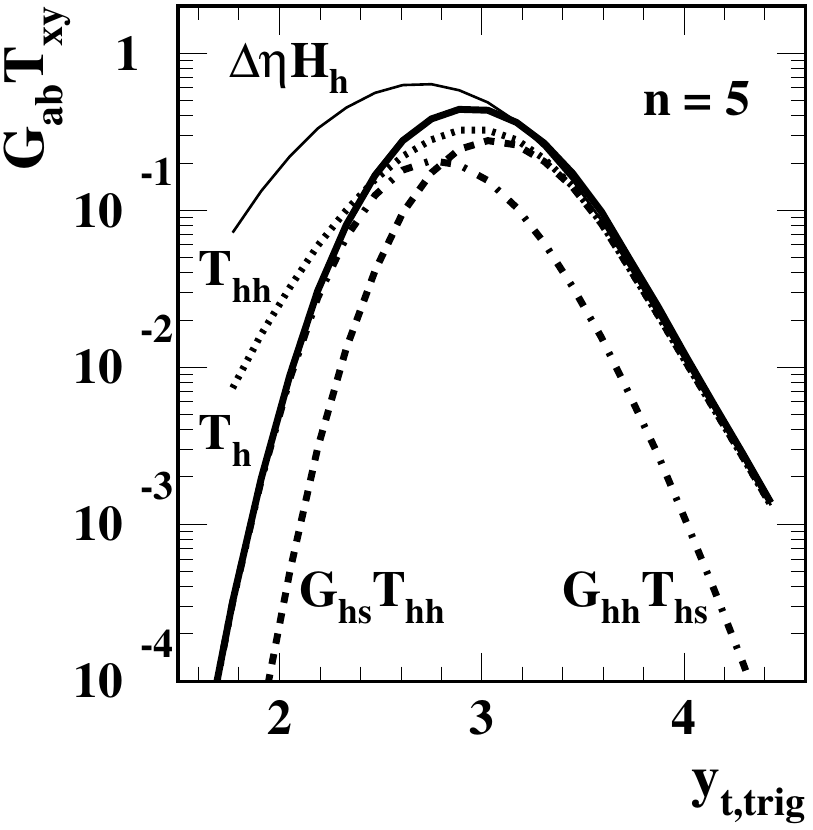}
\caption{\label{trigdemo}
The terms of Eq.~(\ref{thhh}) for two multiplicity classes and for $\Delta \eta = 2$. Unit-normal hard-event trigger spectrum $\hat T_h$ is the bold solid curve. Its soft and hard components are denoted by dashed and dash-dotted curves. The dotted curve is $T_{hh}$ without the $G_{hs}$ factor. The related hard-event spectrum hard component $\Delta \eta H_h$ is the thin solid curve.
 } 
 \end{figure}

Figure~\ref{trigdemo}  shows $\hat T_h$ (bold solid curve, all triggers from hard events), $\hat T_{hs}$ (dash-dotted curve, soft triggers from hard events) and $\hat T_{hh}$ (dashed curve, hard triggers from hard events) for two multiplicity classes compared to the spectrum hard component of hard events $H_h(y_{tt})$ (thin solid curve) integrated over acceptance $\Delta \eta = 2$. The dotted curve is $\hat T_{hh}$ without void probability factor $G_{hs}$.

Figure~\ref{hardcol2} (a) shows predicted unit-normal total trigger spectrum $\hat T(y_{tt})$ (bold solid curve) from Eq.~(\ref{trigspec}) compared to preliminary data (points) from 200 GeV \pp\ collisions for multiplicity class $n = 5$~\cite{ismd13,duncan}. Parameter $\kappa \approx 1.5$ has been adjusted to match the data below the spectrum mode. 
The bold dashed curve is $P_h \hat T_{h}$ and the light dashed curve is spectrum hard-component model $P_h n_{hh} \hat H_0(y_{tt}) = n_h \hat H_0(y_{tt})$. 
The bold dotted curve is $P_s \hat T_{s}$ and the light dotted curve is $P_s \tilde n_{ch} \hat S_0(y_{tt})$. The ratio of bold to light curve is in either case the void probability $G_x(y_{tt})$ for the soft or hard event  type.


\subsection{Conditional trigger-associated TCM} \label{contcm}

We next derive a 2D model for TA two-particle correlations based on the chain rule applied to joint distribution $F_{at}(y_{ta},y_{tt},n_{ch})$ formed from all possible trigger-associated hadron pairs {\em excluding self pairs}. We assume that for a given trigger the associated particles are sampled from soft or hard parent distributions approximated by those in the 1D SP spectrum TCM subject to marginal constraints described in Ref.~\cite{pptrig}. For multiplicity class $n_{ch}$ the total number of triggers is $N_{evt}(n_{ch})$, the observed (detected) associated-particle number per event is $\tilde n_{ch} - 1$, 
and the total trigger-associated pair number for the given $n_{ch}$ class (with self pairs excluded) is $N_{evt}(n_{ch})(\tilde n_{ch}-1)$. 

 \begin{figure}[t]
  \includegraphics[width=1.65in]{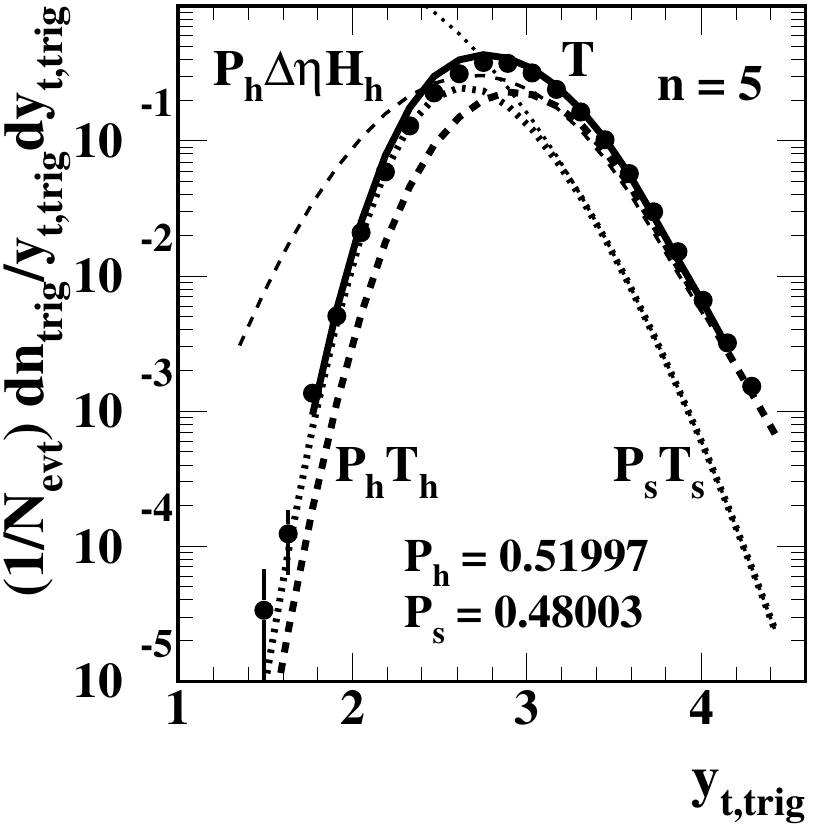}
 \put(-87,100) {\bf (a)}
 \includegraphics[width=1.65in]{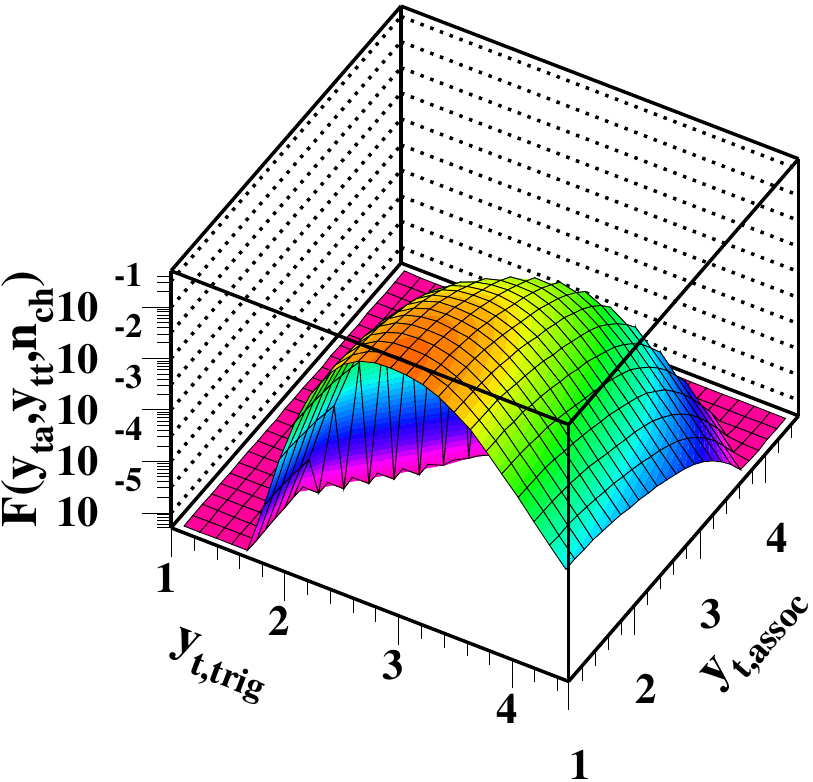}
  \put(-87,100) {\bf (b)} \\
 \includegraphics[width=1.65in]{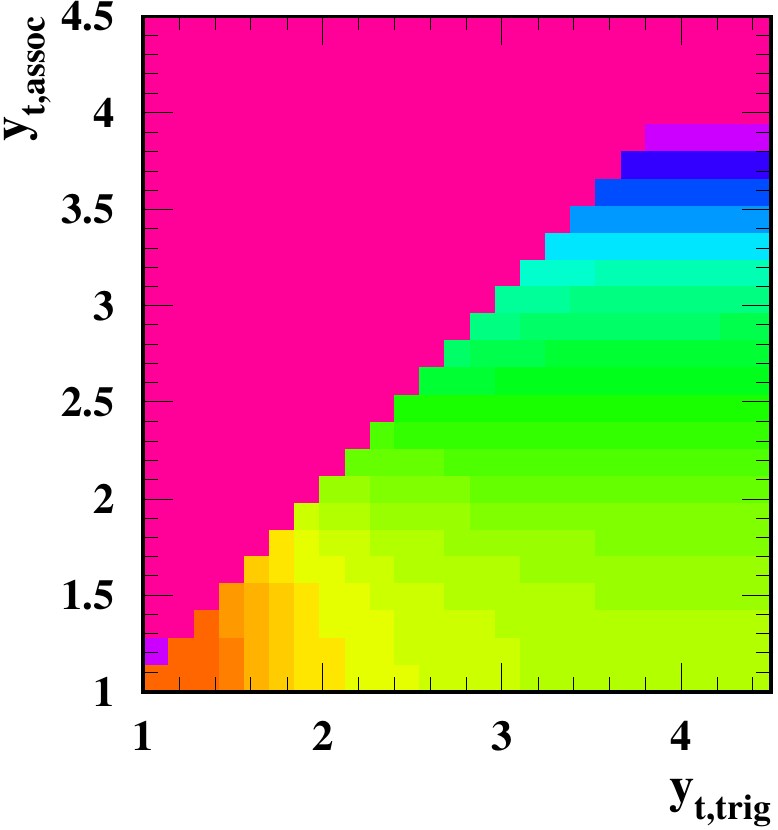}
 \put(-87,109) {\bf (c)}
  \includegraphics[width=1.65in]{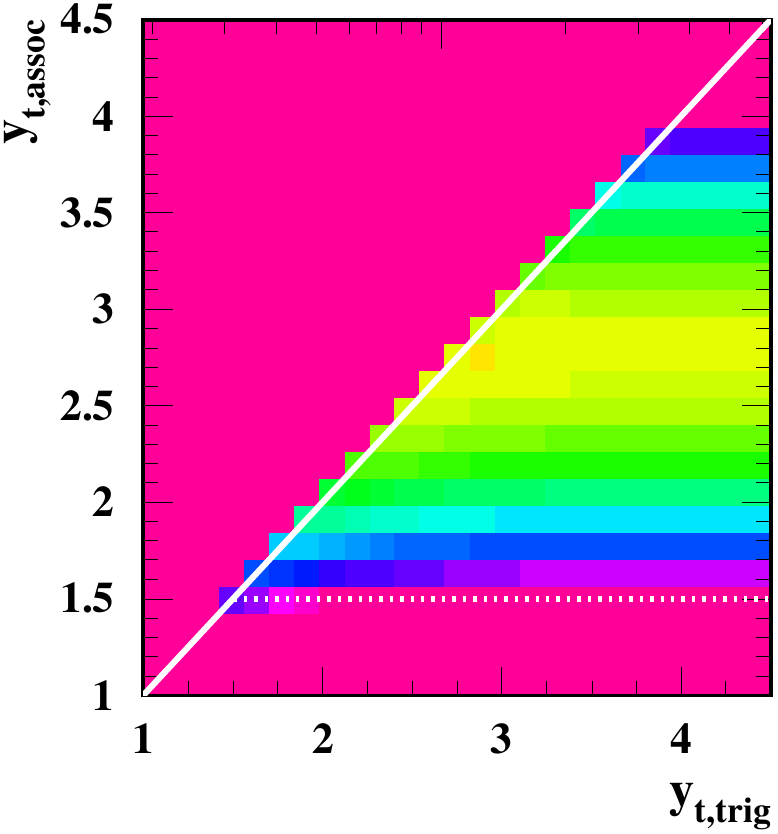}
 \put(-87,109) {\bf (d)}
\caption{\label{hardcol2}
(Color online) 
(a) Unit-normal TCM trigger spectrum $\hat T$ (bold solid curve) for multiplicity class $n = 5$ compared with preliminary TA data from Ref.~\cite{ismd13,duncan} (points). Also shown are separate soft- (bold dotted) and hard- (bold dashed) event spectra.
(b) TCM for 2D TA density $\hat F_{at}(y_{ta},y_{tt})$.
(c) TCM for 2D conditional associated-particle spectrum ensemble $\hat A = \hat F_{at} / \hat T$.
(d) The hard component $y_{ta}\hat A_{hh} \approx y_{ta}\hat H'_0(y_{ta}|y_{tt})$ of TCM $\hat A$ isolated with Eq.~(\ref{hardcomp}).
For (c) and (d) the z-axis limits (log scale) are 3 and 0.03; the acceptance is $\Delta \eta = 2$.
 }  
\end{figure}

We assume that the expressions for soft and hard event types are linearly independent so that $F_{at}(y_{ta},y_{tt}) = P_s F_s(y_{ta},y_{tt}) + P_h F_h(y_{ta},y_{tt})$. According to the chain rule the probability of a TA pair $F_{x}(y_{ta},y_{tt})$ can be written as the product of trigger probability $\hat T_x(y_{tt},n_{ch})$ and $A_x(y_{ta}|y_{tt},n_{ch})$, proportional to the conditional probability of an associated particle at $y_{ta}$ given a trigger at $y_{tt}$.
The model for $A_x(y_{ta}|y_{tt})$ is also based on the $F_x(y_t)$ from the 1D SP TCM but is set to zero above $y_{tt}$ (the sample void). 
We then obtain a 2D TCM for  $F_{at}(y_{ta},y_{tt},n_{ch})$
\bea \label{tadist}
F_{at}(y_{ta},y_{tt}, n_{ch})
&=& 
\frac{1}{N_{evt}(n_{ch})}\frac{d^2n_{ch}}{y_{tt}dy_{tt}y_{ta}dy_{ta}}~~  \\ \nonumber
&& \hspace{-.2in} =  P_s (n_{ch}) \hat T_s(y_{tt},n_{ch})A_s(y_{ta}|y_{tt},n_{ch}) \\ \nonumber
&& \hspace{-.2in} + P_h (n_{ch}) \hat T_h(y_{tt},n_{ch}) A_h(y_{ta}|y_{tt},n_{ch}),
\eea
where $A_s(y_{ta}|y_{tt},n_{ch}) = (\tilde n_{ch} - 1) \hat S_0''(y_{ta}|y_{tt},n_{ch})$ for soft events and $A_h(y_{ta}|y_{tt},n_{ch}) = \tilde n'_{hs}(n_{ch})\hat  S_0'(y_{ta}|y_{tt},n_{ch}) + n'_{hh}(n_{ch}) \hat H_0'(y_{ta}|y_{tt},n_{ch}) \equiv A_{hs} + A_{hh}$ for hard events.  The primes on the spectrum model components  indicate that conditional probabilities may deviate from corresponding 1D SP spectrum models because of imposed marginal constraints (the soft component includes different distortions for the two event types). 

The primes on multiplicities $\tilde n_{xy}$ indicate the effect of the ``missing'' trigger hadron. We want $\tilde n_{hs} + n_{hh} = \tilde n_{ch}$ for $F_{hh}$ and $\tilde n'_{hs} + n'_{hh} = \tilde n_{ch}-1$ for $A_{hh}$. For soft triggers $(\tilde n_{hs} -1) + n_{hh} = \tilde n_{ch}-1$, but for hard triggers $\tilde n_{hs} + (n_{hh} -1) = \tilde n_{ch}-1$. The primed quantities represent averages over  trigger distributions approximated in this case by scaling the estimated $\tilde n_{hs}$ and $n_{hh}$ by factor $(\tilde n_{ch} - 1) / \tilde n_{ch}$.

Figure~\ref{hardcol2} (b) shows the predicted TCM \pp\ TA distribution $\hat F_{at} = F_{at} / (\tilde n_{ch} - 1)$ from Eq.~(\ref{tadist}) for multiplicity class $n = 5$ that compares well with preliminary data  from Refs.~\cite{ismd13,duncan}.


The chain rule can also be applied to the full TA distribution to define the complementary associated-particle conditional distribution $A(y_{ta}|y_{tt},n_{ch})$. 
We divide $F_{at}(y_{ta},y_{tt},n_{nch})$ from Eq.~(\ref{tadist}) by trigger spectrum $\hat T(y_{tt},n_{ch})$ from Eq.~(\ref{trigspec}) to define a TCM for $A(y_{ta}|y_{tt},n_{ch})$. The result is an ensemble of associated-particle distributions on $y_{ta}$ conditional on $y_{tt}$
\bea \label{ftrat}
A(y_{ta}|y_{tt},n_{ch}) &=&   \frac{dn_{ch}(y_{ta}|y_{tt},n_{ch})}{y_{ta}dy_{ta}} \\ \nonumber
&=&   P_s(n_{ch}) R_{s}(y_{tt},n_{ch}) A_s(y_{ta}|y_{tt},n_{ch}) \\ \nonumber
&+&    P_h(n_{ch}) R_{h}(y_{tt},n_{ch}) A_h(y_{ta}|y_{tt},n_{ch}).
\eea
each integrating to $\tilde n_{ch}-1$, with trigger fractions $R_x \equiv\hat T_x / \hat T$. The TA TCM can be compared with data in the forms $\hat T(y_{tt},n_{ch})$, $A(y_{ta}|y_{tt},n_{ch})$ and  $F_{at}(y_{ta},y_{tt}, n_{ch})$. 


Figure~\ref{hardcol2} (c) shows predicted \pp\ TA conditional distribution $A$ from Eq.~(\ref{ftrat}) for multiplicity class $n = 5$ that also compares well with preliminary data  from Refs.~\cite{ismd13,duncan}. The increased amplitude below $y_{tt} \approx 2.5$ is a marginal-constraint distortion. The integrated associated-particle multiplicity is constrained to have the same value $\tilde n_{ch} - 1$ for each $y_{tt}$ condition but the accepted $y_{ta}$ interval decreases with $y_{tt}$.
Such distortions are represented by $O(1)$ weight factor $D_x(y_{tt})$ defined in Ref.~\cite{pptrig}. 


\subsection{Isolating TA data hard component $\bf R_h A_{hh}$} \label{isolate} \label{tahard}

For TCM or real data we can isolate the hard component  of $A$ by subtracting a TCM soft-component  reference from the ratio $A = F/\hat T$ according to the method described in Ref.~\cite{ppprd}.
 We first rearrange Eq.~(\ref{ftrat}) to isolate the product $R_h A_{hh}$ (hard component of hard events)
\bea \label{hardcomp}
P_h R_hA_{hh}(y_{ta}|y_{tt},n_{ch}) \hspace{-.07in} &=& \hspace{-.07in} A - P_s R_s A_s -  P_h R_h A_{hs}.~~
\eea
For real data the product $R_h A_{hh}$ requires further analysis after isolation, as described in Sec.~\ref{ahardcomp}. Note that $A_{hh}$ above is denoted by $H'_h$ in Ref.~\cite{pptrig}.

Figure~\ref{hardcol2} (d) shows TA hard component $A_{hh}$ obtained with Eq.~(\ref{hardcomp}) from the TCM as defined in Sec.~\ref{contcm}. The TA TCM assumes factorization of the hard component $A_{hh}(y_{ta}|y_{tt}) \rightarrow n'_{hh} \hat H_0'(y_{ta}|y_{tt})$, the latter derived from 1D SP model function $\hat H_0(y_t)$, and that structure is returned in this exercise. There is no correlation between $y_{tt}$ and $y_{ta}$ -- the input model is uniform on $y_{tt}$ modulo the effect of the $y_{ta} < y_{tt}$ constraint.  The marginal-constraint distortion is evident along the left edge of $A_{hh}$. 

For real data (as in Refs.~\cite{ismd13,duncan}) we expect to encounter nontrivial jet-related correlation structure in data hard-component product $R_h A_{hh}(y_{ta}|y_{tt})$ that is not factorizable. The associated-particle spectrum should change shape with varying trigger condition $y_{tt}$. Correlation structure should correspond to jet fragment distributions derived from measured FFs and may reveal systematic details of low-energy parton fragmentation.

\section{Partitioning FF ensembles} \label{partition}

We want to predict the dijet-related hard component of TA hadron correlations from \pp\ collisions by constructing trigger-associated conditional fragment distributions derived from measured FF ensembles. The parametrizations of FF ensembles  from \ee\ and \ppbar\ collisions used in  this study are summarized in App.~\ref{fragfunc}. In this section we partition FF ensembles into trigger and associated components via the method described in Sec.~\ref{trigtcm}.


An FF ensemble $D_{u}(y|y_{max})$ can be partitioned into trigger and associated components by defining a void probability that depends on the absolute number of jet fragments detected in some angular acceptance. A trigger particle appearing at rapidity $y_{trig}$ implies a sample void above that point with void probability $G_{t}(y_{trig})$ derived from the mean sample number per dijet within the void interval and detector angular acceptance. The number of detected fragments per dijet above $y_{trig}$ within acceptance $\Delta \eta$ is
\bea
n_{\Sigma}(y_{trig}|y_{max}) &=& \epsilon(\Delta \eta) \int_{y_{trig}}^\infty dy D_{u}(y|y_{max}),
\eea
where factor $\epsilon(\Delta \eta)$ represents the effect of the detector $\eta$ acceptance~\cite{fragevo,jetspec}. The void probability is
defined by $G_{t}(y_{trig}|y_{max}) = \exp[-\kappa n_{\Sigma}(y_{trig}|y_{max})]$ where $\kappa \approx 1$ represents the effects of non-Poisson correlations. Given void (trigger) probability $G_{t}$ we define complementary associated probability $G_{a} = 1 - G_{t}$. We then obtain the FF trigger component as $\hat S_{t}(y_{trig}|y_{max})= G_t(y_{trig}|y_{max}) \epsilon(\Delta \eta) D_{u}(y_{trig}|y_{max})$ and the associated component $D_{a}(y_{trig}|y_{max})$ by replacing $G_t$ with $G_a$.

 \begin{figure}[h]
  \includegraphics[width=3.3in]{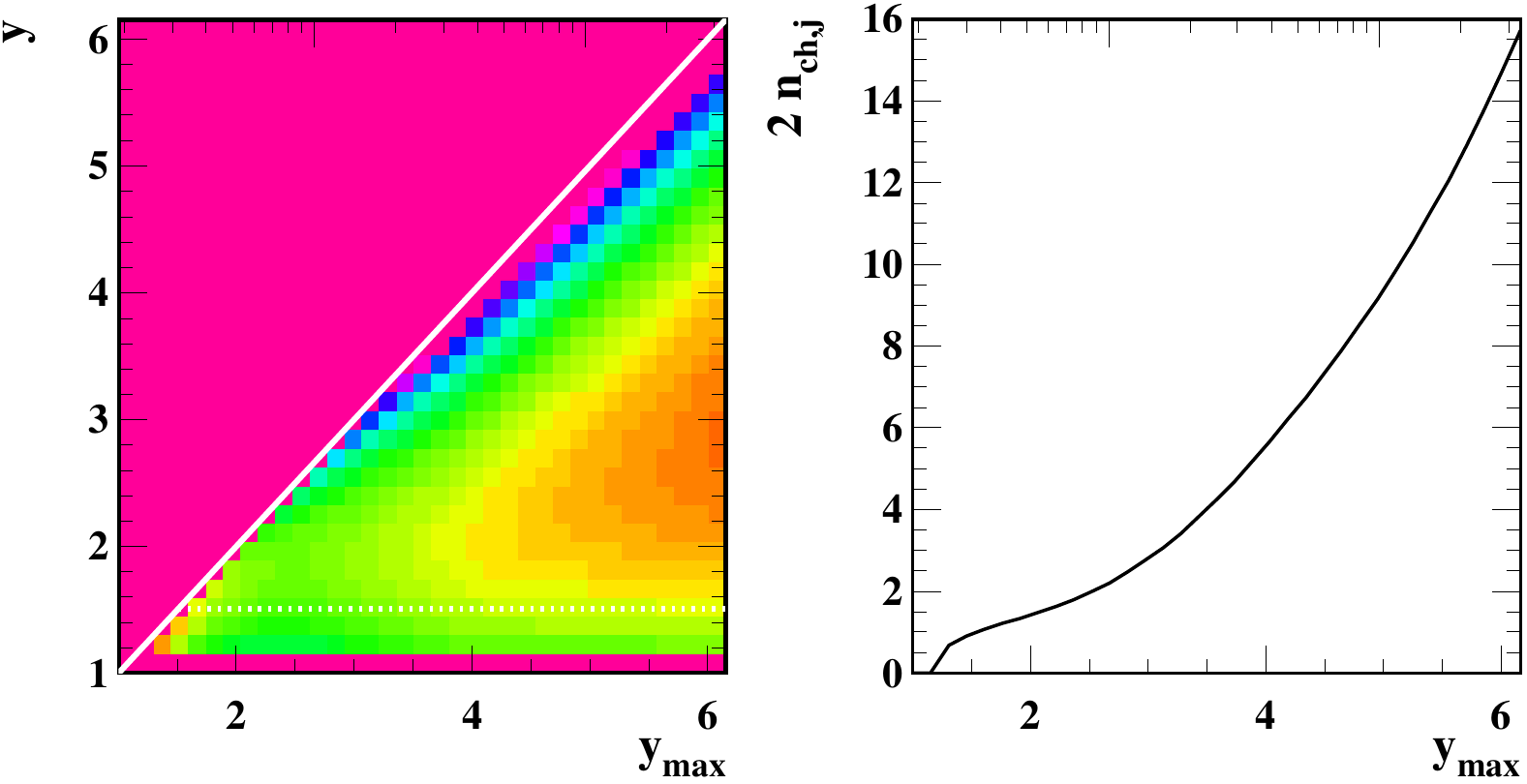}
  \put(-67,108) {\scriptsize \bf 1}
  \put(-28.,108) {\scriptsize \bf 10}
  \put(-62,100) {\scriptsize $\bf E_{jet} (GeV)$}
 \put(-212,100) {\bf (a)}
 \put(-87,100) {\bf (b)}\\
 \includegraphics[width=3.3in]{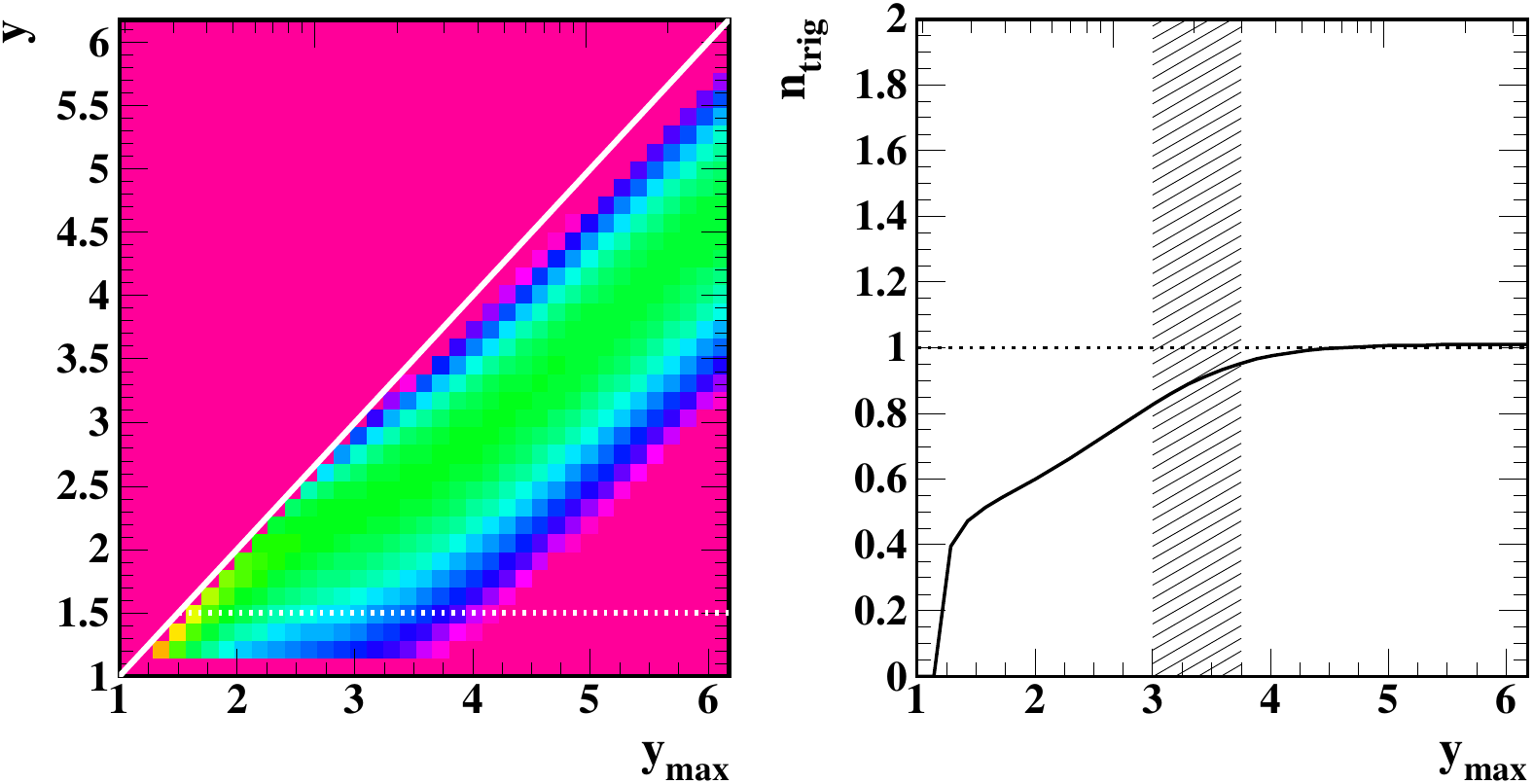}
 \put(-212,100) {\bf (c)}
 \put(-87,100) {\bf (d)}\\
 \includegraphics[width=3.3in]{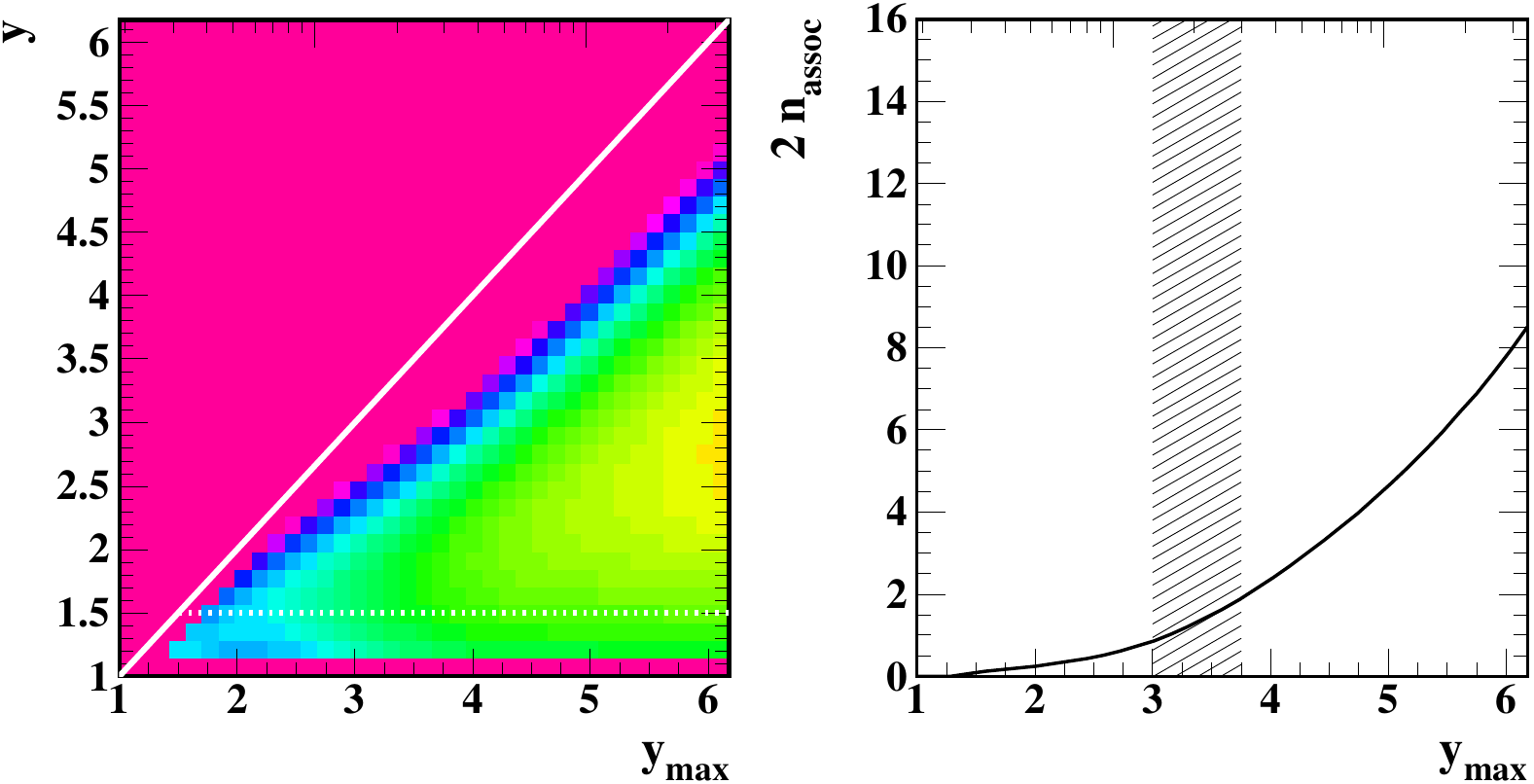}
 \put(-212,100) {\bf (e)}
 \put(-87,100) {\bf (f)}\\
\caption{\label{void}
(Color online) Trigger/associate partitioning of the FF ensemble for light (uds) quarks from \ee\ collisions.
Upper panels: (a) parametrized FF ensemble $D_u(y|y_{max})$~\cite{eeprd} and (b) its projection onto parton rapidity $y_{max}$.
Middle panels: (c) Conditional trigger-spectrum ensemble $\hat S_t(y_{}|y_{max})$ and (d) its projection.
Bottom: (e) Conditional associated-fragment-spectrum ensemble $D_a(y_{}|y_{max})$ and (f) its projection. The trigger and associated spectra sum as $\hat S_t + D_s = \epsilon D_u$ (see text). The z-axis (log scale) limits are 6 and 0.06 for (a) and (e) and  4 and 0.04 for (c). The hatched bands indicate the estimated lower kinematic limit for MB  dijet production.
 }  
\end{figure}
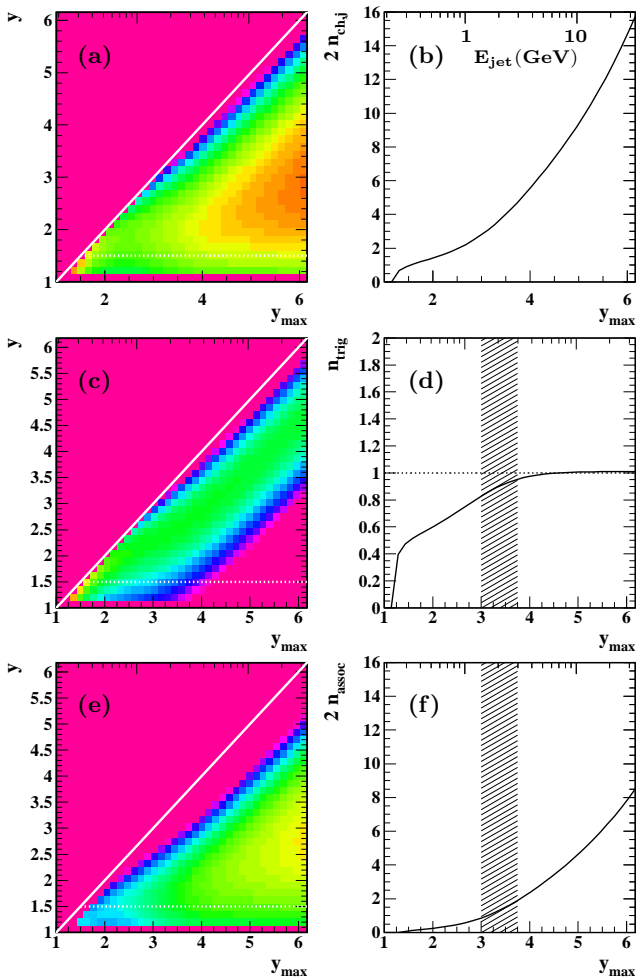

Figure~\ref{void} (top row) shows (a) the FF ensemble $D_{u}(y|y_{max})$ for quark jets from \ee\ collisions within $4\pi$ and (b) its projection onto $y_{max}$ $2 n_{ch,j}(y_{max})$. Figure~\ref{void} (middle row) shows (c) trigger component $\hat S_{t}(y_{trig}|y_{max})$ and (d) its projection $n_{trig}(y_{max})$ (one trigger per dijet). Figure~\ref{void} (bottom row) shows (e) associated component $D_{a}(y_{assoc}|y_{max})$ and (f) its projection $2 n_{assoc}(y_{max}) \approx \epsilon 2 n_{ch,j}(y_{max})-1$. Trigger multiplicity $n_{trig}(y_{max})$ should integrate approximately to 1 and does so (with $\kappa = 1.1$) for \ee\ FFs with $y_{max} > 3.75$ ($E_{jet} >3$ GeV) corresponding to observable charged-hadron jets to the right of the hatched band. 
Note that $\hat S_t + D_a = \epsilon(\Delta \eta) D_u$ sums to the fragment distribution {\em appearing within  the detector angular acceptance} determined by acceptance factor $\epsilon$.
We next consider FF 1D trigger-hadron and  2D TA distributions in isolation and then relate them to physical \pp\ collisions including soft components and multiple dijets.


 \section{1D Trigger spectrum from FF$\text{s}$} \label{fftrig}

To derive a prediction for the trigger-spectrum hard component from FFs
we can combine a measured unit-normal MB jet/parton spectrum $\hat S_p(y_{max})$ with the inferred FF 2D trigger component $\hat S_t(y_{trig}|y_{max})$ from Fig.~\ref{void} (c) to obtain the required FF trigger spectrum $\hat S_t(y_{trig})$ as a 1D projection.  
The jet spectrum is represented by $S_p(y_{max}) = p_{jet} d^2\sigma_j/ dp_{jet} d\eta \ = (d\sigma_j/d\eta) \hat S_p(y_{max})$ as described in App.~\ref{parspec}. 
The full FF joint hadron-parton distribution is $F_{up}(y,y_{max}) = \hat S_p(y_{max}) D_u(y|y_{max}) $.  Its marginal projection is $D_u(y)$, the average of $D_u(y|y_{max})$ weighted by the unit-normal parton spectrum, with integral  $2\bar n_{ch,j}$ the MB mean dijet fragment multiplicity within $4\pi$. 
In the previous section we separated the FF ensemble into components as  $\epsilon D_u(y|y_{max}) = \hat S_t(y|y_{max}) + D_a(y|y_{max})$ based on void probability $G_t(y|y_{max})$, with 
 \bea \label{stdu}
 \hat S_t(y_{}|y_{max}) \hspace{-.07in} &=&\hspace{-.07in} G_t(y_{}|y_{max}) \epsilon(\Delta \eta) D_u(y_{}|y_{max})~~~
 \eea
The FF trigger-parton joint distribution is then $\hat F_{tp}(y_{trig},y_{max}) = \hat S_p(y_{max}) \hat S_t(y_{trig}|y_{max})$, the marginal projection being trigger spectrum $\hat S_t(y_{trig})$  
 \bea \label{sthat}
\hat S_t(y_{trig}) &=& \int_0^\infty dy_{max}   \hat S_p(y_{max}) \hat  S_t(y_{trig}|y_{max})
\\ \nonumber
&\equiv& G_t(y_{trig})\epsilon(\Delta \eta) D_u(y_{trig})
 \eea
defining marginal void probability $G_t(y_{trig})$ as a spectrum-weighted average of the $G_t(y_{trig}|y_{max})$.  Similarly, $F_{ap}(y_{assoc},y_{max}) = \hat S_p(y_{max}) D_a(y_{assoc}|y_{max})$ with marginal projection $D_a(y_{assoc})$ integrating to $\epsilon 2 \bar n_{ch,j} - 1$. 

 \begin{figure}[h]
  \includegraphics[width=1.65in,height=1.65in]{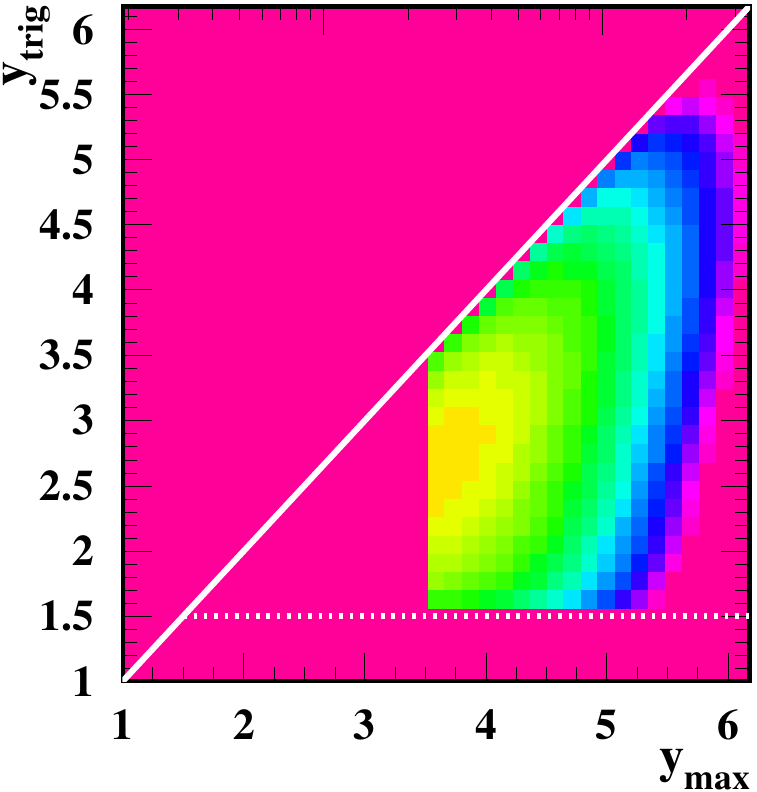}
  \includegraphics[width=1.6in,height=1.65in]{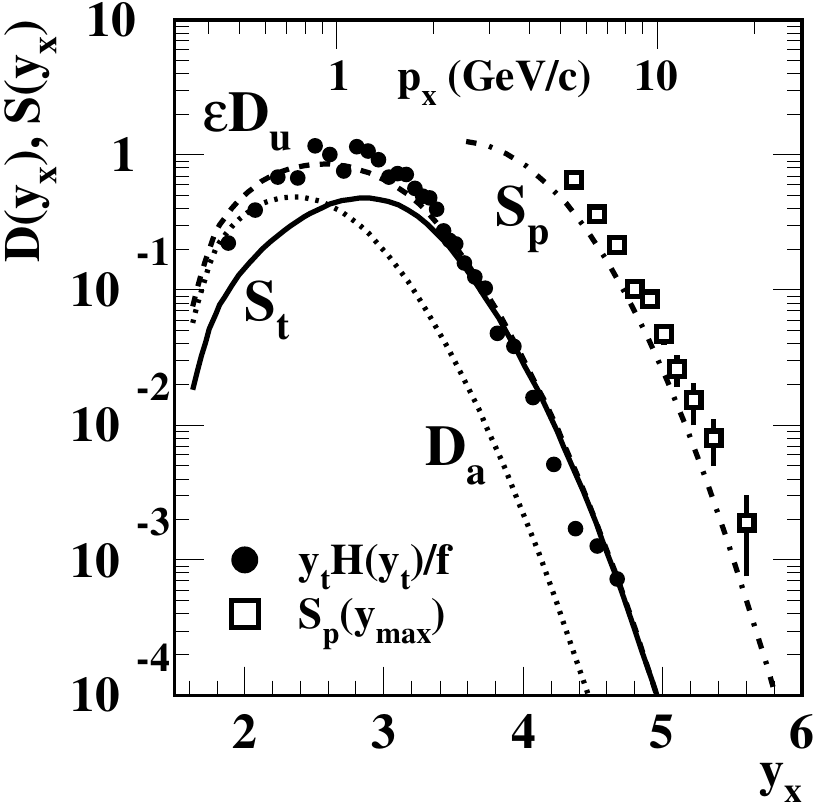}
\caption{\label{trigs}
(Color online) Conditional fragment distributions for light (uds) quarks from \pp\ collisions.
Left: Joint distribution $\hat F_{tp}(y_{},y_{max})$ relating hadron trigger and parton rapidities with z-axis (log scale) limits 4 and $1 \times 10^{-6}$.
Right: Trigger-fragment spectrum $\hat S_t(y_{})$ (solid curve), associated-fragment distribution $D_a(y_{})$ and their sum (dashed curve), the jet fragment distribution $\epsilon D_u(y)$ within acceptance $\Delta \eta = 1$, compared to the spectrum hard component from NSD \pp\ collisions (solid points)~\cite{ppprd}. The MB parton spectrum $S_p(y_{max})$ (dash-dotted curve) used to construct $\hat F_{tp}(y_{trig},y_{max})$ is compared to a jet spectrum from 200 GeV \ppbar\ collisions (open squares)~\cite{ua1}.
 } 
 \end{figure}   
 




Figure~\ref{trigs} (left panel) shows the joint trigger-parton distribution $\hat F_{tp}(y_{trig},y_{max})$ 
obtained by combining FF trigger component $\hat S_t(y_{trig}|y_{max})$ from Fig.~\ref{void} (c) with pQCD parton spectrum $\hat S_p(y_{max})$. The 2D mode corresponds to parton (gluon) energies $\approx 3$ GeV and hadron trigger momenta $\approx 1$ GeV/c. In this case the FFs represent light-quark jets from \pp\ collisions within $\Delta \eta = 1$. The locus of modes on $y_{trig}$  for larger $y_{max}$ is then considerably higher than for gluon jets, but ($\approx$ valence) quark jets dominate at the larger parton energies. Near the 2D distribution mode quark and gluon FFs are equivalent.

Figure~\ref{trigs} (right panel) shows the marginal projections of $\hat F_{tp}(y_{trig},y_{max})$ in the left panel compared with other results.  The projection onto $y_{max}$ is the parton spectrum $\hat S_p(y_{max})$ introduced to construct $\hat F_{tp}(y_{trig},y_{max})$, compared here with 200 GeV jet cross section data (open squares, \cite{ua1}) (the dash-dotted curve includes factor $d\sigma_j/d\eta \approx 1$ mb). The projection onto $y_{trig}$ (solid curve)  is the sought-after FF trigger spectrum $\hat S_t(y_{trig})$. The dotted curve is the projection of $F_{ap}(y_{assoc},y_{max})$, the associated-particle distribution $D_a(y_{assoc})$. The dashed curve is the sum $\epsilon D_u(y) = \hat S_t(y) +D_a(y)$ with $\epsilon = 0.6$ for $\Delta \eta = 1$ integrating to 1.2. Hard-component data from 200 GeV \pp\ collisions (solid points) are scaled as $y_t H(y_t) = d^2 n_h/dy_td\eta$ divided by $f_{NSD} = 0.025$ for comparison with $\epsilon D_u(y)$.  The hard component includes trigger and associated components (see Sec.~\ref{datacomp}) and integrates to $\rho_h \approx 0.03$ consistent with the text below Eq.~(\ref{f}).  For NSD \pp\ collisions $\Delta n_j \ll 1$ (the probability of a second dijet in a hard event is negligible).

 

    

 \section{2D TA Distribution from FF$\text{s}$} \label{fftcm}

From trigger $\hat S_t(y|y_{max})$ and associated $D_a(y|y_{max})$ FF components conditional on parton rapidity $y_{max}$ we want to construct TA fragment-pair distribution $D_{a}(y_{assoc}|y_{trig})$ conditional on hadron trigger rapidity that can be compared with measured TA hard component $A_{hh}(y_{ta}|y_{tt})$ as defined in Sec.~\ref{tatcm}. Fragment TA distributions can be obtained from the two FF components by introducing a convolution  integral over $y_{max}$ that eliminates the parton degree of freedom. The result describes single dijets within a \pp\ hard event. For direct comparisons with data we must accommodate the \pp\ soft components and  effects of multiple dijets per event.

\subsection{Deriving the TA convolution integral}

We begin with a schematic description of the method. From a minimum-bias dijet ensemble we can construct a three-particle joint distribution on rapidity space ($y_{max},y_{trig},y_{assoc}$) for partons (p) and for trigger (t) and associated (a) hadron fragments. Introducing a compact notation the joint distribution (on rapidities) is indexed by $(p,t,a)$. The marginal projections appear on $(a,p)$, $(t,p)$, $(a,t)$, $(p)$, $(t)$ and $(a)$. The sum of $(t,p)$ and $(a,p)$ is $(u,p)$, the measured FF ensemble (describing partons fragmenting to unidentified hadrons).  The chain rule can be applied to that system in various ways: $f(p,t,a) = g(a,t|p)h(p)$, $f(t,p) = g(t|p) h(p) =  g'(p|t) h'(t)$, etc. The measured conditional distributions are FFs on $(u|p)$ and TA hadron correlations on $(a|t)$. Measured FFs can be decomposed into trigger on $(t|p)$ and associated on $(a|p)$ components using void probabilities as described in Sec.~\ref{partition} and illustrated in Fig.~\ref{void}.
We assume an approximate factorization to marginal projections in the form $f(a,t,p) \approx f(a,p)f(t,p) / f(p)$.  Note that marginal projections onto $(t,p)$ and $(a,p)$ on both sides are equal. Projection onto $(a,t)$ gives $f(a,t) \approx \int dp f(a,p) f(t,p) / f(p)$. 



We now restore part of the conventional FF notation. The full integral of joint distribution $F_{atp}(a,t,p)$ per dijet over the 3D rapidity space is $2\bar n_a \equiv 2\bar n_{ch,j} - 1$, the mean number of associated fragments per dijet in $4\pi$. The relation among $F_{atp}(a,t,p)$ marginals is  then
\bea \label{foldd}
F_{at}(a,t) &\approx& \int dy_{max} F_{ap}(a,p) F_{tp}(t,p) / F_p(p) \\ \nonumber
&\rightarrow& \int dy_{max}  F_{ap}(a,p) \hat F_{pt}(p,t) / \hat S_p(p) \\ \nonumber
D_a(a|t) &\approx& \int dy_{max}  D_{a}(a|p) \hat S_{p}(p|t), \\ \nonumber
\eea
where the second line is obtained by canceling a common factor $2\bar n_a$ on the RHS and permuting $p$ and $t$. The third line is the required convolution integral obtained by applying  the chain rule to each function $F_{\alpha \beta}$ and canceling common factors $\hat S_t(t)$ and $\hat S_p(p)$. To proceed we require the parton-trigger conditional distribution $\hat S_p(p|t) \rightarrow \hat S_p(y_{max}|y_{trig})$.




\subsection{Parton-trigger conditional distribution}
 

For a given parton--collision-system combination we partition FFs $D_u(y|y_{max})$ into trigger $\hat  S_t(y_{trig}|y_{max})$ and associated $D_a(y_{assoc}|y_{max})$ components as in Sec.~\ref{partition}. We obtain the required parton-trigger conditional distribution $\hat S_p(y_{max}|y_{trig})$ from the trigger component of the FFs as follows. The first line of
\bea \label{ftpp}
\hat F_{tp}(y_{trig},y_{max}) &=& \hat S_t(y_{trig}|y_{max}) \hat S_p(y_{max}) \\ \nonumber
 &=& \hat S_p(y_{max}|y_{trig}) \hat S_t(y_{trig})
\eea
is constructed from the FF trigger component as in Fig.~\ref{void} (c) and parton spectrum $\hat S_p(y_{max})$ derived from jet data (App.~\ref{parspec}). The second line is an alternative application of the chain rule. $ \hat S_t(y_{trig})$ is the projection of $\hat F_{tp}$ onto $y_{trig}$ obtained in Sec.~\ref{fftrig}. We then have
\bea
\hat S_{p}(y_{max}|y_{trig})  \approx \frac{\hat S_t(y_{trig}|y_{max})\hat S_p(y_{max})}{\hat S_t(y_{trig})}
\eea
as an application of Bayes' theorem. 

 \begin{figure}[h]
  \includegraphics[width=1.6in]{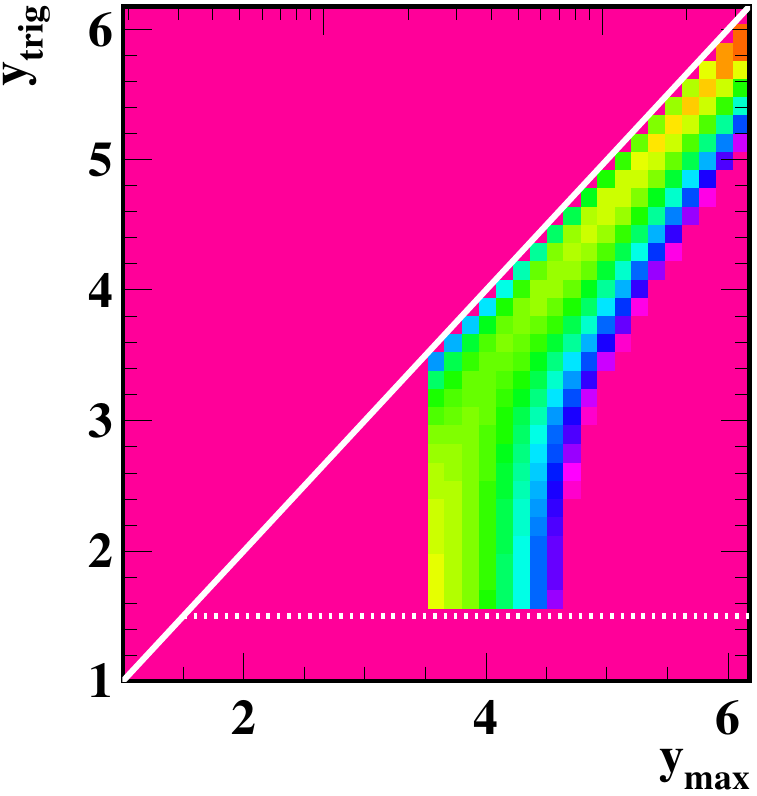}
\put(-67.8,110) {\scriptsize \bf 1}
 \put(-26.9,110) {\scriptsize \bf 10}
 \put(-68,102) {\scriptsize $\bf E_{jet} (GeV)$}
 \includegraphics[width=1.6in]{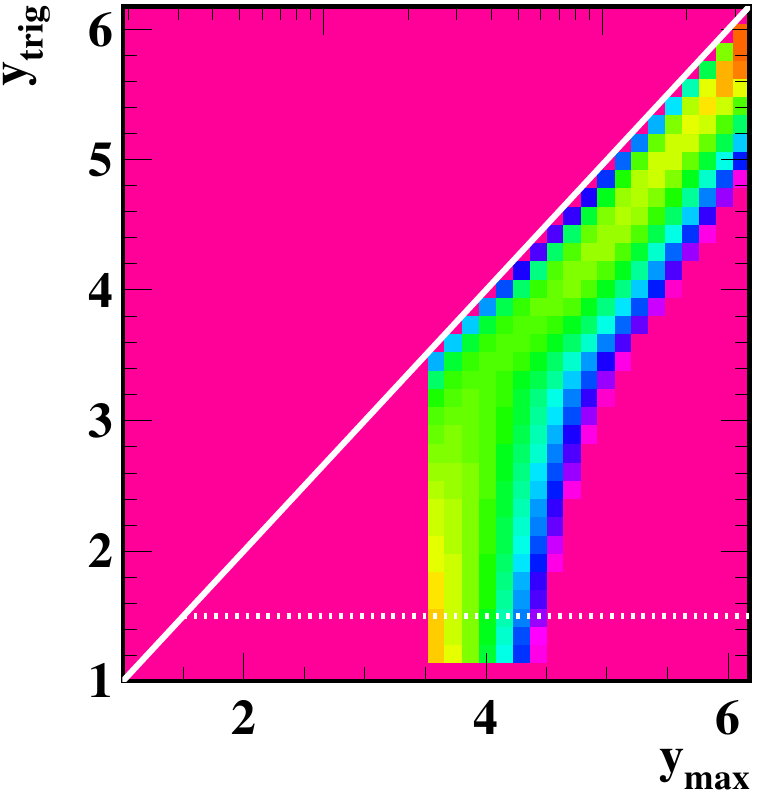}
\caption{\label{demo}
(Color online)  Conditional parton spectra $\hat S_{p}(y_{max}|y_{trig})$ for quark jets from \pp\ collisions (left) and gluon jets from \ee\ collisions (right). The z-axis limits (log scale) are 4 and 0.04. Although the FF ensembles for the two systems are quite different these joint distributions are similar. Measured jet-related TA structure does not extend below trigger rapidity $y_{trig} \approx 2.5$ ($p_{trig} \approx 0.85$ GeV/c).
 } 
 \end{figure}

Figure~\ref{demo} shows parton spectra $\hat S_p(y_{max}|y_{trig})$, unit normal on $y_{max}$ for each $y_{trig}$ condition, for quark jets from \pp\ collisions (left panel) and gluon jets from \ee\ collisions (right panel). Those result can be compared with Fig.~\ref{void} (c) where $\hat S_t(y_{trig}|y_{max})$ is approximately unit normal on $y_{trig}$ for each $y_{max}$ condition. The z-axis limits are the same in the two panels. The large difference between Figs.~\ref{void} (c) and \ref{demo} corresponds to the difference in marginal distributions. The trigger marginal is broad, with mode near $y_{trig} = 3$, whereas the parton marginal (jet spectrum) is steeply falling above a mode near $y_{max} = 3.75$. For given parton rapidity the trigger mode lies substantially below the kinematic upper bound, whereas for given trigger rapidity the parton mode lies just above the kinematic lower bound. 

 
Figure~\ref{demo} illustrates the relation between hadron trigger momentum (or $y_{trig}$) and conditional mean parton/jet energy (or $y_{max}$). The \pp\ collision energy imposes kinematic constraints on $y_{max}$ and therefore $y_{trig}$. For a trigger hadron with $y_{trig} \approx 5.5$ ($p_t \approx 15$ GeV/c) the most probable parton rapidity is $y_{max} \approx 6$ ($E_{jet} \approx 30$ GeV), implying valence quarks near the kinematic upper bound for $\sqrt{s} = 200$ GeV \pp\ collisions. For trigger rapidity $y_{trig} < 3.3$  ($p_t < 2$ GeV/c -- most triggers) the most-probable parton rapidity is $y_{max} \approx 3.75$ ($E_{jet} \approx 3$ GeV, minijets) near the lower bound of the parton spectrum. Over the typical  trigger  rapidity range considered in this study we encounter the extremes of parton flavor and hadronization: from valence-quark jets for 15 GeV/c triggers  to small-$x$ gluon minijets for 1-2 GeV/c triggers.

\subsection{Obtaining TA distributions from FFs} \label{taff}

The conditional distribution $\hat S_p(y_{max}|y_{trig})$ from Fig.~\ref{demo}, an ensemble of parton spectra approximately unit-normal on $y_{max}$, can be inserted into Eq.~(\ref{foldd}) (third line) to obtain (with full notation)
\bea \label{fold1}
D_a(y_{assoc}|y_{trig}) &=& \\ \nonumber
&& \hspace{-.75in}  \int_{y_{trig}}^\infty dy_{max} D_a(y_{assoc}|y_{max}) \hat S_p(y_{max}|y_{trig}),
\eea
an ensemble of associated-hadron FF components averaged over a subset of parton energies determined by the trigger hadron. For a given $y_{trig}$ condition all associated FFs $D_a(y_{assoc}|y_{max})$ are truncated so that $y_{assoc} < y_{trig}$. The 2D TA distribution then extends up to the diagonal, whereas $D_a(y_{assoc}|y_{max})$ in Fig.~\ref{void} (e) does not. Calculations of Eq.~(\ref{fold1}) are extended in this case to $y_x = 8$ for all rapidities. The result is then cropped to $y_x < 4.5$ to minimize the distorting effects at larger $y_x$ of an imposed upper bound on $y_{max}$. The fragment $y_x$ acceptance [1,4.5] corresponds to detected-hadron \pt\ interval [0.15,6] GeV/c, consistent with typical detector $p_t$ acceptance and data volumes (statistics limits at larger \pt). 








 \begin{figure}[h]
 \includegraphics[width=3.3in]{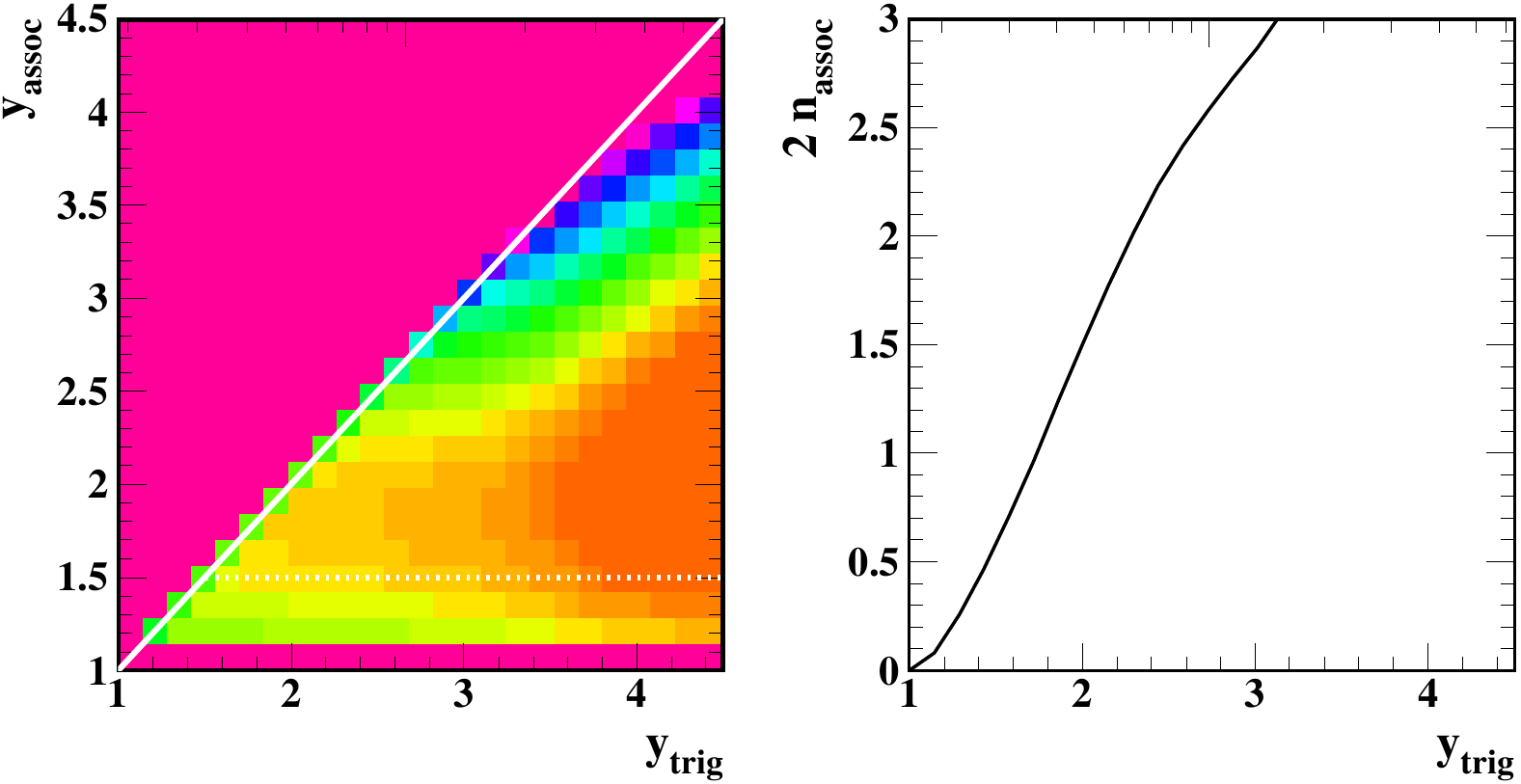}
  \put(-212,100) {\bf (a)}
 \put(-87,100) {\bf (b)}\\
\includegraphics[width=3.3in]{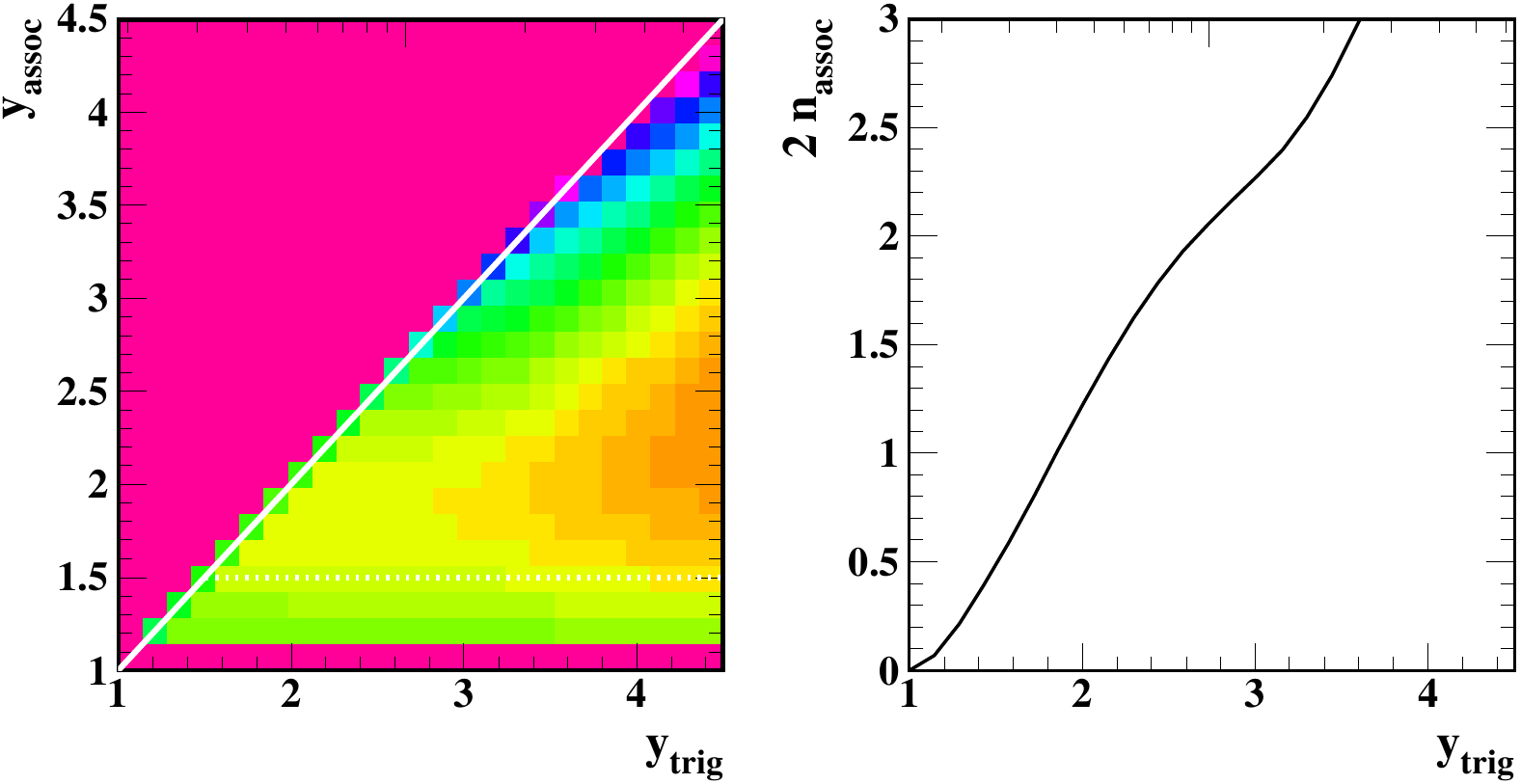}
 \put(-212,100) {\bf (c)}
\put(-51,110) {\scriptsize \bf 1}
 \put(-15,110) {\scriptsize \bf 4}
 \put(-56,102) {\scriptsize $\bf p_{trig} (GeV/c)$}
 \put(-87,100) {\bf (d)}
\caption{\label{tcrossa1}
(Color online) 
Associated-hadron distributions $D_a(y_{assoc}|y_{trig})$ conditional on trigger-hadron rapidity $y_{trig}$ for gluon jets (a) and quark jets (c) from \ee\ collisions (App.~\ref{ppeerat}). Right panels show projections onto $y_{trig}$ to obtain associated multiplicities $2 n_{assoc}(y_{trig}) \approx 2 n_{ch,j}(y_{trig}) - 1$. The left-panel z-axis limits (log scale) are 3 and 0.03.
 }  
 \end{figure}

 \begin{figure}[t]
 \includegraphics[width=3.3in]{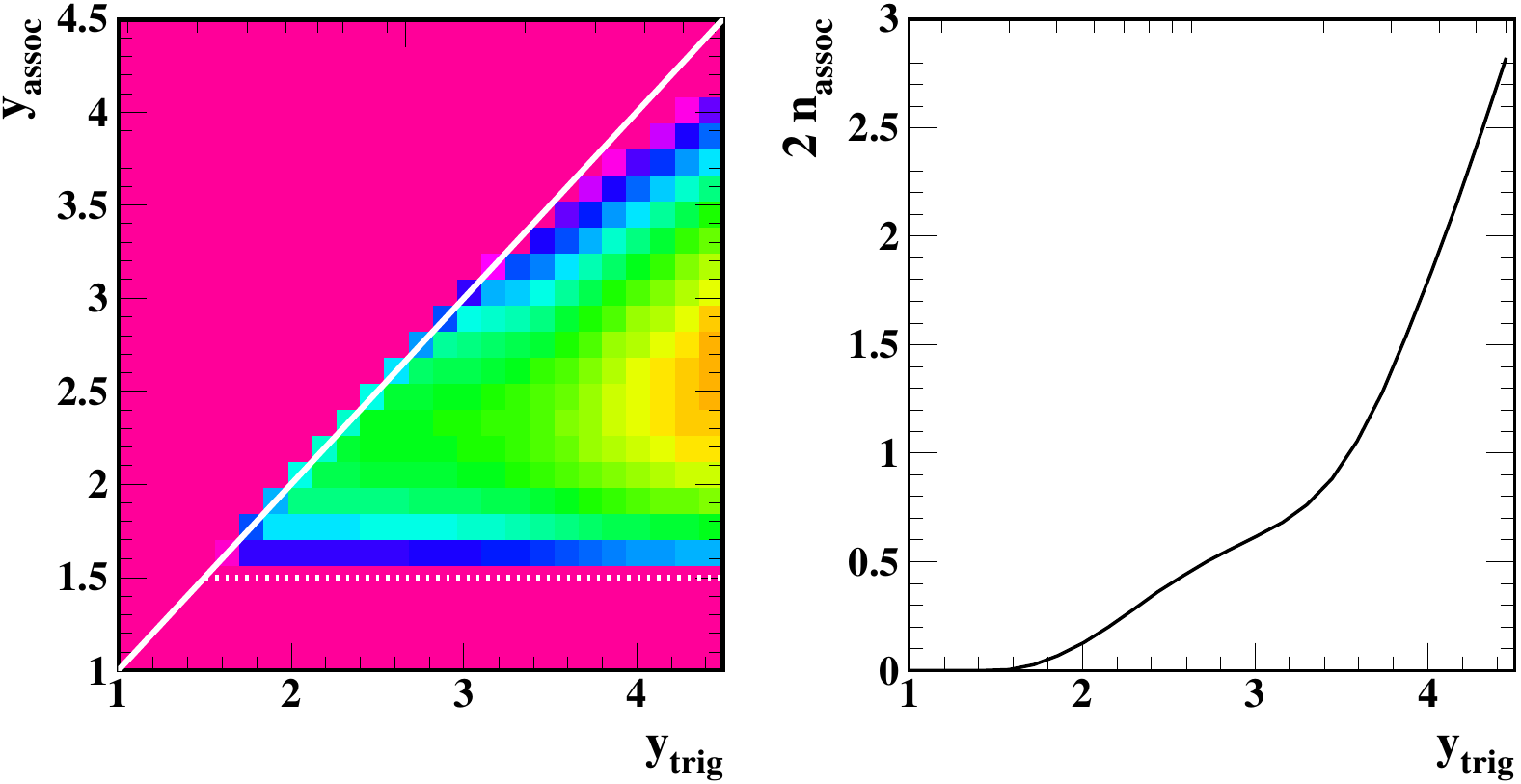}
 \put(-212,100) {\bf (a)}
 \put(-87,100) {\bf (b)}\\
 \includegraphics[width=3.3in]{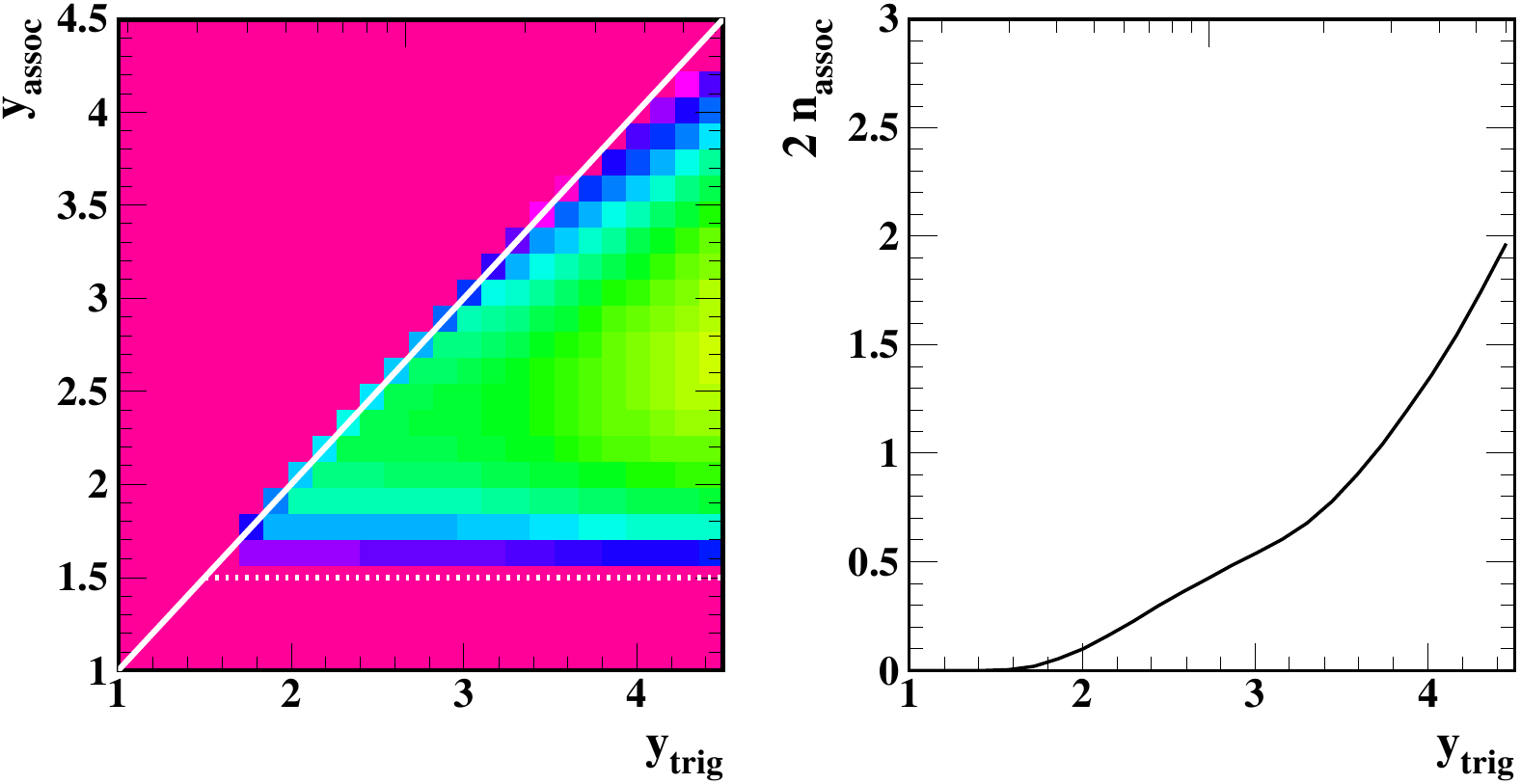}
 \put(-212,100) {\bf (c)}
 \put(-87,100) {\bf (d)}
\put(-51,110) {\scriptsize \bf 1}
 \put(-15,110) {\scriptsize \bf 4}
 \put(-56,102) {\scriptsize $\bf p_{trig} (GeV/c)$}
\caption{\label{tcrossa2}
(Color online) 
The same as Fig.~\ref{tcrossa1} except  from \ppbar\ collisions (App.~\ref{ppeerat}). Note the lower bound near $y_{assoc} = 1.5$ in comparison to $y_{assoc} = 1$ for \ee\ FFs. The left-panel z-axis limits (log scale) are 3 and 0.03.
}  
\end{figure}
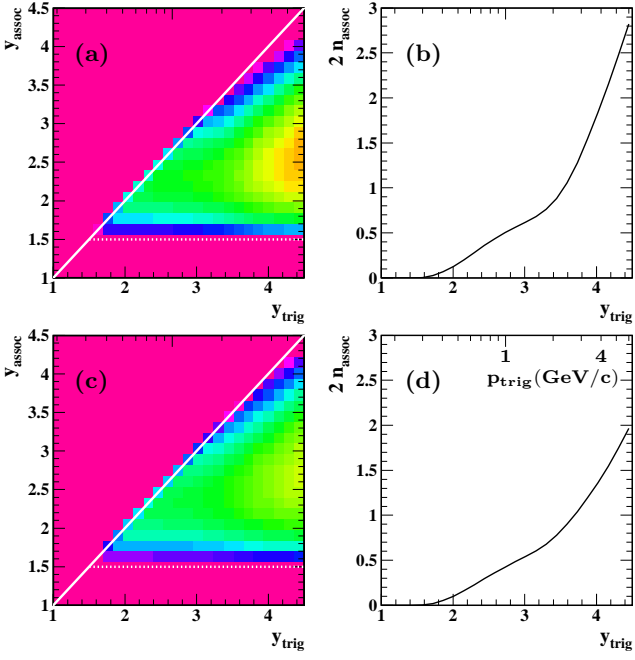


Figures~\ref{tcrossa1} and \ref{tcrossa2} show examples of FF TA distributions for four combinations of partons (light quarks or gluons) and collision systems (\ee\ or \pp\ collisions).  
 Figure~\ref{tcrossa1} shows results for gluon (a) and (b) and light-quark (c) and (d) FFs from \ee\ collisions, with FFs extending down to $y_{assoc} = 1$ (detector acceptance limit).  Panels (b) and (d) show projections onto $y_{trig}$ giving the associated-fragment yield per dijet $2n_{assoc}(y_{trig}) \approx 2 n_{ch,j}(y_{trig})-1$.
Figure~\ref{tcrossa2} shows equivalent results from \ppbar\ collisions. The effect of the FF cutoff near $y = 1.5$ (white dotted line) observed in \ppbar\ FF data is apparent. 
Generally, gluon FFs manifest substantially larger fragment multiplicities and lower modes on $y_{assoc}$ than light-quark FFs, as reflected in the structure of these 2D TA distributions.
  
We can check the consistency of the algebra leading to Eq.~(\ref{fold1}) by the following exercise. From Eq.~(\ref{ftpp}) we have
\bea \label{spst}
\hat S_p(y_{max}) &=& \int_0^{y_{max}} dy_{trig} \hat S_p(y_{max}|y_{trig}) \hat S_t(y_{trig})~~
\eea
as one marginal projection of $\hat F_{tp}(y_{trig},y_{max})$. Multiplying Eq.~(\ref{fold1}) through by factor $\hat S_t(y_{trig})$ and integrating over $y_{trig}$ establishes with Eq.~(\ref{spst}) the following relations
\bea \label{dayassoc}
D_a(y_{assoc}) 
&=& \int dy_{trig} D_a(y_{assoc}|y_{trig}) \hat S_t(y_{trig}) 
\\ \nonumber
&=& \int dy_{max} D_a(y_{assoc}|y_{max})\hat S_p(y_{max}).
\eea
The second line is the equivalent of Eq.~(\ref{sthat}) for FF associated-hadron component $D_a$, the common marginal projection from joint distributions  $F_{at}$ and $F_{ap}$.

\begin{figure}[h]
  \includegraphics[width=3.3in]{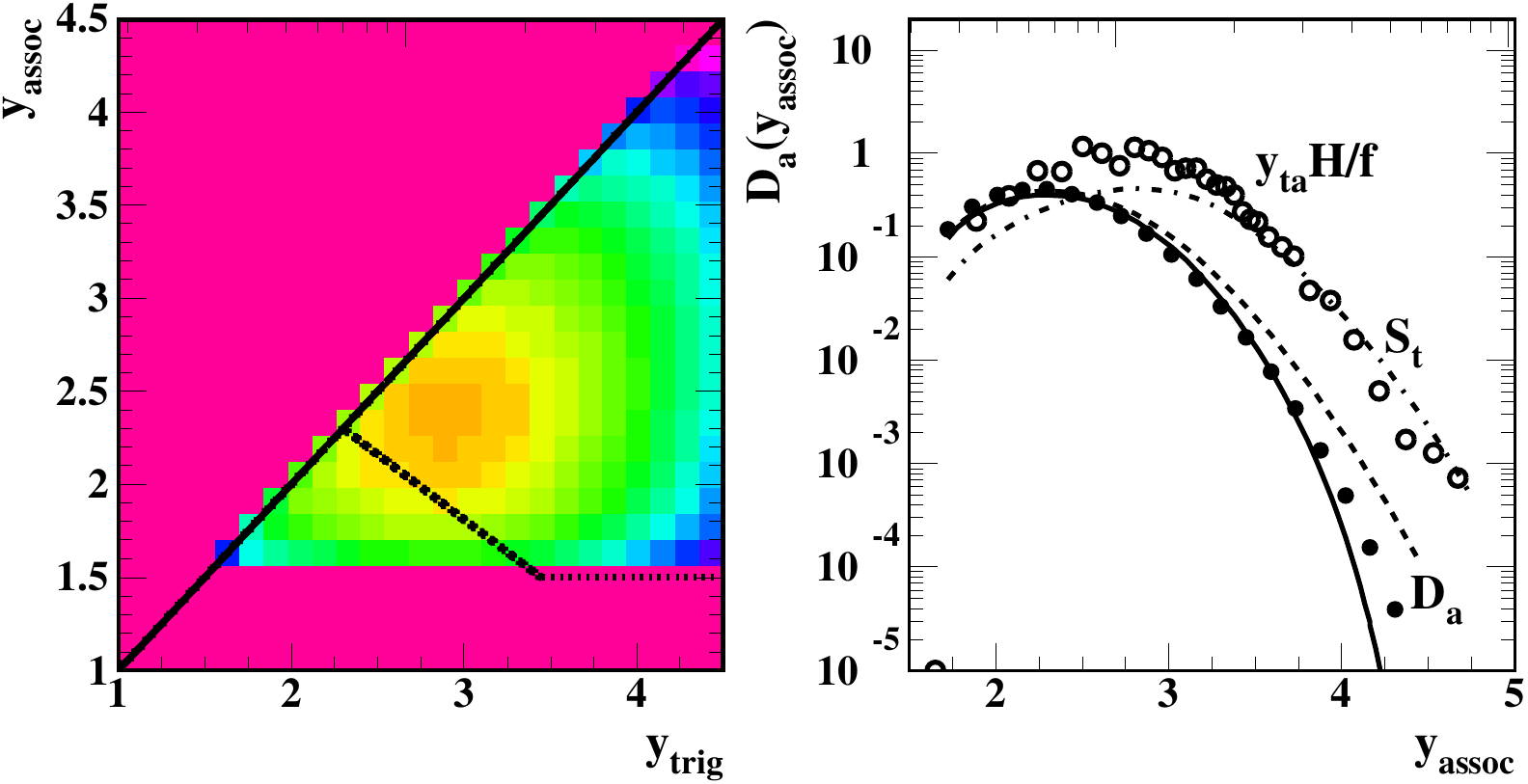}
\put(-67.2,109) {\scriptsize \bf 1}
 \put(-30,109) {\scriptsize \bf 4}
 \put(-78,103) {\scriptsize $\bf p_{assoc} (GeV/c)$}\\
\caption{\label{alephgg}
(Color online) 
Left panel:  Distribution $D_a(y_{assoc}|y_{trig})$ from Fig.~\ref{tcrossa2} (c) multiplied by  trigger fragment spectrum $\hat S_t(y_{trig})$ from Fig.~\ref{trigs} (right panel) to form the FF prediction for TA joint distribution $F_{at}(y_{assoc},y_{trig})$. The z-axis limits (log scale) are  0.5 and $5 \times 10^{-5}$.
Right panel: Projection of the histogram in the left panel onto $y_{assoc}$ (solid points) compared to $D_{a}(y_{assoc})$ from Eq.~(\ref{dayassoc}) (dashed and solid curves), the spectrum hard component from NSD \pp\ collisions (open circles) and the trigger spectrum $\hat S_t$ derived from \ppbar\ FFs and jet spectrum (dash-dotted curve).
 } 
 \end{figure}

Figure~\ref{alephgg} (left panel) shows FF TA joint distribution $F_{at}(y_{assoc},y_{trig})$ as the product of 
$D_a(y_{assoc}|y_{trig})$ from Fig.~\ref{tcrossa2} (c) and FF  trigger-fragment spectrum $\hat S_t(y_{trig})$ from Fig.~\ref{trigs} (right panel, solid curve). The predicted structure indicates that the great majority of jet fragments appears within $p = 0.5$ - 2 GeV/c ($y \approx 2$ - 3.3).

Figure~\ref{alephgg} (right panel) shows the projection $D_a(y_{assoc})$ of the histogram in the left panel $\hat S_t(y_{trig}) D_a(y_{assoc}|y_{trig})$ onto $y_{assoc}$ compared to two analytic expressions.  The solid points (2D histogram projection) represent the upper line of Eq.~(\ref{dayassoc}). The solid curve represents the lower line integrated within the same $y_{trig}$ interval confirming the approximate equivalence of the two forms.  The dashed curve is the same projection with $y_{trig}$ extended to 8 to reduce the marginal distortion at larger $y_{assoc}$ resulting from the limited $y_{trig}$ interval.

The open circles are the 200 GeV NSD \pp\ spectrum hard component in the form $y_{ta} H(y_{ta}) / f_{NSD}$ for comparison with the FF results. The spectrum hard component for hard events from Ref.~\cite{ppprd} can be expressed as
\bea
\Delta \eta H_h(y_{ta}) 
&=& G_{hs}  \hat T_{hh}(y_{ta}) + A_{hh}(y_{ta})
\\ \nonumber
 &\approx&(1 + \Delta n_j) \epsilon D_u(y_{assoc}) / y_{assoc}.
\eea
Thus $y_{ta} H / f_{NSD} = y_{ta} \Delta \eta P_h H_h/n_{j,NSD}  \approx y_{ta} \Delta \eta H_h \approx \epsilon D_u(y_{assoc}) = \hat S_t(y_{assoc}) + D_a(y_{assoc})$, since for NSD \pp\ collisions $P_h / n_{j} \approx 1$ and $\Delta n_j \approx 0$.

\section{FF Comparisons with  $\bf p$-$\bf p$ data} \label{datacomp}

Predictions derived from fragmentation functions can be compared with TA data from \pp\ collisions  in several ways, including 2D joint distributions and 1D marginal projections onto $y_{trig}$ and $y_{assoc}$.   
Such comparisons require a quantitative relation between data from \pp\ collisions and predictions from FFs as established in Sec.~\ref{dijet}.
Sections VI and VII assume single dijets within a specific detector acceptance with no soft component. Effects of soft components, limited \yt\ and angular acceptances and additional (untriggered) dijets on the inferred TA hard component must be accommodated in comparisons of FF predictions with real \pp\ data.

Hard events include both soft and hard components. At least one but possibly several dijets may contribute to the hard component, the number represented by quantity $1 + \Delta n_j(n_{ch})$. Each event includes a single trigger hadron which may come from the soft or hard component. If the trigger is soft there is no correlation with the dijet contribution (per the TCM). If the trigger is hard the associated hadrons from one dijet should be correlated with the trigger but not those from other dijets. Those several possibilities are considered in each subsection below.

\subsection{Jet correlations and p-p data}

In Sec.~\ref{tatcm} a two-component model was developed for trigger spectra and TA correlations based on the TCM for SP spectra. By construction the TA TCM includes no 2D correlations. For comparison of FF predictions with real data we must enumerate the possible jet-related correlations in data compared to other contributions, specifically for hard events that include a dijet contribution.

To relate the measured trigger-spectrum hard component to FF jet triggers we must account for multiple dijets in some hard events. The \pp\ spectrum hard component $H$ is related to the FF fragment distribution $D_u$ by 
%
\bea \label{yh}
y H(y,n_{ch}) &=& P_h yH_h \approx f(n_{ch}) \epsilon(\Delta \eta) D_u(y).
\eea
The possibility of multiple dijets in the same \pp\ hard event is represented by factor $\Delta \eta f (n_{ch})/ P_h(n_{ch}) = 1 + \Delta n_j$.  We then have
\bea \label{deletahh}
\Delta \eta y H_h(y,n_{ch}) &\approx& [1  + \Delta n_j(n_{ch})] \epsilon D_u(y),
\eea
where the second term on the right presents a contribution from secondary dijets not correlated with the single hard trigger in an event. The hadron spectrum hard component for event multiplicity $n_{ch}$ is thus related to FF trigger and associated components for a single dijet. 


For soft triggers in hard events the ensemble-average fragment number should be $[1 + \Delta  n_j(n_{ch})] \epsilon 2 \bar n_{ch,j} \approx n_{hh}$ uncorrelated with the trigger. For hard triggers with rapidity $y_{tt}$ the fragment number per event should be $\epsilon 2 n_{ch,j}(y_{tt}) - 1$ associated hadrons correlated with the trigger  and $\Delta  n_j(n_{ch}) \epsilon 2 \bar n_{ch,j}$ hadrons uncorrelated with  the trigger, where the triggered-dijet fragment multiplicity depends on the trigger rapidity (or parton energy).

\subsection{Hard component of 1D trigger spectrum $\bf \hat T(y_{tt})$}

Section~\ref{fftrig} describes a 1D trigger hard component derived from FFs based on fragments from single dijets appearing within acceptance $\Delta \eta$ and with no soft component. We now derive the additional factors required to relate FF predictions to  \pp\ data. From Eq.~(\ref{thhh}) $\hat T_{hh} = G_{hh} \Delta \eta H_h$ represents all potential triggers from dijets. From Eq.~(\ref{deletahh})  we have $G_{hh} = G_t G_t^{\Delta n_j}$, and from Eq.~(\ref{stdu}) we have $\hat S_t = G_t \epsilon D_u$. Given Eq.~(\ref{yh}) we conclude that
\bea \label{phytt}
P_h y_{tt} \hat T_{hh}(y_{tt},n_{ch}) / n_j(n_{ch}) &\approx& G_t(y)^{\Delta n_j(n_{ch})} \hat S_t(y)~~
\eea
is the required relation between the  hadron trigger hard component $y_{tt} \hat T_{hh}$ and the predicted fragment trigger spectrum $\hat S_t$ based on FFs. They should coincide approximately for a single dijet ($\Delta n_j \approx 0$). 

\begin{figure}[h]
  \includegraphics[width=1.6in,height=1.65in]{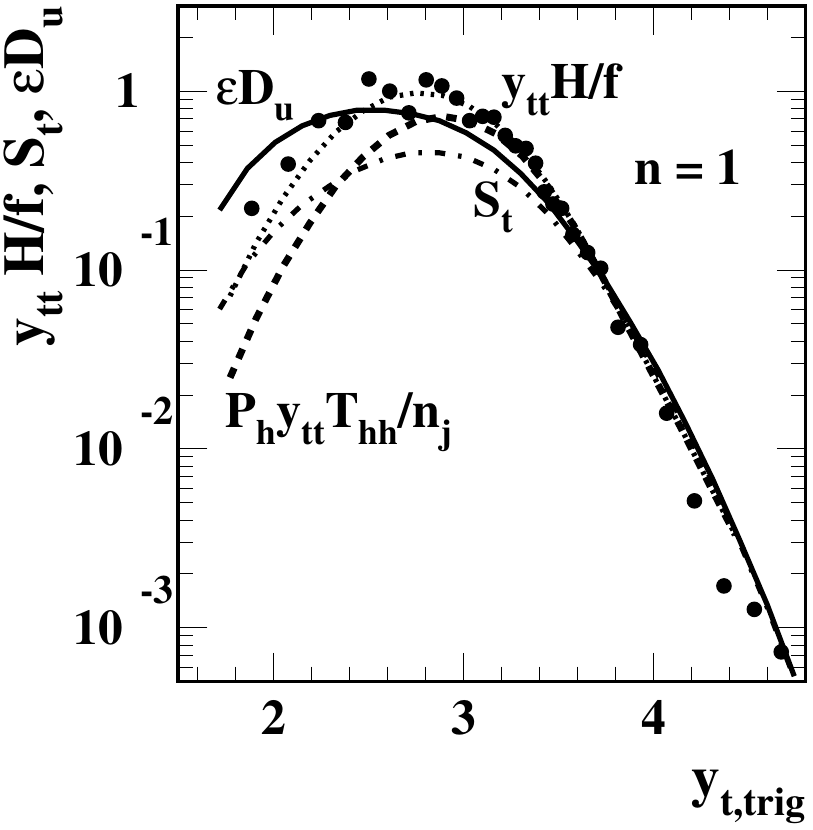}
  \includegraphics[width=1.6in,height=1.65in]{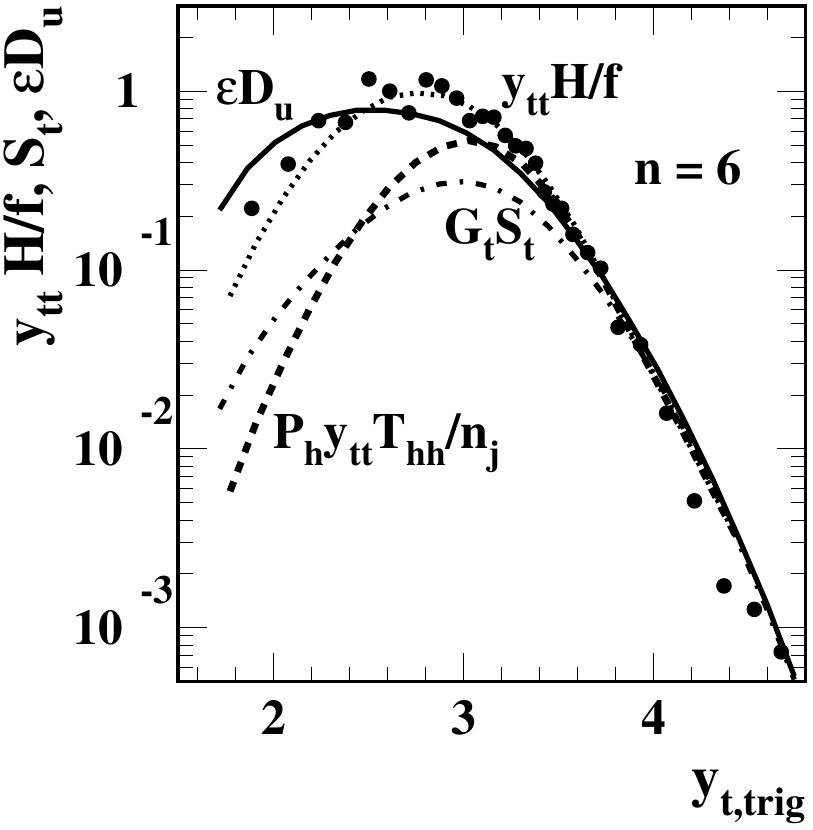}
\caption{\label{1dtrig}
TCM $\hat T_{hh}$ (dashed curve) and FF $\hat S_t$ (dash-dotted curve) trigger-spectrum models for two \pp\ multiplicity classes. Full spectrum hard-component models from the TCM (dotted curve) and FFs (solid curve) are compared with the spectrum hard component from NSD \pp\ collisions in the form $y_{tt}H/f$ (solid points)~\cite{ppprd}. The trigger-spectrum models differ in form for smaller $y_{t,trig}$ but vary together with increasing $n_{ch}$ in a manner consistent with Eq.~(\ref{phytt}).
 }  
 \end{figure}

Figure~\ref{1dtrig} (left panel) shows a comparison between the accepted fragment distribution $\epsilon D_u$ and fragment trigger spectrum $\hat S_t$ predicted from FF distributions for isolated dijets and the hadron- and trigger-spectrum hard components derived from 200 GeV \pp\ collisions for multiplicity class $n = 1$. The solid curve is the dijet total fragment distribution $\epsilon D_u$ within acceptance $\Delta \eta = 1$. The dash-dotted curve is the corresponding fragment trigger spectrum $\hat S_t$. The dotted curve is the hadron spectrum hard-component model $y_{tt}H(y_{tt},n_{ch})$ divided by dijet frequency $f(n_s)$ for a given multiplicity class, and  the dashed curve is the hadron trigger hard component in the form $P_h y_{tt} \hat T_{hh} / n_j$. The points are NSD \pp\ hard-component data from Ref.~\cite{ppprd}. For multiplicity class $n = 1$ the probability of a second dijet in a hard event is negligible ($\Delta n_j \approx 0$). The difference in shape between the predicted fragment-trigger and hadron-trigger spectra corresponds to the difference in the ``parent'' distributions $D_u$ and $y_{tt} H$ below $y_{tt} = 3.3$ ($p_t \approx 2$ GeV/c).

Figure~\ref{1dtrig} (right panel) shows  the same curves for multiplicity class $n = 6$. From Fig.~\ref{njph} we determine that $\Delta n_j \approx 1$. Based on Eq.~(\ref{phytt}) we plot $G_t^1 \hat S_t$ as the dash-dotted curve and find that the relation to $P_h y_{tt} T_{hh} / n_j$ is equivalent to that in the left panel modulo the differences in the parent distributions. Thus, Eq.~(\ref{phytt}) is a good description of the relation between the predicted and measured trigger spectra. Comparison of left and right panels indicates that as the number of dijets in a hard event increases significantly above 1 the void probability for smaller $y_{tt}$ values decreases accordingly.

\subsection{Hard component of 2D conditional $\bf A(y_{ta}|y_{tt})$} \label{ahardcomp}

Section~\ref{fftcm} describes a 2D TA hard component derived from FFs and is also based solely on fragments from single dijets appearing within acceptance $\Delta \eta$ with no soft spectrum component. We now derive additional factors and terms that relate FF TA predictions to \pp\ TA data, taking into account the presence of  soft components and the possibility of multiple dijets in hard events.

From the measured TA conditional distribution $A$ we first subtract the TCM  soft-component models
\bea \label{phrhahh}
P_h R_h A_{hh} &=& A - P_s R_s A_s - P_h R_h A_{hs}.
\eea
Because the hard-event trigger ratio has the structure
\bea
R_h &=& G_{hh} R_{hs} + G_{hs} R_{hh}
\eea
with soft and hard trigger components we obtain
\bea \label{rhahh}
R_h(y_{tt}) A_{hh}(y_{ta}|y_{tt})  &=& G_{hh}(y_{tt}) R_{hs}(y_{tt}) A'_{hh}(y_{ta}|y_{tt}) 
 \nonumber\\
&& \hspace{-.35in} +~ G_{hs}(y_{tt}) R_{hh}(y_{tt}) A^\dagger_{hh}(y_{ta}|y_{tt}),
\eea
where $A'_{hh}(y_{ta}|y_{tt},n_{ch}) \approx n_{hh} \hat H'_0(y_{ta}|y_{tt})$ with $n_{hh} = (1 + \Delta n_j)\epsilon 2\bar n_{ch,j} $ since there is no jet correlation with a soft trigger. 
$A^\dagger_{hh}$ in the second term representing hard triggers in hard events can be further decomposed as
\bea \label{yahh}
A^\dagger_{hh}(y_{ta}|y_{tt},n_{ch}) &=&A^*_{hh}(y_{ta}|y_{tt},n_{ch}) 
+ A''_{hh}(y_{ta}|y_{tt})~~~~~
\eea
with $A''_{hh}(y_{ta}|y_{tt})  \approx   \Delta n_j(n_{ch}) \epsilon 2 \bar n_{ch,j} \hat H_0'(y_{ta}|y_{tt})$ since the second term (from one or more MB secondary dijets) is not correlated with the trigger hadron. The first term $A^*_{hh}$ with mean fragment number $\epsilon 2 n_{ch,j}(y_{tt}) - 1$ corresponds to a single dijet correlated with the hard  trigger at $y_{tt}$. Introducing the ratios in the first line of
 \bea \label{xhs}
 X_{hs} &=& G_{hh} T_{hs} / T_h~~~~ X_{hh} = G_{hs} T_{hh} / T_h
 \nonumber \\
 A^*_{hh}&=& (A_{hh} - X_{hs} A_{hh}')/X_{hh} -  A_{hh}''
 \eea
gives the triggered-jet component in the second line.

To summarize, combining the TA TCM soft components with measured $\hat T$ and $A$ we isolate $R_h A_{hh}$ in Eq.~(\ref{phrhahh}). Further application of the TCM in  the form of $R_{hs}A'_{hh}$ (soft-trigger--dijet component) and  $A''_{hh}$ (hard-trigger--secondary-dijet component) in Eqs.~(\ref{rhahh}) and (\ref{yahh}) isolates $y_{ta}A^*_{hh}(y_{ta}|y_{tt},n_{ch})$ in Eq.~(\ref{xhs}) (hard-trigger--primary-dijet component), the data equivalent of $D_a(y_{assoc}|y_{trig})$ from FFs (hard-trigger--triggered-dijet component) that is the principal object of TA analysis.

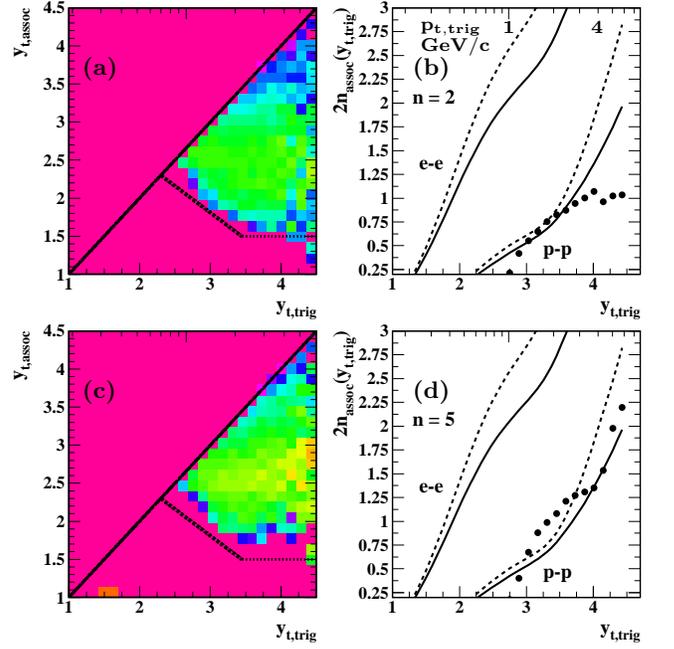
\begin{figure}[h]
   \includegraphics[width=3.3in]{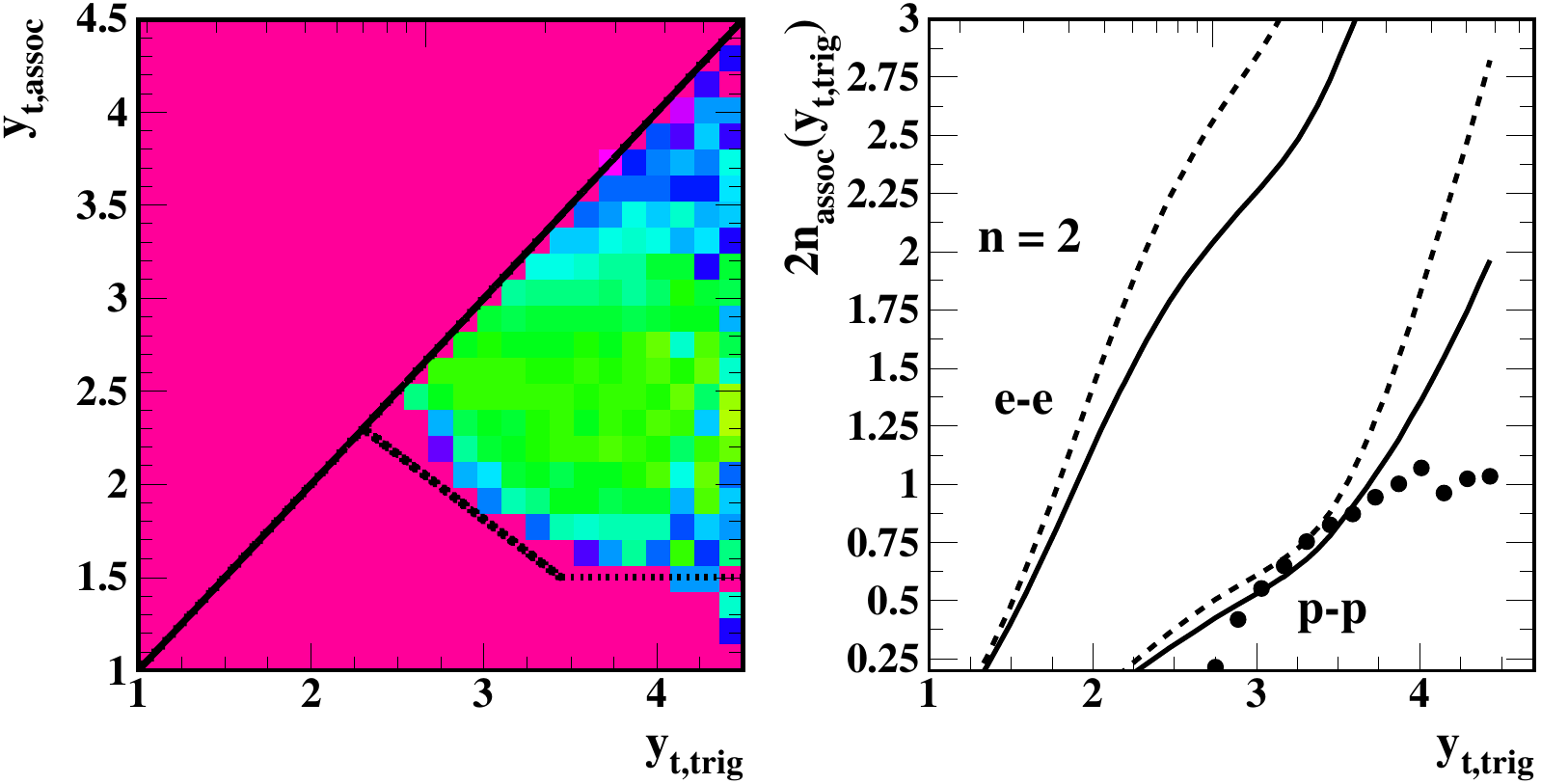}
\put(-212,93) {\bf (a)}
 \put(-87,93) {\bf (b)}
 \put(-52.9,109) {\scriptsize \bf 1}
  \put(-19.5,109) {\scriptsize \bf 4}
  \put(-84,110) {\scriptsize $\bf p_{t,trig}$}
 \put(-84,102) {\scriptsize $\bf GeV/c$}
\\
  \includegraphics[width=3.3in]{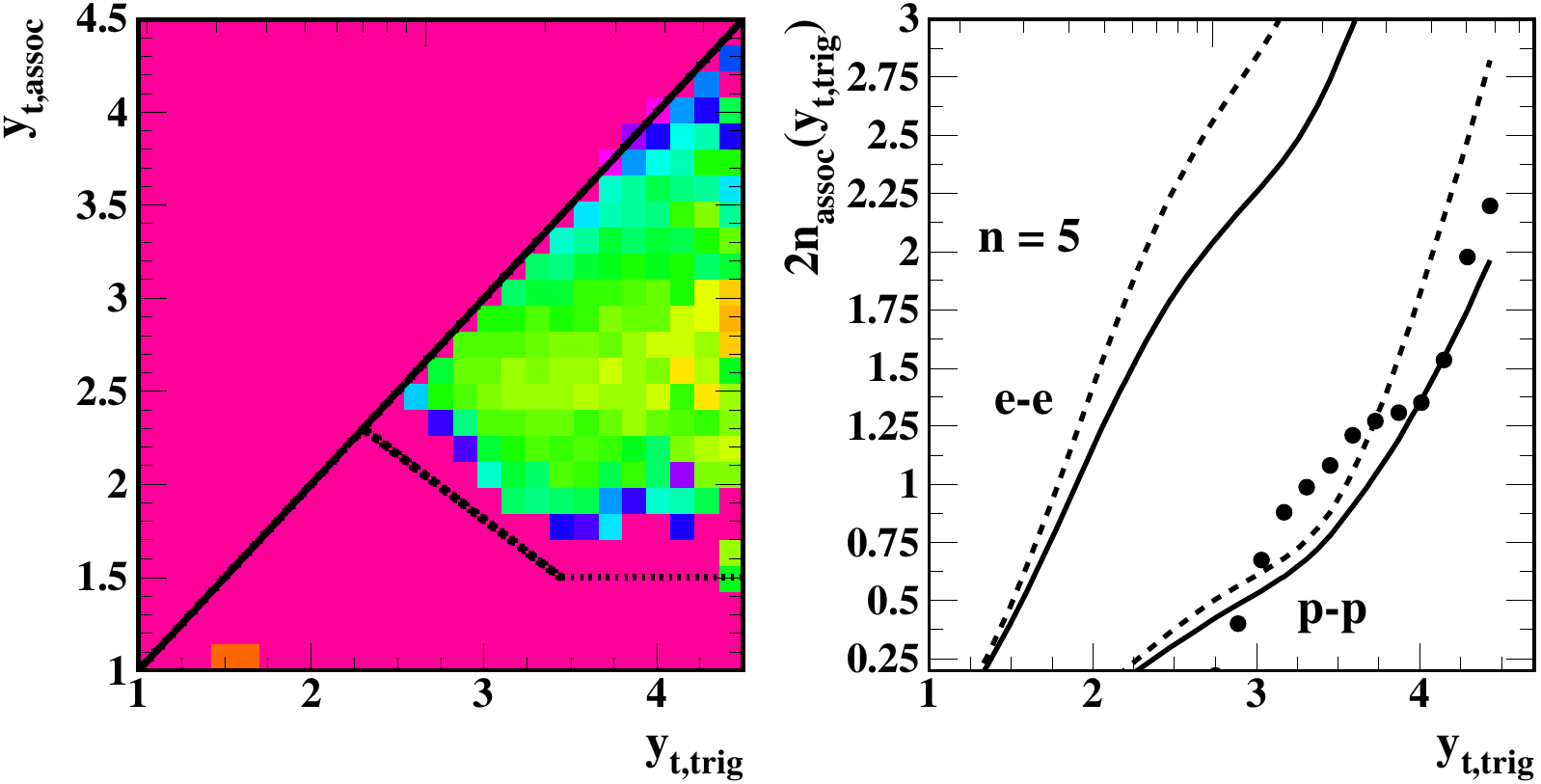} 
 \put(-212,93) {\bf (c)}
 \put(-87,93) {\bf (d)}
  \caption{\label{compare1}
 (Color online) 
Left panels:  
Associated-hadron conditional distribution $A^*_{hh}$ defined in Eq.~(\ref{xhs}) in the form $y_{ta}A^*_{hh}(y_{ta}|y_{tt},n_{ch})$ from 200 GeV \pp\ collisions for multiplicity classes n = 2, 5. The z-axis limits (log scale) are 3 and 0.03. The distributions are corrected for marginal-constraint distortions~\cite{pptrig}. The dotted reference lines are explained in the text.
Right panels: 
Projections $2n_{assoc}$ of the 2D histograms onto trigger rapidity $y_{t,trig}$ (points) compared to FF predictions from \pp\ and \ee\ collisions and for quark jets (solid curves) and gluon jets (dashed curves). Multiplicity classes 2 and 5 represent a factor 9 increase in dijet frequency per \pp\ collision.
  }  
  \end{figure}

 
Figure~\ref{compare1} (left panels) shows data distributions $y_{ta}A^*_{hh}(y_{ta}|y_{tt},n_{ch})$ from 200 GeV \pp\ collisions for  multiplicity classes $n = 2$, 5~\cite{ismd13,duncan} that can be compared with FF equivalents in Figs.~\ref{tcrossa1} and \ref{tcrossa2}.   The z-axis (log scale) limits are the same in the two figures. The dotted reference lines represent lower bounds typically observed for data TA hard components.

Figure~\ref{compare1} (right panels) shows 2D histograms in the left panels projected onto $y_{tt}$ to obtain (points)
\bea
2 n_{assoc}(y_{tt}) &\approx& \epsilon 2 n_{ch,j}(y_{tt}) - 1
\eea
corresponding to single triggered dijets within the angular acceptance.  Also shown for comparison are the FF projections from the right panels of Figs.~\ref{tcrossa1} and \ref{tcrossa2} for four combinations of parton type and collision system. For each of \ee\ and \pp\ systems the dashed curve is for gluons, the solid curve for light (e.g.\ valence) quarks. The comparison favors FFs from \ppbar\ collisions.
 
As noted above, for 6 GeV triggers ($y_{t,trig} \approx 4.5$) the most probable jet energy is approximately 10 GeV (Fig.~\ref{demo} -- left panel) or $x \approx 0.1$.  Given proton PDF structure we expect ($\approx$ valence) quark jets to make a substantial contribution in that region. 
For 1 GeV triggers ($y_{t,trig} \approx 2.7$) the most probable jet energy is near 3 GeV, and from the proton PDF structure we expect dominance of gluon jets for energy fraction $x \approx 0.03$, but the FF difference between gluon and quark jets is negligible for such low parton energies~\cite{eeprd}. 

For small $n_{ch}$ the dijet contribution (hard component) to hard events is comparable to and may even exceed the soft contribution (low-multiplicity hard events are very hard) in which case the dijet fragment multiplicity may be strongly biased by the imposed  \nch\ condition, as in panel (b). Since the hard fraction of hard events scales approximately as $1/n_{ch}$ the dijet contribution for larger \nch\ is a smaller fraction and may be less biased, consistent with panel (d) where TA data appear to favor FFs from \ppbar\ collisions and follow that $y_{t,trig}$ trend closely.

\subsection{Hard component of 2D joint $\bf F_{at}(y_{ta},y_{tt})$}

\begin{figure}[h]
  \includegraphics[width=3.3in]{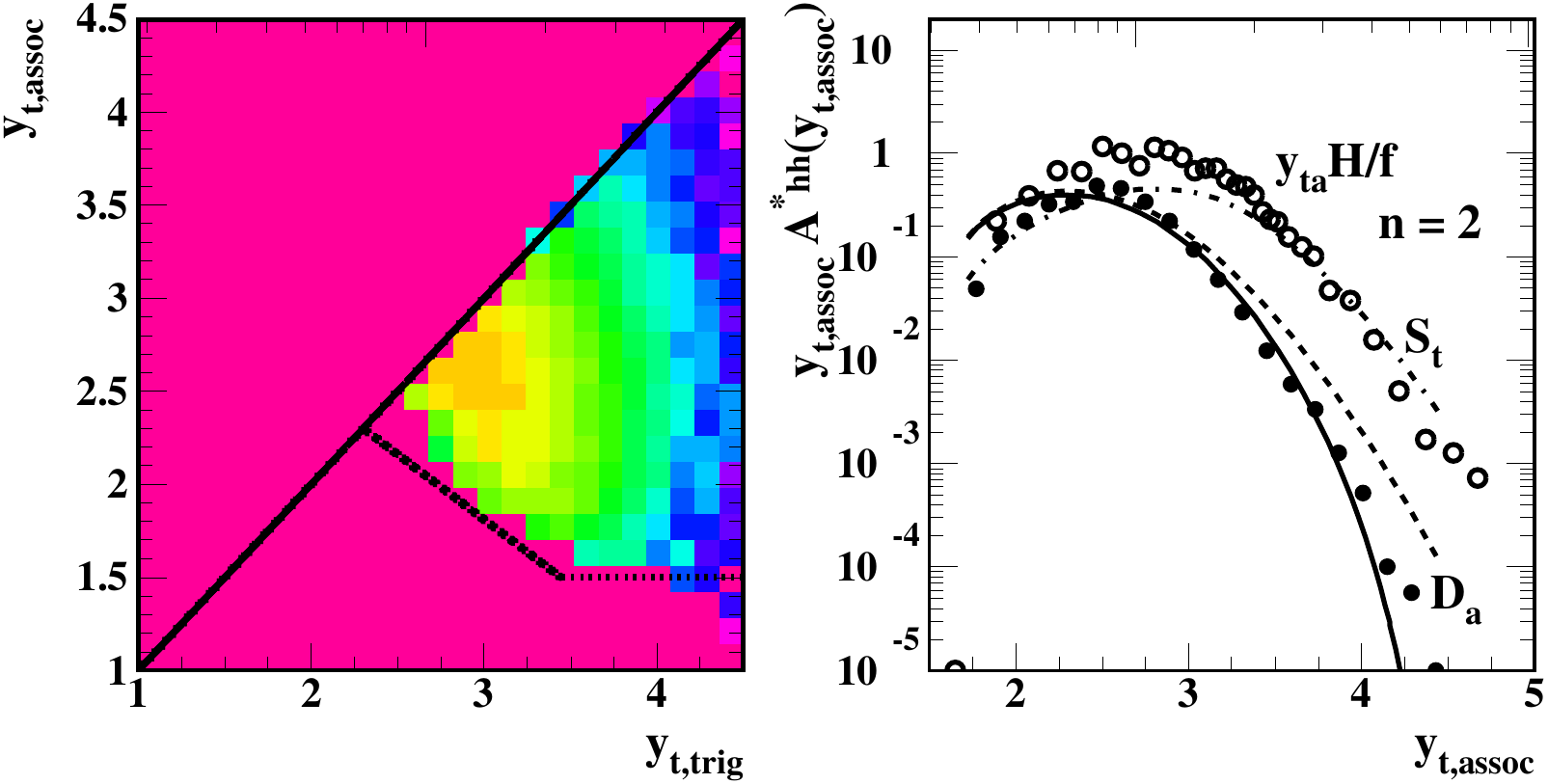}
\put(-212,108) {\bf (a)}
 \put(-90,108) {\bf (b)}
\put(-65.6,109) {\scriptsize \bf 1}
 \put(-30,109) {\scriptsize \bf 4}
 \put(-78,103) {\scriptsize $\bf p_{t,assoc} (GeV/c)$}\\
  \includegraphics[width=3.3in]{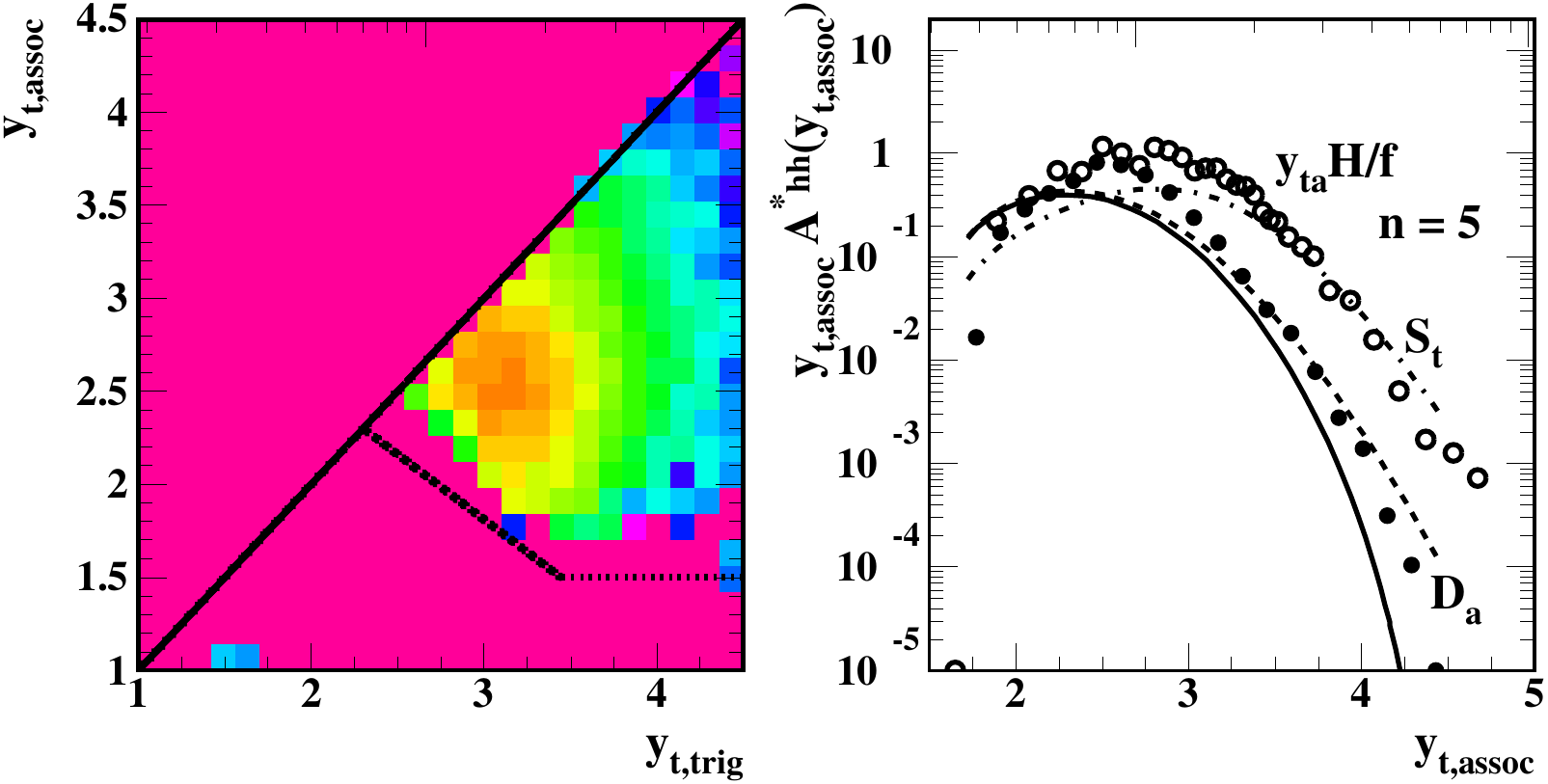}
\put(-212,108) {\bf (c)}
 \put(-90,108) {\bf (d)}
\caption{\label{compare2}
(Color online) 
Left panels:  
The per-dijet hard component of TA joint distribution $F_{at}$ in the form $\hat T_{h}(y_{tt},n_{ch}) y_{ta}A^*_{hh}(y_{ta}|y_{tt},n_{ch})$ from 200 GeV \pp\ collisions for multiplicity classes $n = 2$, 5. The z-axis limits (log scale) are 0.5 and $5\times 10^{-5}$, the same as Fig.~\ref{alephgg} (left panel). The distributions are corrected for marginal-constraint distortions~\cite{pptrig}.
Right panels: Projections of histograms in the left panels onto $y_{t,assoc}$ (solid points) compared to equivalent projections (solid and dashed curves) of $D_a(y_assoc|y_{trig})$ derived from measured FFs~\cite{eeprd,fragevo} and reconstructed-jet spectra~\cite{jetspec}. The 200 GeV \pp\ \yt\ spectrum hard component (open circles) and FF trigger spectrum $\hat S_t(y_{trig})$ are included for comparison.
 }  
 \end{figure}
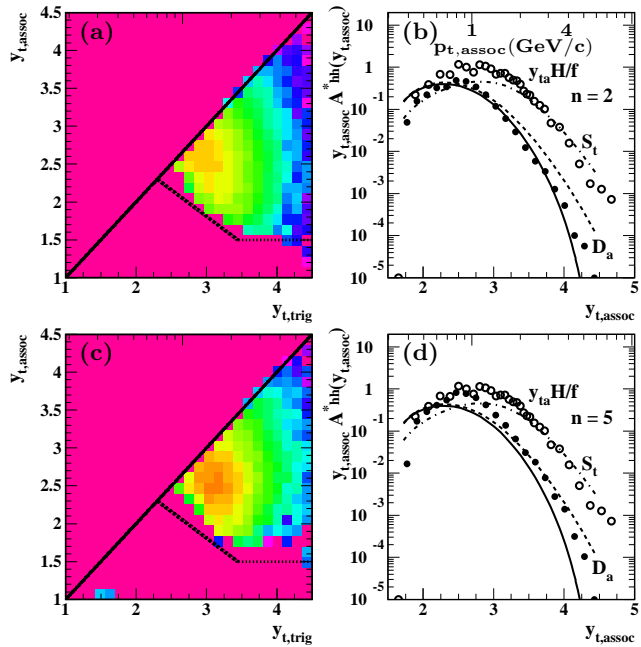

Figure~\ref{compare2} (left panels) shows the product $\hat T_{h}(y_{tt},n_{ch}) y_{ta} A^*_{hh}(y_{ta}|y_{tt},n_{ch})$ representing the {\em triggered-dijet} TA hard component of hard events from $F_{at}(y_{ta},y_{tt},n_{ch})$ in Fig.~\ref{hardcol2} (b). The TCM $P_h \hat T_{h}$ [bold dashed curve in Fig.~\ref{hardcol2} (a)] represents all hard-event triggers. From Eq.~(\ref{tadist}) we can define the $F_{at}$ hard component of hard events as $F_{hh} = P_h \hat T_h A_{hh}$. To some approximation $A^*_{hh} \leftrightarrow P_h A_{hh} / n_j$ so that $\hat T_{h} A^*_{hh} \approx  F_{hh}/ n_j$, the $F_{at}$ hard-event hard component per dijet. Whereas $F_{hh}/n_j$ may represent an average over multiple dijets per hard event $\hat T_h A^*_{hh}$ represents single triggered dijets in hard events.

Figure~\ref{compare2} (right panels) shows corresponding projections onto $y_{ta}$ (solid points) compared to the FF-predicted projection $D_a(y_{assoc})$ (solid, dashed curves) for quark jets from \ppbar\ collisions. As described in the text below Eq.~(\ref{fold1}), to reduce distortions from finite integration domains 2D TA predictions from FFs described in previous sections are calculated within rapidity spaces extended to $y_x = 8$, and the 2D histograms are then cropped to smaller intervals as required. The dashed curves in Fig.~\ref{compare2} (right panels) were produced with that procedure, but the data domains in the left panels are limited to $y_{tt} < 4.5$. The solid curves in the right panels were obtained by projecting across the reduced integration domain, and $D_a$ thus falls to zero near $y_{ta} = 4.5$.


Comparison of Fig.~\ref{compare2} (left panels) with Figure~\ref{alephgg} (left panel) demonstrates general similarity, including 2D mode positions and mean amplitudes (the z-axis limits are the same).  However, the 2D FF TA model does not match the observed data lower bounds (dotted lines in Figs.~\ref{compare1} and \ref{compare2}, left) that may reflect kinematic limits on low-energy jet formation.

The \pp\ TA data distributions are significantly more peaked near the 2D modes suggesting that the shape of the underlying parton spectrum model $\hat S_p(y_{max})$ near its lower bound could be more sharply peaked than the Gaussian model described in App.~\ref{parspec} , closer to the power-law spectrum model used in Ref.~\cite{fragevo} that terminates sharply near 3 GeV.  The jet (parton) spectrum appearing in Fig.~\ref{trigs} (right panel) as the dash-dotted curve and defined as a Gaussian shape on parton rapidity $y_{max}$ in App.~\ref{parspec} provides an accurate and comprehensive spectrum model for measured jet cross sections at jet energies above 5 GeV. However, there are no jet cross-section data below that energy where most MB jets are produced in high-energy nuclear collisions. Comparison of high-statistics \pp\ TA correlation data and FF predictions may help to refine the MB jet spectrum model near its lower bound.

Figure~\ref{compare1} emphasizes the $y_{tt}$ dependence of TA associated-particle distributions $A^*_{hh}$ compared to FF predictions, whereas Fig.~\ref{compare2} confirms that most jet-related triggers appear near $y_{tt} \approx 3$ ($p_{tt} \approx 1.6$ GeV/c) corresponding to 3 GeV minijets, and most associated fragments appear near $y_{ta} = 2.5$ ($p_{ta} \approx 0.85$ GeV/c).

 \section{Systematic uncertainties} \label{syserr}

\subsection{Internal consistency}

The accuracy of the FF-based TA TCM is indicated by the extent of its internal consistency. The TA TCM introduced in Ref.~\cite{pptrig} was based only on the \pp\ SP spectrum \nch\ dependence whose systematic uncertainties are discussed in the next subsection. In that case the application of probability analysis was rather simple. A consistency check is provided by various marginal projections compared to 1D SP spectrum components. The present study extends the TCM by including information derived from parton FFs and a jet spectrum via the probability chain rule and Bayes' theorem.

Decomposition of  the FF ensemble into trigger and associated components can be checked by comparison of the trigger+associated sum with the SP spectrum hard component as in Fig.~\ref{trigs} (right panel). The significant differences near the distribution mode are probably due to inaccuracy of the assumed jet spectrum near its lower bound, for which there are no jet cross-section data. The same discrepancy is apparent in Fig.~\ref{1dtrig} where nevertheless the relation in Eq.~(\ref{phytt}) is confirmed as a reasonable approximation. A check on the internal consistency of application of the probability chain rule and Bayes' theorem is provided by comparison of the two forms of Eq.~(\ref{dayassoc}) in Fig.~\ref{alephgg} (right panel). Finally, comparison of the FF-based TA TCM with data in Figs.~\ref{compare1} and \ref{compare2}  shows good agreement with the data for the \ppbar\ FFs.

\subsection{Implications from SP spectrum uncertainties}

The 1D TCM inferred from \nch\ dependence of SP spectra from 200 GeV \pp\ collisions~\cite{ppprd} provides the basis for the 2D TA TCM described in Sec.~\ref{tatcm}. Thus, some 2D systematic uncertainties arise from the 1D system.

The main source of systematic uncertainty in the inferred 1D SP spectrum hard component 
is the definition of the soft-component model $\hat S_0$ as a limiting case of  spectrum \nch\ variation. 
$\hat S_0$ is a rapidly-decreasing function in the interval \yt\  = 1.5 - 2.5 where the hard component becomes significant. The main effect of varying $\hat S_0$ model parameters is to change the magnitude of
$\hat S_0$ in that interval, shape changes being secondary. 
Certain limit criteria described in Ref.~\cite{ppprd} establish
stringent constraints on $\hat S_0$ already in the \yt\ interval 1.5-2, limiting
systematic uncertainty at \yt\ = 2 to $\pm$0.002 (1/3 of $\hat H_0$ at that point). The uncertainty range rapidly decreases above that point. Uncertainty in the inferred hard component is therefore greatest in that interval. Above \yt\ = 2.5 ($p_t \approx 1$ GeV/c) SP spectra are dominated by $H$ and the hard component  is  accurately defined in  that interval.

Some confirmation of the SP uncertainty estimate arises from comparison of the inferred SP spectrum hard component for NSD \pp\ collisions and a pQCD prediction as established in Ref.~\cite{fragevo}. Figure~\ref{trigs} (right panel) shows the SP spectrum hard component (solid points) and the corresponding pQCD prediction $\epsilon D_u$ (dashed curve) defined as a convolution of measured FFs and measured jet spectrum. The agreement is good. Below 2 GeV/c the pQCD calculation uncertainty is determined by uncertainty in the jet spectrum lower bound and uncertainty in the low-momentum structure of \ppbar\ FFs.

Implications for the  TA analysis are that 2D structure for $y_{ta} > 2.5$ is well defined whereas structure below $y_{ta} = 2$ is increasingly uncertain with decreasing $y_{ta}$ due to uncertainty in the subtracted TA soft-component model.

\subsection{Ratio comparison of TA TCM and TA data}

Direct comparison of TA data with the TA TCM as a ratio provides an indication of the quality of  the model over the entire kinematic domain whereas data-TCM differences such as those appearing in Fig.~\ref{compare1} do not.

\begin{figure}[h]
  \includegraphics[width=3.3in]{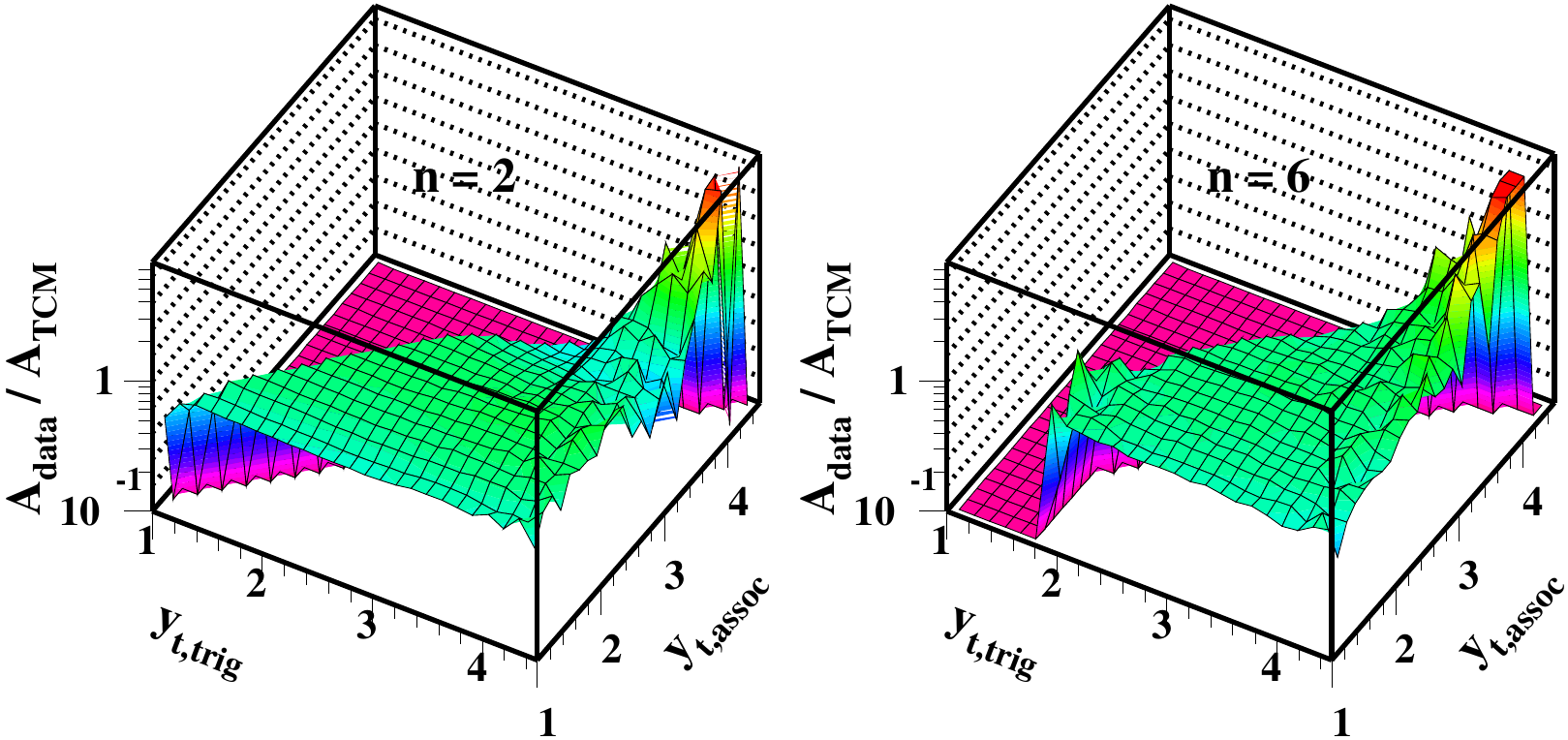}
\caption{\label{116bdub}
Ratio $A_\text{data} / A_\text{TCM}$ for two multiplicity classes. The soft components are well matched leading to a ratio $\approx 1$ for $y_{ta} < 2.5$. Above that point triggered-dijet correlations are apparent as strong deviations from the TA TCM.
 } 
 \end{figure}

Figure~\ref{116bdub} shows ratio $A_\text{data} / A_\text{TCM}$ for two multiplicity classes $n = 2$, 6 representing a  factor 4 multiplicity increase (factor 16 dijet rate increase). $A_\text{TCM}$ for $n = 5$ is shown in Fig.~\ref{hardcol2} (c). The ratio is approximately 1 (within 10\%) for all $y_{tt}$ and for $y_{ta} < 2.5$ indicating  that the TCM soft components provide a reasonable data model. The same soft reference applies to all event classes. For larger \nch\ (right panel) the empty bins at small $y_{tt}$ arise from lack of statistics for smaller event and trigger numbers.

If the TA data system were trivial (no correlations) we should expect the 2D ratio to be uniform across the entire kinematic domain. The deviations at larger $y_{ta}$ and $y_{tt}$ represent the desired nonfactorizable jet structure that is the object of TA analysis. The TCM hard component appearing in Fig.~\ref{hardcol2} (d) is uniform on $y_{tt}$ (modulo the marginal-constraint distortion at the left edge) and represents a marginal projection of the 2D TA hard component. The dijet contribution {\em correlated with a trigger particle} is less than the average for smaller $y_{tt}$ and greater than the average for larger $y_{tt}$ as shown in Fig.~\ref{116bdub}.



\subsection{Implications from $\bf n_{ch}$ dependence}

The TA hard component increases relative to the soft component linearly $\propto n_{ch}$. If there were a systematic bias arising from subtraction of TCM soft components to isolate the TA data hard component, as in Sec.~\ref{datacomp}, any such bias should scale approximately with the soft/hard ratio and should be most apparent near the hard-component  kinematic boundaries where that ratio is largest. No significant \nch\ bias is apparent in Figs.~\ref{compare1} and \ref{compare2} (left panels).


\subsection{Marginal-constraint distortions}

For a given multiplicity class \nch\ the associated particles for each trigger class $y_{tt}$ are constrained to sum to $\tilde n_{ch} - 1$. That constraint biases the data in at least two ways. First, the TCM model is distorted as illustrated in Fig.~\ref{hardcol2} (d) (increased amplitude along the left edge) compared to the input hard-component model that is uniform on $y_{tt}$ modulo the condition $y_{ta} < y_{tt}$. Comparison of the TCM hard component with the input generates a correction function $D_h(y_{tt})$ as noted in Ref.~\cite{pptrig}. The correction derived from the TCM has been applied to the $A^*_{hh}$ data in Figs.~\ref{compare1} and \ref{compare2} (left panels).

A second issue is the bias imposed on {\em triggered} dijets. Dijet structure should depend strongly on $y_{tt}$ as a proxy for the parton/jet energy. However, to some extent requirement of a fixed number of associated fragments must bias the fragmentation cascade (for instance to one higher-momentum trigger and a compensating reduced number of  lower-momentum fragments if $n_{ch}$ is small). That bias is suggested by comparing panels (b) and (d) of Fig.~\ref{compare1}. The bias should be most severe where the dijet contributes a major fraction of the event multiplicity, as for smaller event multiplicities (b), and the data seem consistent with that expectation.




 \section{Discussion} \label{disc}
 
 In this study we have established an algebraic connection between dijet TA systematics (via measured FFs and jet spectrum) and the hard component of hadron TA correlations from high-energy \pp\ collisions. A major goal of the study is a test of the hypothesis that the observed 2D TA hard component does represent dijet production, as already strongly suggested by 1D SP spectrum analysis.

Several other aspects of dijet production in high-energy \pp\ collisions can be addressed by comparing FF TA predictions with TA data: 
(a) What are the kinematic lower limits on dijet production (parton energy and fragment momentum) in nuclear collisions? 
(b) Are FFs universal as assumed in some treatments?
(c) What are the implications for underlying event (UE) studies and multiple parton interactions (MPI)?


\subsection{Kinematic limits on dijets in p-p collisions}

Dijet contributions to yields, spectra and correlations at lower fragment momenta and parton (jet) energies have been dismissed in the context of heavy ion collisions. It is argued that since a pQCD description is not well established in that kinematic domain jet interpretations are not supported by theory. In the absence of definitive theoretical predictions hadrons  below 2 GeV/c are conventionally attributed to emission from a thermalized flowing bulk medium. Such arguments are questionable for two reasons: (a) the systematics of reconstructed MB dijets and (b) accurate description of measured hadron hard components based on the same  jet systematics.

Appendix~\ref{fragfunc} presents FFs from isolated dijets measured accurately down to 0.1 GeV/c (\ee\ FFs) or 0.35 GeV/c (\ppbar\ FFs) fragment momentum. Complete FFs have been measured accurately down to $E_{jet} = 7$ GeV~\cite{tasso}, and jet fragment multiplicities have been measured down to $E_{jet} = 2.25$ GeV~\cite{cleo}. A simple parametrization describes FF evolution with parton energy down to 3 GeV~\cite{eeprd}. Although pQCD in the form of DGLAP equations~\cite{dglap} or the MLLA~\cite{mlla} describes FFs only down to the FF mode (0.7 - 2 GeV/c) the FF data are fully represented over the entire momentum acceptance of typical particle detectors. Appendix~\ref{parspec} presents jet spectra measured down to 4-5 GeV jet energy~\cite{ua1} and described accurately over a large range of jet and \ppbar\ collision energies by a simple QCD-inspired parametrization~\cite{jetspec}. The absence of a pQCD description over some part of that kinematic domain does not preclude real jet fragment production down to low hadron momenta.

In Ref.~\cite{fragevo} measured \ppbar\ FFs and a jet spectrum model derived from measured jet cross sections accurately describes spectrum hard components down to 0.35 GeV/c hadron (fragment) momentum from 200 GeV NSD \pp\ collisions~\cite{ppprd} and from \auau\ collisions over a range of centralities~\cite{hardspec}. In Sec.~\ref{datacomp} measured FFs and a MB jet spectrum are combined to predict hadron TA correlations  quantitatively over a large fraction of the 2D kinematic domain. Disagreements for small hadron momenta may reveal kinematic limits to jet production near 3 GeV jet energy and 0.35-0.75 GeV/c fragment (trigger and associated) momentum, providing unprecedented access to the details of low-energy jet formation and confirming the importance of dijet production at low hadron momenta..

\subsection{e-e vs p-\=p FFs and FF universality} \label{ffcompare}

QCD factorization allows the separation of long-range (soft) and short-range (hard) processes in pQCD calculations of cross sections~\cite{factor}. Soft processes are measured and hard processes are predicted. In pQCD calculations of hadron production via hard parton scattering to jets measured fragmentation functions represent (mostly soft) parton fragmentation, and it is typically assumed that the FFs are universal, independent of the particular context in which they are applied. Universality has been tested in various contexts including \ppbar\ collisions as in Ref.~\cite{kniehl} where it is concluded that ``...our global analysis of inclusive charged-hadron production provides evidence that both the predicted scaling violations and the universality of the FFs are realized in nature.''

However, universality tests in the context of \pp\ or \ppbar\ collisions involve comparisons between predictions based on FF ensembles inferred from \ee\ collisions and hadron \pp\ SP spectra where application of a factorization theorem and assumed FF universality applies only  to the jet-related spectrum hard component. The sensitivity of such tests is then reduced at lower \pt\ by the dominant spectrum soft component resulting from projectile proton dissociation~\cite{ppprd,hardspec}. Comparisons are typically not extended below hadron $p_t \approx 1$ GeV/c where discrepancies of a factor 2 or more may already be apparent.

In Fig.~\ref{ppffs} a direct comparison between \ee\ and \ppbar\ FFs inferred from event-wise jet reconstruction shows large deviations below 3-4 GeV/c where the \ppbar\ FFs fall well below the \ee\ FFs~\cite{eeprd,fragevo}. One could argue that jet reconstruction in \ppbar\ collisions may lead to inefficiencies for smaller hadron momenta and larger angular distances from the jet axis due to a limited cone radius.  However, direct comparisons between fragment distributions based on \ppbar\ FFs and hadron spectrum hard components, where jet reconstruction does not play a role, appear to confirm the discrepancy~\cite{ppprd,hardspec,fragevo}. Differential study of hadron SP spectrum structure via a two-component analysis and direct comparison of measured FFs from \ee\ and \ppbar\ collisions with the isolated spectrum hard component suggests that FF universality is strongly violated below fragment momentum 2 GeV/c. 

Universality may be tested  more differentially with TA correlation analysis of \pp\ collision data. Figure~\ref{compare1} (right panels) indicates that the ratio of dijet-associated yields for \ee\ vs \ppbar\ FFs is approximately a factor 4-5 near $y_{tt} = 3$ where the majority of jet fragments emerge, as indicated in Fig.~\ref{compare2} (left panels). The preliminary \pp\ TA data in Fig.~\ref{compare1} strongly favor  the \ppbar\ FFs that closely follow the data trends. Universality might be argued for larger parton energies and fragment momenta, but near  the kinematic limits of jet production FF universality is strongly violated. 2D TA results  are consistent with similar indications from previous 1D spectrum analysis~\cite{fragevo}.

\subsection{Relation to UE studies and interpretations}

The underlying event (UE) in hard \pp\ collisions (including at least one dijet within the acceptance) has been studied extensively in the context of searches for multiple-parton interactions (MPI)~\cite{mpi}. It is assumed that relative to the triggered-jet axis the azimuth {\em transverse region} (TR, $||\phi - \phi_{trigger}| - \pi/2| < \pi/6$) includes no contribution from the triggered dijet and therefore provides diagnostic measures for UE contributions thought to represent beam-beam remnants (BBR, fragments from projectile-nucleon dissociation) and MPI~\cite{rick,cdfue,cmsue}. 

TR properties exhibit characteristic variations with changing  trigger condition over a range of \pp\ or \ppbar\ collision energies. Some TR trends have been interpreted in terms of a growing probability of MPI resulting from increased \pp\ centrality. With increasing trigger \pt\ the TR hadron multiplicity (or \pt\ sum) first increases rapidly and then saturates, the saturation value depending strongly on collision energy. 
A \pp\ centrality increase is inferred from the TR multiplicity increase  and associated with increased dijet production~\cite{mpi,frankfurt}.  The TR \pt\ spectrum includes a hard component~\cite{cdfue}.  Given conventional UE assumptions and inferred centrality trend the TR  hard component is attributed to non-triggered dijets (MPI). 

However, conventional assumptions that the UE does not include a contribution from the trigger dijet can be questioned~\cite{pptheory}. Analysis of dijet 2D angular correlations reveals that MB (mainly low-energy) dijets contribute substantially to the TR and that higher-energy dijets include the same MB low-momentum, large-angle base structure that may be excluded from conventional event-wise jet reconstruction. The TR region should therefore include a hard contribution from the triggered dijet. 

Analysis of \pp\ \nch\ dependence indicates that a \pt-based jet trigger probably selects low-multiplicity hard events where MPI are  unlikely, whereas imposition of a large-\nch\ condition would make MPI quite probable (dijet frequency per \pp\ collision increases  as $n_{ch}^2$~\cite{ppprd,pptheory}).

The TA analysis introduced in Ref.~\cite{pptrig} may contribute to UE studies in several ways: (a) Given a TA azimuth reference the jet-related TA correlation structure within the TR becomes accessible. Is there a triggered-dijet contribution to the TR, especially in low-multiplicity events with negligible secondary dijets? (b) In higher-multiplicity events is there a nontriggered-dijet contribution to the TR uncorrelated with the trigger? (c) Are jet-related charge correlations consistent with MPI? Triggered-jet same-side TA correlations should exhibit strong charge correlations, whereas untriggered secondary dijets in the same azimuth interval should not.

FF universality also relates to UE studies. If \ppbar\ FFs were equivalent to \ee\ FFs with their much larger lower-momentum (and hence larger-angle) fragment density the TR should include a much larger contribution from the triggered dijet, as suggested by Fig.~\ref{compare1}. Even for \ppbar\ FFs the triggered-dijet contribution is substantial.

\subsection{Triggered dihadron correlations}

So-called triggered dihadron correlations~\cite{starta} have certain features in common with the TA analysis described in Ref.~\cite{pptrig} and the present study. For a specified class of collision events the highest-momentum hadron in each event falling within a restricted {\em trigger} \pt\ interval is paired with each one of that fraction of the other hadrons falling within an {\em associated} \pt\ interval. The hadron pairs meeting those trigger-associated \pt\ conditions (cuts) are then used to construct angular correlations on 1D azimuth $\phi$ or 2D $(\eta,\phi)$. The trigger hadron is assumed (with some probability) to be a proxy for the leading parton of a jet, and some of the associated hadrons may be fragments of that jet. Based on certain strong assumptions a combinatoric background model function is subtracted from the sibling (same-event) pair distribution to isolate the nominal jet-related correlation structure~\cite{tzyam}.

Although the terminology is similar and the goals are related (identify jet-related correlation structure) the details and results are quite different. The object of TA correlation analysis as described in Ref.~\cite{pptrig} and the present study is MB jet-related correlation structure distributed on trigger and associated \pt\ or \yt\ with no restrictions on the hadron momenta. Thus, {\em all MB jet structure} is identified. The subtracted soft-component model is based on 1D spectrum analysis of \pp\ collision data where the complementary hard component has been quantitatively  confirmed as representing all MB QCD jet structure.

In contrast, triggered-dihadron correlations include restrictive \pt\ cuts that select {\em only a small fraction} of the MB jet structure. For instance, a typical cut combination is $p_{t,trig}\in [4,6]$ GeV/c ($y_{t,trig} \in [4,4.5]$) and $p_{t,assoc}\in [2,4]$ GeV/c ($y_{t,assoc} \in [3.3,4]$)~\cite{starta} corresponding to a small rectangle at upper right in Fig.~\ref{compare2} (left panels) including a tiny fraction of the total MB jet fragments and corresponding to jet energies near 7 GeV according to Fig.~\ref{demo} (left panel). Angular correlations from MB jet fragments that survive such cuts are then subjected to so-called ZYAM (zero yield at minimum) background subtraction 
(assuming that there is no overlap of intrajet and interjet azimuth correlation peaks) that may remove a further substantial fraction of the surviving jet structure, biasing and distorting the result~\cite{tzyam}.

\subsection{Hard-component stability with varying $\bf n_{ch}$}

In Sec.~\ref{dijet} the dijet production rate is  $n_j \propto n_s^2 \approx n_{ch}^2$. Thus, charge multiplicity provides strong control of dijet rates and MPI. Typical \pp\ data volumes insure a usable factor 10 increase of \nch\ relative to NSD \pp\ collisions and therefore a factor 100 increase in the dijet production rate, from an average few-percent probability per NSD collision (within $\Delta \eta \approx 2$) to two or more dijets in each collision. In Figs.~\ref{compare1} and \ref{compare2} there is a factor 9 increase in dijet rate between multiplicity classes $n = 2$ and 5. 

If multiple dijets were coupled in some way (as some MPI scenarios suggest) we might expect to observe a quadratic dependence of coupling effects corresponding to the dijet production rate per event. No such dependence is evident in  data. The TA hard (dijet) component is remarkably stable over a large dijet frequency range.

In Fig.~\ref{compare1} (right panels)  we do see possible indications of selection bias due to the event multiplicity. For the lowest-multiplicity events the hard-component multiplicity of hard events is comparable to the soft-component multiplicity. The imposition of an event multiplicity constraint may then  bias the hard component, and preliminary data in Fig.~\ref{compare1}  (b) suggest that is the case: $2 n_{assoc}$ is nearly independent of $y_{tt}$.  For larger event multiplicities the hard-component fraction falls toward 10\% and bias from the multiplicity constraint may be substantially reduced, as suggested by Fig.~\ref{compare1}  (d) where $2 n_{assoc}(y_{tt})$  follows the TA prediction from \ppbar\ FFs.

\subsection{p-p TA data as a reference for dijets in A-A}

Analysis of yields, spectra and correlations from \pp\ collisions provides a testing ground for any theoretical description of  soft and hard QCD processes. Improved understanding of QCD in \pp\ collisions should provide a more accurate reference for novel physics in \aa\ collisions and LHC searches for physics beyond the Standard Model.  Current \pp\ issues include UE analysis and interpretation, dijet production and angular structure, the role of MPI, the relevance of \pp\ centrality, possible partonic collectivity in \pp\ and \pa\ collisions, Monte Carlo modeling and the TCM for hadron production.  Although a major effort has been devoted to \pp\ measurements and theoretical analysis a number of issues remain unresolved.

Conventional \pp\ vs \aa\  comparisons of dijet production have been quite  limited, for instance spectrum ratio $R_{AA}$ or dihadron azimuth correlations with trigger-associated \pt\ cuts and ZYAM subtraction that typically access a small fraction of all  jet fragments within a small fraction of momentum space~\cite{tzyam,fragevo}. Such restricted comparisons are contrasted with MB analysis (no \pt\ conditions) of yields, spectra and correlations in the TCM context where almost all jet fragments within a detector acceptance are addressed quantitatively~\cite{ppprd,hardspec,fragevo,porter2,porter3,anomalous,jetspec}.

TA analysis as described in Ref.~\cite{pptrig} further extends the  \pp\ reference system for comparisons with similar analysis of \aa\ collisions. In previous work we established a pQCD context for the TCM of spectra and angular correlations in \pp\ and \aa\ collisions~\cite{ppprd,hardspec,fragevo,jetspec}. In the present study we extend the pQCD description to jet-related hard components of TA correlations from \pp\ collisions. Follow-up TA analysis of \pa, \da\ and \aa\ collisions may reveal  changes in the TA TCM (and pQCD) description required to accommodate modified dijet production in larger collision systems.

\section{Summary}\label{summ}


The extent of  dijet contributions to hadron production in high-energy nuclear collisions has been strongly questioned recently. Nominal jet manifestations in small as well as large collision systems have been reinterpreted as representing collective flows, the paradigm shift motivated in part by {\em a priori} assumptions about kinematic and theoretical limits on jet production as a QCD phenomenon. Resolving the apparent conflict between opposing interpretations requires a more complete reference for minimum-bias dijet production in \pp\ collisions---directly linked to measured in-vacuum jet properties---that can be applied to \pa, \da\ and \aa\ collisions.


In previous studies certain features of hadron spectra and angular correlations from 200 GeV \pp\ collisions were related quantitatively to measured minimum-bias QCD jet systematics. A hadron spectrum hard component was predicted by convoluting measured parton fragmentation functions (FFs) with a measured minimum-bias jet spectrum. A hard component of 2D angular correlations was in turn related to the 1D spectrum hard component and thereby to QCD dijets. Complementary  transverse-momentum or -rapidity (symmetrized $y_t \times y_t$) correlations also include a jet-related hard component, but a quantitative connection to QCD jets has not been established. 


In a more-recent study we developed a two-component model (TCM) for   trigger-associated (TA) correlations.  Asymmetric TA rapidity correlations on $y_{ta}\times y_{tt}$ are distinct from but closely related to symmetrized correlations on $y_t \times y_t$.  Subtracting the TCM soft component from measured TA correlations for \pp\ collisions should reveal a hard component representing all fragments from all jets appearing within some detector acceptance.

In the present study we distinguish a TA hard component representing triggered dijets from that for secondary dijets accompanying the triggered dijet (corresponding to multiple parton interactions (MPI). We derive a quantitative relation among TA hadron correlations, measured FFs and a MB jet spectrum. To establish a connection between  TA hard components inferred from \pp\ collisions and jet measurements we first partition measured FFs  from \ee\ and \ppbar\ collisions into trigger and associated components. We  combine the trigger and associated FF components with a measured MB jet spectrum according to the probability chain rule and Bayes' theorem to define theoretical QCD (FF) predictions for  \pp\ TA hard components. We then compare the FF predictions with preliminary TA data from 200 GeV \pp\ collisions.


We find quantitative agreement between measured TA correlations and  TA predictions derived from measured \ppbar\ FFs and a MB jet spectrum inferred from \ppbar\ collisions. Predictions based on \ee\ FFs strongly disagree with the \pp\ TA data, challenging assumptions about FF universality. The discrepancy cannot be attributed to differences in jet-finding algorithms or to the \ppbar\ underlying event, since TA correlations include all fragments from all dijets within the acceptance. Kinematic lower bounds on dijet energy and fragment momentum inferred from these comparisons are substantially lower than conventional assumptions about dijet production in nuclear collisions and confirm a jet-spectrum lower bound near 3 GeV. From these TA results we confirm that {\em most jet fragments are produced with} $p_t < 2$ GeV/c ($y_t \approx 3.3$).


TA correlations from \pp\ collisions combined with QCD predictions from the present study may further clarify the quantitative connection between QCD jets and measured MB spectrum and correlation structures. The TA hard component represents all jet contributions, not just a biased sample determined by imposed \pt\ cuts and background subtractions. Relative to the trigger, distinct structures from secondary jets (MPI) may be compared with FF predictions. And, conventional assumptions about azimuth dependence invoked in underlying-event studies may be tested.
TA correlation analysis can be applied to \pa, \da\ and \aa\ collisions to access MB dijet structure in those systems and test claims of collective motion (flows) as an alternative to dijet production.

This work was supported in part by the Office of Science of the U.S.\ DOE under grant DE-FG03-97ER41020. 

\begin{appendix}

\section{Fragmentation functions} \label{fragfunc}

Reconstruction of isolated dijets from \ee\  collisions has provided accurate determination of in-vacuum jet properties for specific jet energies. Nonperturbative fragmentation functions (FFs) have been measured down to small hadron momenta~\cite{opal,tasso} and have been parametrized simply and precisely over a large jet energy range (3 to 200 GeV)~\cite{eeprd}. Jet systematics in elementary \pp\ and composite \aa\ collisions are less certain. 

Fragmentation functions derived from in-vacuum dijets have been described in terms of several kinematic variables, including scalar total momentum $p$~\cite{opal} and longitudinal (along the dijet axis) momentum $p_z$ and transverse momentum $p_t$~\cite{aleph}. The relation to the jet energy has been represented by momentum or energy fraction $x_p = p/p_{jet}$ or $x_E = E/E_{jet}$ and logarithmic measures $\xi = \log(1/x)$.

Measured FFs are derived from isolated (di)jets reconstructed within high-energy elementary collisions (e.g.\ \ee, \pp, \ppbar). Although the higher-momentum portions of high-energy FFs can be described by pQCD much of the distribution is not amenable to theory and must be measured. FFs are conventionally represented by quantity $D_\alpha^\beta(x|Q^2)$ where $\alpha$ and $\beta$ represent hadron and parton types, $x$ is the fragment momentum or energy fraction (of jet energy $E_{jet}$) and $Q$ is the energy scale (dijet total energy $2 E_{jet}$).

In this study we employ rapidity variables $y$ and $y_{max}$ as defined in Sec.~\ref{kine} to describe FFs, with $D_\alpha(y|y_{max}) \equiv 2dn_{ch,j}(y_{max})/dy$, the fragment-multiplicity rapidity density per dijet into $4\pi$ acceptance. The explicit factor 2 reminds that this quantity represents a dijet fragment multiplicity. 
As noted, subscript $\alpha$ in this study has values $u$ (all unidentified-hadron fragments), $t$ (trigger fragments) or $a$ (associated fragments). The leading-parton type (light quark or gluon) is noted in the text as required.

\subsection{FF parametrization of $\bf e^+$-$\bf e^-$ data}

\begin{figure}[h]
  \includegraphics[width=1.65in,height=1.6in]{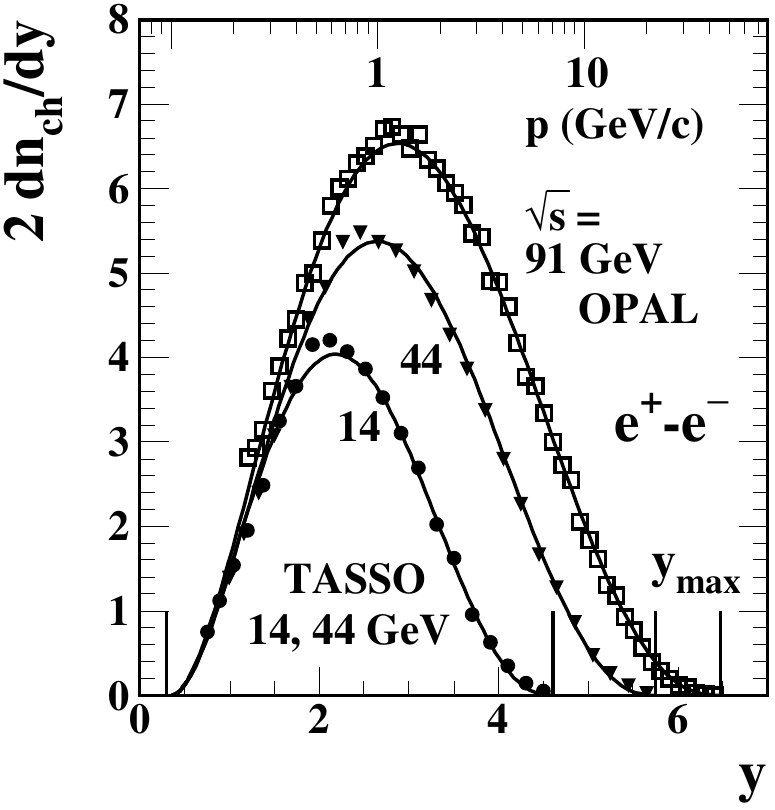}
 \includegraphics[width=1.65in,height=1.6in]{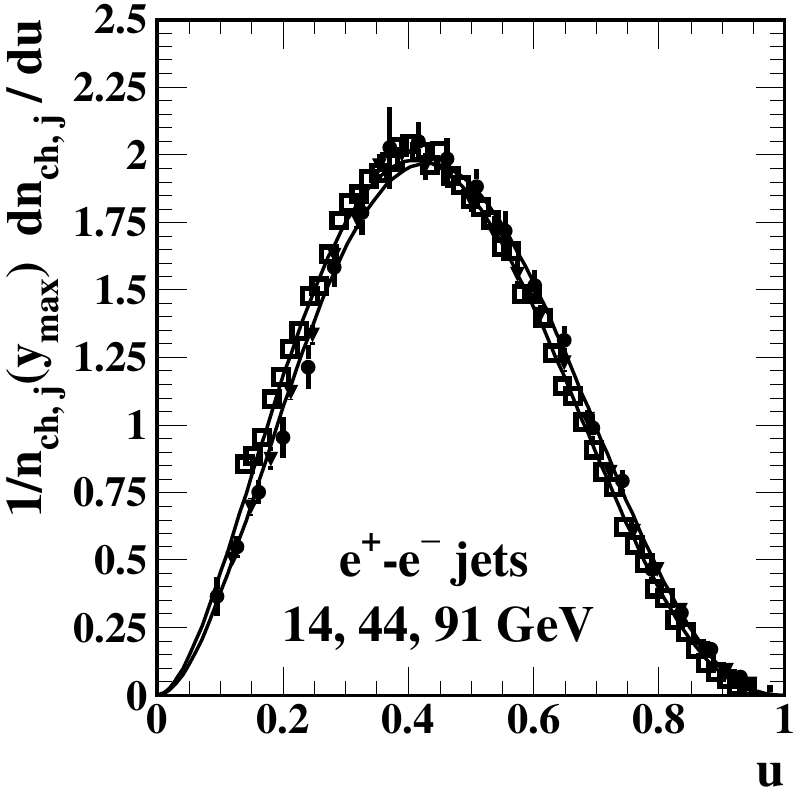}
\caption{\label{ffs}
Left: Fragmentation functions from \ee\ collisions for three energies~\cite{opal,tasso} plotted vs rapidity $y = \ln[(E + p)/m_\pi]$ as in Ref.~\cite{eeprd}. Dijet energies are specified. The left vertical line indicates a common $y_{min} \approx 0.35$ ($p_t \approx 0.05$ GeV/c). Other vertical lines indicate $y_{max}$ values.
Right: The same FFs normalized to unit integral and plotted vs normalized rapidity $u = (y - y_{min}) / (y_{max} - y_{min})$. The data over a large jet energy range and the full fragment momentum range are well described by a single beta distribution.
 } 
 \end{figure}

Figure~\ref{ffs} (left panel) shows measured FFs (points) for three dijet energies derived from \ee\ collisions by TASSO~\cite{tasso} and OPAL~\cite{opal}. The data are of exceptional quality and extend down to low fragment momentum. When plotted on fragment rapidity $y$ the FFs exhibit self-similar evolution with jet energy ($y_{max}$). The solid curves show the FF parametrization used in this study.

Figure~\ref{ffs} (right panel) shows the self-similar data in the left panel plotted on scaled rapidity $u = (y - y_{min}) / (y_{max} - y_{min})$ with $y_{min} \approx 0.35 $ ($p \approx 50$ MeV/c) rescaled to unit integral. The solid curves are beta distributions with parameters $p$ and $q$ nearly constant over the large jet energy interval. The simple two-parameter description is accurate to a few percent within the jet energy interval 3 GeV ($y_{max} \approx 3.75$) to 200 GeV ($y_{max} \approx 8$).~\cite{eeprd}.  FF data for light-quark and gluon jets are separately parametrized, but the parametrizations for gluon and quark jets converge near $E_{jet} = $ 3 GeV.

\subsection{$\bf p$-$\bf p~ vs~  e^+$-$\bf e^-$ FF comparison} \label{ppeerat}

There are substantial differences between in-vacuum dijets from \ee\ collisions  and event-wise reconstructed jets from \pp\ or \ppbar\ collisions. Accommodation of those differences is critical for quantitative comparisons between FFs and \pp\ spectra and correlations.

\begin{figure}[h]
  \includegraphics[width=1.65in]{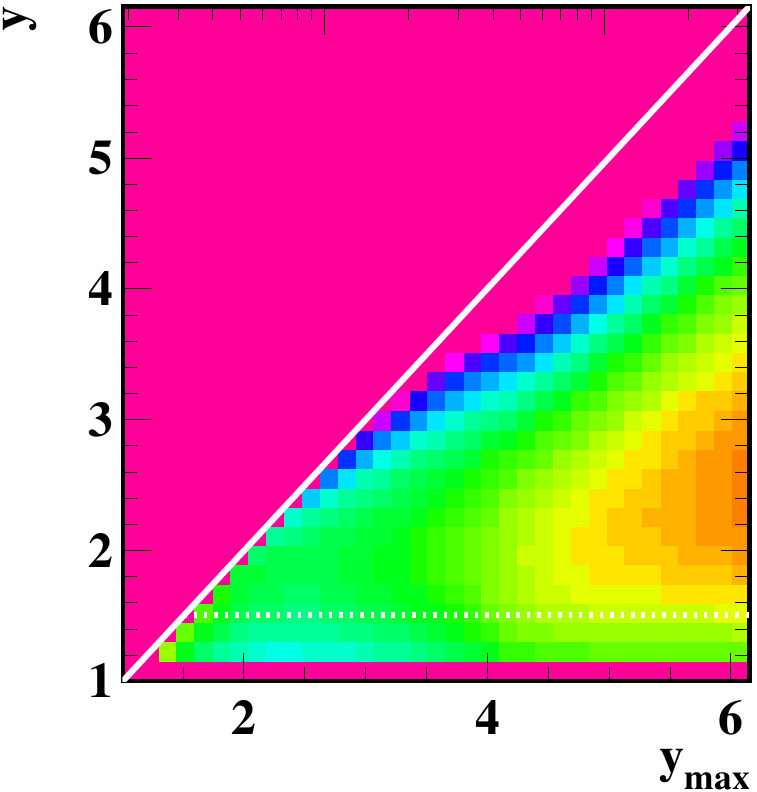} 
 \put(-70.4,112) {\scriptsize \bf 1}
 \put(-28.9,112) {\scriptsize \bf 10}
 \put(-68,104) {\scriptsize $\bf E_{jet} (GeV)$}
  \includegraphics[width=1.65in]{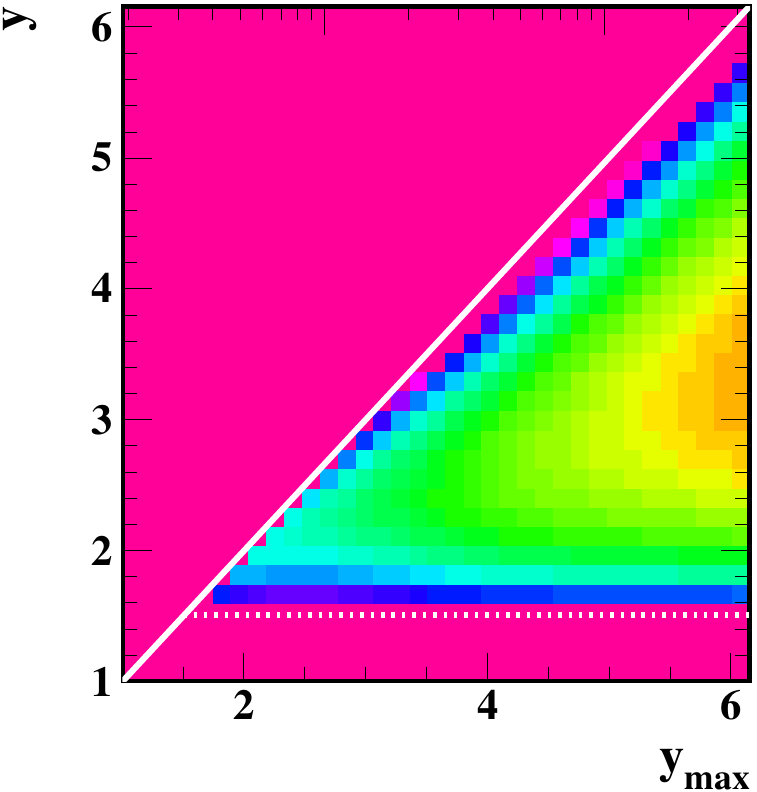}
\caption{\label{qcd}
(Color online) Left: Parametrized dijet fragmentation functions $D_{u}(y|y_{max})$ for gluons from \ee\ collisions fragmenting to unidentified hadrons derived from LEP~\cite{opal} and HERA~\cite{tasso} data with fragment $y = [(E + p)/m_\pi]$ and parton $y_{max} = \ln(2E_{jet}/m_\pi)$~\cite{eeprd}. The z-axis limits (log scale) are 12 and 0.12.
Right: The parametrization on the left has been altered (see text) to describe FFs from 1.8 TeV \ppbar\ collisions. The z-axis limits (log scale) are 6 and 0.06.
 } 
 \end{figure}

Figure~\ref{qcd} shows parametrizations of an ensemble of FFs for gluons from \ee\ collisions (left panel) compared to quarks from \ppbar\ collisions (right panel)~\cite{eeprd}. The two systems represent limiting cases for this study. The gluon FF modes on $y$  are significantly lower and the fragment yields substantially larger because of the larger color charge. The FF mode on momentum $p$ for high-energy jets shifts by more than a factor 2 between the two cases. The measured FFs for \pp\ or \ppbar\ collisions appear to be cut off near $y = 1.5$ ($\approx 0.3$ GeV/c) whereas those from \ee\ collisions follow the parametrization down to a much smaller cutoff at $y_{min} \approx 0.35$. 

\begin{figure}[h]
  \includegraphics[width=1.65in,height=1.63in]{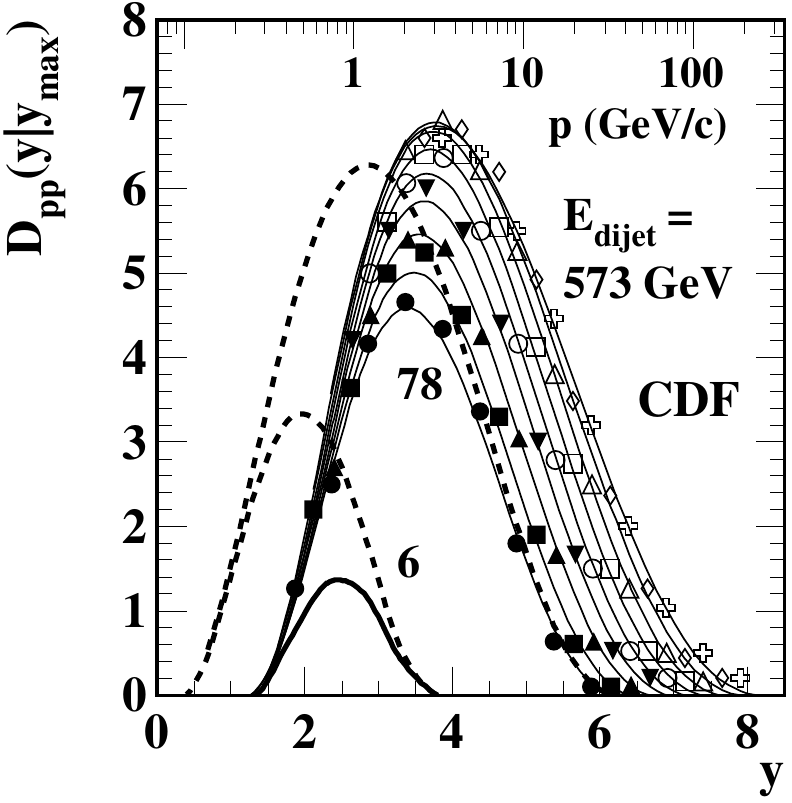}
   \includegraphics[width=1.65in,height=1.63in]{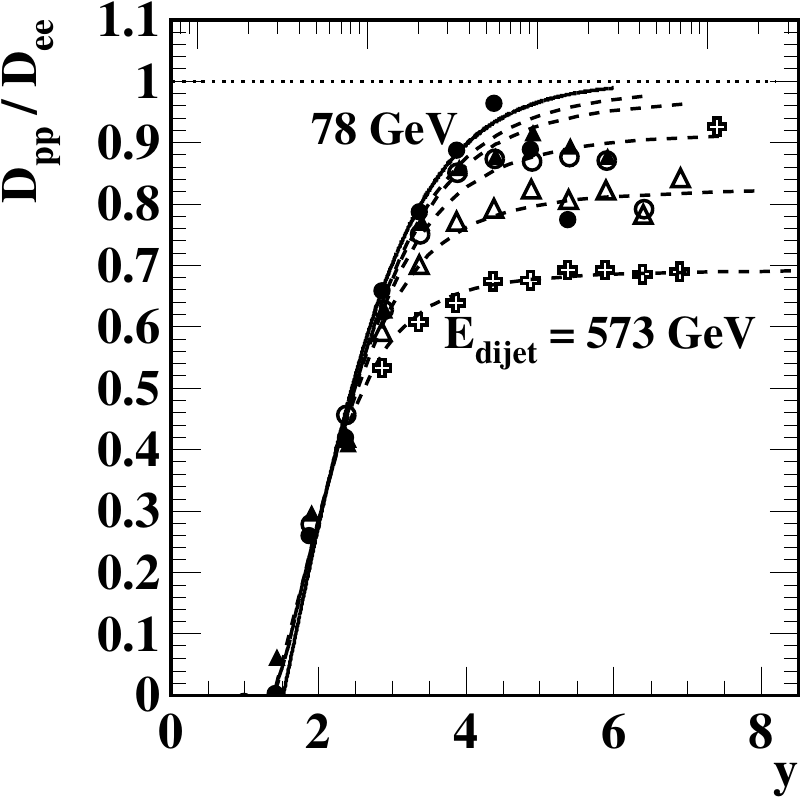}
\caption{\label{ppffs}
Left: FFs from 1.8 TeV \ppbar\ collisions (points) from Ref.~\cite{cdf}. The solid curves represent the \pp\ FF parametrization from Ref.~\cite{fragevo}. Dijet energies are specified. The dashed curves are from the \ee\ parametrization of Ref.~\cite{eeprd} for comparison.
Right: The ratio of \pp\ to \ee\ parametrizations (curves) and \ppbar\ data to \ee\ parametrization (points) from the left panel. The solid curve $\tanh[(y-1.5)/1.7]$ describing dijets below 70 GeV is used in the present study.
 } 
 \end{figure}

Figure~\ref{ppffs} (left panel) shows FFs derived from \ppbar\ collisions by the CDF collaboration (points) using an event-wise jet-finder method~\cite{cdf}. Comparison with the  \ee\ parametrization (dashed curves) indicates that a substantial fraction of dijets may be missing from \ppbar\ FFs at lower fragment momentum.  We conjecture that some low-momentum part of the \pp\ dijets may be excluded from  the mid-rapidity angular acceptance due to longitudinal transport, as discussed in Ref.~\cite{fragevo}  Sec. XIII-C.

Figure~\ref{ppffs} (right panel) shows ratios of \ppbar\ FFs in the left panel to equivalent \ee\ parametrizations illustrating the differences. The solid curve is $\tanh[(y-1.5)/1.7]$ which describes measured  \ppbar\ FFs relative to \ee\ FFs for dijet energies below 70 GeV. That combination is used to represent \pp\ FFs for this study. 

Observed jet fragment yields from 200 GeV NSD \pp\ collisions~\cite{ppprd} can be compared with expectations from \ee\ FFs~\cite{eeprd}. The most probable jets in \pp\ collisions have $E_{jet} \approx 3$ GeV (minijets). For NSD collisions we observe $dn_h /d\eta = 0.005 \times 2.5^2 \approx 0.03$ from Eq.~(\ref{freq}) and $f_{NSD} \approx 0.025$ leading to mean dijet fragment multiplicity $2 \bar n_{ch,j} \approx 2$ from Eq.~(\ref{f}). That value can be compared with  $2n_{ch,j}(y_{max}) \approx 5$ for \ee\ dijets in Fig.~\ref{void} (b) with $y_{max} = 3.75$ ($E_{jet} = 3$ GeV). Thus, measured 3 GeV \pp\ dijets include approximately 40\% of \ee\ dijet fragments. As noted in Sec.~\ref{ffcompare}  such results call into question conventional assumptions about FF universality.

\section{jet/Parton spectra} \label{parspec}

The MB scattered-parton spectrum for a given collision energy $\sqrt{s_{NN}}$ is denoted by  $d^2\sigma_j /dy_{max}d\eta \equiv S_p(y_{max}|y_{beam})$ with $y_{beam} \rightarrow y_{b}$ defined below. Systematic analysis of available jet production data from the ISR and Sp\=pS below 1 TeV beam energy has lead to a simple parametrization based on rapidities~\cite{jetspec2}. The beam rapidity relative to pion mass is $y_b \equiv \ln(\sqrt{s} / \text{0.14 GeV})$, $y_{b0} \equiv \ln(Q_0 / 0.14 )$ with $Q_0 \approx 10$ GeV determined by jet-related correlation trends and $y_{m0} =  \ln(2E_{cut} / 0.14 )$. We then define $\Delta y_b = y_b - y_{b0}$, $\Delta y_{max} = y_{b} - y_{m0}$ and  normalized rapidity $u =  (y_{max} - y_{m0}) /\Delta y_{max}$. 
The resulting parametrized parton spectrum conditional on beam rapidity is
 \bea \label{jetspecc}
\frac{d^2\sigma_j}{dy_{max} d\eta} 
&=& 
p_t \frac{d^2 \sigma_j}{dp_t d\eta} 
\\ \nonumber
&=& 0.026 \Delta y_b^2  \frac{1}{\sqrt{2\pi \sigma^2_u}} e^{-u^2 / 2 \sigma^2_u} 
 \eea
 with $\sigma_u \approx 1/7$ and $E_{cut} \approx 3$ GeV determined by data.

\begin{figure}[h]
   \includegraphics[width=1.65in,height=1.63in]{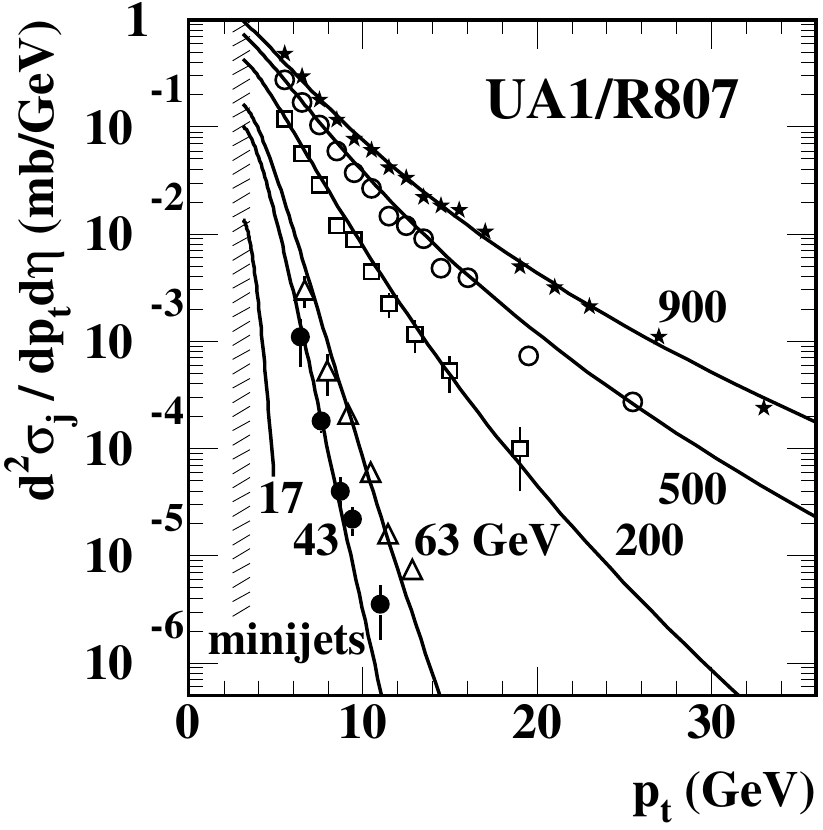}
  \includegraphics[width=1.65in,height=1.63in]{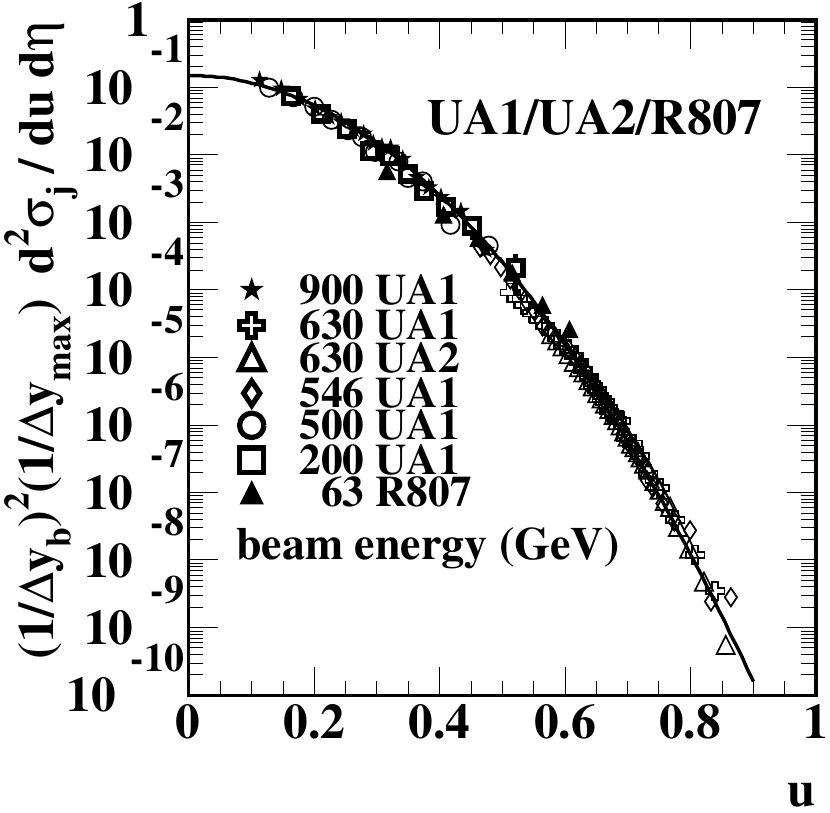}
\caption{\label{specapp}
Left: Measured jet spectra from \ppbar\ collisions (points) for several collision energies below 1 TeV. The solid curves through data were generated by Eq.~(\ref{jetspecc}).
Right: The data from the left panel normalized to unit integral and plotted on normalized rapidity $u = (y_{max} - y_{m0}) / (y_{beam} - y_{m0})$.
 } 
 \end{figure}

Figure~\ref{specapp} (left panel) shows jet production data for several beam energies (points) compared to Eq.~(\ref{jetspecc}) (solid curves). The cross-section data for a broad range of collision energies are described accurately down to $E_{jet} \approx 3$ GeV where jet production via charged hadrons apparently terminates due to kinematic constraints.

Figure~\ref{specapp} (right panel) shows the data from the left panel normalized according to the $y$-axis label and plotted vs normalized rapidity $u$. The data fall on a common Gaussian locus as in Eq.~(\ref{jetspecc}), a manifestation of spectrum self-similarity when plotted on a rapidity variable. That result can be compared with Fig.~\ref{ffs} (right panel) where a similar self-similarity is exhibited by fragmentation functions.

The jet frequency $f_{NSD} \equiv (1/\sigma_{NSD}) d\sigma_j /d\eta$ required to relate jet manifestations in  \pp\ collisions to FF data is determined as follows. The jet spectrum integral is
\bea
 \frac{d\sigma_j}{d\eta} &=& 0.026  \Delta y_b^2 \Delta y_{max},
 \eea
and the NSD cross section at 200 GeV is $\sigma_{NSD} \approx 34$ mb~\cite{nsdsig,jetspec}. For 200 GeV \pp\ collisions the ratio corresponds to $f_{NSD} \approx 0.029$. Given the stated systematic uncertainties for the Sp\=pS jet cross sections we adopt 200 GeV $f_{NSD} \approx 0.025 \pm 0.005$ for this study.

We have used two parton spectrum models to relate measured dijet FFs to the spectrum hard component from \pp\ collisions. In Ref.~\cite{fragevo} the power-law spectrum model varied as $1/ E_{jet}^{5.75}$ with a lower bound at  $E_{jet,cut} \approx 3$ GeV ($y_{max,cut} \approx 3.75$) integrating to $d\sigma_j/d\eta \approx$ 1.2 mb. In the present study the Gaussian model of Eq.~(\ref{jetspecc}) developed in Ref.~\cite{jetspec} is cut off at $E_{jet,cut} \approx 2.5$ GeV ($y_{max,cut} \approx 3.55$) and integrates to $d\sigma_j/d\eta \approx$ 0.85 mb. While the Gaussian model describes jet spectra over a large range of jet and \pp--collision energies the power-law shape near  the spectrum lower bound may be more appropriate based on results in Sec.~\ref{datacomp}.
   
\end{appendix}



\begin{thebibliography}{99}

\bibitem{hepjets}  S.~M.~Berman, J.~D.~Bjorken and J.~B.~Kogut,
  Phys.\ Rev.\ D {\bf 4}, 3388 (1971).

\bibitem{rick1}  R.~D.~Field and R.~P.~Feynman,
  Phys.\ Rev.\ D {\bf 15}, 2590 (1977).
  
  \bibitem{rick2}  R.~P.~Feynman, R.~D.~Field and G.~C.~Fox,
    Nucl.\ Phys.\ B {\bf 128}, 1 (1977).
 
\bibitem{atlas}  G.~Aad {\it et al.}  (ATLAS Collaboration),
  Nature Commun.\  {\bf 2}, 463 (2011).

\bibitem{bec}  F.~Becattini, M.~Bleicher, T.~Kollegger, M.~Mitrovski, T.~Schuster and R.~Stock,
  Phys.\ Rev.\ C {\bf 85}, 044921 (2012).
 
\bibitem{bron} W.~Broniowski and P.~Bozek,
  arXiv:1311.6412.

\bibitem{perfliq1}  M.~Gyulassy and L.~McLerran,
  Nucl.\ Phys.\ A {\bf 750}, 30 (2005).

\bibitem{perfliq2}  B.~Muller,
  Acta Phys.\ Polon.\ B {\bf 38}, 3705 (2007).

\bibitem{2gev}  J.~Adams {\it et al.}  (STAR Collaboration),
  Phys.\ Rev.\ C {\bf 72}, 014904 (2005).

\bibitem{eeprd}   T.~A.~Trainor and D.~T.~Kettler,
  Phys.\ Rev.\ D {\bf 74}, 034012 (2006).

\bibitem{hardspec}  T.~A.~Trainor,
  Int.\ J.\ Mod.\ Phys.\  E {\bf 17}, 1499 (2008).

\bibitem{jetspec}   T.~A.~Trainor and D.~T.~Kettler,
  Phys.\ Rev.\ C {\bf 83}, 034903 (2011).

\bibitem{fragevo}    T.~A.~Trainor,
  Phys.\ Rev.\  C {\bf 80}, 044901 (2009).

\bibitem{anomalous}  G.\ Agakishiev, {\it et al.} (STAR Collaboration),
  Phys.\ Rev.\ C {\bf 86}, 064902 (2012).

\bibitem{axialci}   J.~Adams {\it et al.}  (STAR Collaboration),
  Phys.\ Rev.\  C {\bf 73}, 064907 (2006).

\bibitem{ppprd} J.~Adams {\it et al.}  (STAR Collaboration),
  Phys.\ Rev.\  D {\bf 74}, 032006 (2006).

\bibitem{porter2} R.~J.~Porter and T.~A.~Trainor  (STAR Collaboration),
  J.\ Phys.\ Conf.\ Ser.\  {\bf 27}, 98 (2005).

\bibitem{porter3}  R.~J.~Porter and T.~A.~Trainor  (STAR Collaboration),
  PoS {\bf CFRNC2006}, 004 (2006).

\bibitem{jetspec2}  T.~A.~Trainor,
  Phys.\ Rev.\ D  {\bf 89}, 094011 (2014).

\bibitem{pptheory}  T.~A.~Trainor,
Phys.\ Rev.\ D {\bf 87}, 054005 (2013).

\bibitem{pptrig} T.~A.~Trainor and D.~J.~Prindle,
Phys.~Rev.~D {\bf 88}, 094018 (2013).

\bibitem{axialcd} J.~Adams {\it et al.}  (STAR Collaboration),
  Phys.\ Lett.\  B {\bf 634}, 347 (2006).

\bibitem{inverse} T.~A.~Trainor, R.~J.~Porter and D.~J.~Prindle,
  J.\ Phys.\ G {\bf 31}, 809 (2005).

\bibitem{davidhq}  D.~T.~Kettler  (STAR collaboration),
  Eur.\ Phys.\ J.\  C {\bf 62}, 175 (2009).

\bibitem{davidhq2}  D.~Kettler ( STAR Collaboration),
  J.\ Phys.\ Conf.\ Ser.\  {\bf 270}, 012058 (2011).

\bibitem{porter1}  R.~J.~Porter and T.~A.~Trainor  (STAR Collaboration),
  Acta Phys.\ Polon.\  B {\bf 36}, 353 (2005).

\bibitem{mtxmt}  J.~Adams {\it et al.}  (STAR Collaboration),
  J.\ Phys.\ G {\bf 34}, 799 (2007).

\bibitem{ytxyt} E.~W.~Oldag (STAR Collaboration),
  J.\ Phys.\ Conf.\ Ser.\  {\bf 446}, 012023 (2013).

\bibitem{ismd13} T.~A.~Trainor and D.~J.~Prindle,
 arXiv:1310.0408.

\bibitem{duncan} D.~J.~Prindle (STAR Collaboration),
  arXiv:1406.5225.

\bibitem{ua1} C.~Albajar {\it et al.}  (UA1 Collaboration),
  Nucl.\ Phys.\  B {\bf 309}, 405 (1988).

\bibitem{tasso} W.\ Braunschweig {\em et al.} (TASSO Collaboration), Z.\ Phys.\
C {\bf 47}, 187 (1990).

\bibitem{cleo} M.\ S.\ Alam {\em et al.} (CLEO Collaboration), Phys.\ Rev.\ D
{\bf 56}, 17 (1997).

\bibitem{dglap} V.\ N.\ Gribov and L.\ N.\ Lipatov, Sov.\ J.\ Nucl. Phys. {\bf 15}, 438 (1972); 
L.\ N.\ Lipatov, Sov.\ J.\ Nucl.\ Phys.\ {\bf 20}, 95 (1975); 
Yu.\ L.\ Dokshitzer, Sov.\ Phys.\ JETP {\bf 46}, 641 (1977).
G.\ Altarelli and G.\ Parisi, Nucl.\ Phys.\ B {\bf 126} , 298 (1977).

\bibitem{mlla} Ya.\ I.\ Azimov, Yu.\ L.\ Dokshitzer, V.\ A.\ Khoze,
S.\ I.\ Troyan, Z.\ Phys.\ C {\bf 27}, 65 (1985), Z.\ Phys.\ C {\bf 31}, 213 (1986).

\bibitem{starta} J.\ Adams {\em et al.} (STAR Collaboration), Phys.\ Rev.\ Lett. {\bf 95}, 152301 (2005).

\bibitem{tzyam} T.~A.~Trainor,
  Phys.\ Rev.\  C {\bf 81}, 014905 (2010).

\bibitem{factor}  J.~C.~Collins, D.~E.~Soper and G.~F.~Sterman,
  Adv.\ Ser.\ Direct.\ High Energy Phys.\  {\bf 5}, 1 (1988).

\bibitem{kniehl} B.~A.~Kniehl, G.~Kramer and B.~Potter,
  Nucl.\ Phys.\ B {\bf 597}, 337 (2001).

\bibitem{mpi} P.~Bartalini and L.~Fano,
  arXiv:1003.4220.

\bibitem{rick} R.~Field,
  Acta Phys.\ Polon.\ B {\bf 42}, 2631 (2011).

\bibitem{cdfue}   T.~Affolder {\it et al.}  (CDF Collaboration),
  Phys.\ Rev.\ D {\bf 65}, 092002 (2002).

\bibitem{cmsue}  V.~Khachatryan {\it et al.}  (CMS Collaboration),
  Eur.\ Phys.\ J.\ C {\bf 70}, 555 (2010).

\bibitem{frankfurt}  L.~Frankfurt, M.~Strikman and C.~Weiss,
  Phys.\ Rev.\  D {\bf 83}, 054012 (2011).

\bibitem{opal} M.\ Z.\ Akrawy {\em et al.} (OPAL Collaboration) Phys.\ Lett.\ B, {\bf 247}, 617 (1990).

\bibitem{aleph} D.\ Buskulic {\em et al.} (ALEPH Collaboration), Z.\ Phys.\ C {\bf 66}, 355 (1995).

\bibitem{nsdsig} G.\ J.\ Alner {\em et al.} (UA5 Collaboration), Z.\ Phys.\ C {\bf 32}, 153 (1986).

\bibitem{cdf} D.~Acosta {\it et al.}  (CDF Collaboration),
Phys.\ Rev.\ D {\bf 68}, 012003 (2003).

 
\end{thebibliography}
\end{document}